%% file: mumonemt_arxiv.tex
\PassOptionsToPackage{pdftex}{graphicx} 
\documentclass[preprint,11pt]{elsarticle}

\usepackage{lineno,hyperref}  

\usepackage[pdftex]{graphicx}  

\modulolinenumbers[100]  

\usepackage[top=30truemm,bottom=30truemm,left=30truemm,right=30truemm]{geometry}

\journal{arXiv}

\begin{document}

\begin{frontmatter}


\title{{\bf A new electron-multiplier-tube-based beam monitor \\ for muon monitoring
            at the T2K experiment}}
  

\author[kyoto]{Y. Ashida\corref{cor1}}
\ead{assy@scphys.kyoto-u.ac.jp}

\author[kek]{M. Friend}
\author[kyoto]{A. K. Ichikawa}
\author[kek]{T. Ishida}
\author[kyoto]{H. Kubo}
\author[kyoto]{\\ K. G. Nakamura}
\author[kek]{K. Sakashita}
\author[kyoto]{W. Uno}

\cortext[cor1]{Corresponding author}
\address[kyoto]{Department of Physics, Kyoto University, Kyoto 606-8502, Japan}
\address[kek]{High Energy Accelerator Research Organization (KEK), Tsukuba, Ibaraki 305-0801, Japan}

\begin{abstract}
Muon beam monitoring is indispensable for indirectly monitoring
accelerator-produced neutrino beams in real time.
Though Si photodiodes and ionization chambers have been successfully
used as muon monitors at the T2K experiment,
sensors that are more radiation tolerant are desired for
future operation.
We have investigated the electron-multiplier tube (EMT) as a new
sensor for muon monitoring.
Secondary electrons produced by the passage of muons at dynodes are multiplied
in the tube and produce signal.
Two prototype detectors were installed at the T2K muon monitor location,
and various
EMT properties were studied based on in situ data taken with the T2K muon beam.
The signal size is as expected based on calculation, and the
EMTs show a sufficiently fast time response for bunch-by-bunch beam monitoring.
The spill-by-spill intensity resolution is 0.4\%, better than the required value (1\%).
Signal linearity within $\pm$1\% is achieved at proton beam powers up to
460~kW (with +250~kA focusing horn operation).
A gradual signal decrease was observed during the initial exposure,
due to the stabilization
of dynode materials, before the response became stable within $\pm$1\%.
This work demonstrates that EMTs are a good candidate for future muon monitoring
at T2K,
and may also have other more general applications.
\end{abstract}





\end{frontmatter}


\input{introduction}

%
\input{t2kmuonmonitor}

%
\newpage
\input{emt}

%
\input{setup}
\clearpage
\input{performance}

%
\input{conclusion}

%
\clearpage
\section*{Acknowledgements}
The authors are grateful to the J-PARC accelerator group for supplying
a stable beam.
This work was partially supported by MEXT KAKENHI Grant Numbers 25105002, 16H06288,
15J01714, and 17J06141.









\end{document}

%% file: introduction.tex
\section{Introduction}
\label{sec:introduction}

With the steady increase of beam power at high-intensity beam facilities, as well as the
push for higher precision measurements, the importance of beam monitoring is growing, both to protect equipment, 
and to precisely measure beam properties.
A beam intensity can span many orders of magnitude, from one 
to $10^{13}$ particles per hundred nanoseconds.
Counting detectors are applicable when the intensity is up to order 
$10^7$/s, but do not work at higher intensities.  
Rather, current-mode read-out is used at these higher intensities.
When the number of charged particles exceeds around $10^{15}$/s,
a secondary emission monitor, the signal of which is an electrical current
produced by the radiation-induced emission of secondary electrons from a conductor foil,
has sufficiently large signal for 50-$\Omega$-load read-out.
For intermediate intensities still too high for particle counting,
detectors utilizing ionization or excitation of media, such as semiconductors,
scintillators, and gases, are used.

For accelerator neutrino beam experiments, monitoring the muons produced 
along with neutrinos has played a key role in beam control 
(see Refs.~\cite{pattison,kopp} and references therein).
Ionization detectors and Si photodiodes have been used at the T2K experiment, 
where the typical number of muons is around $10^{12}$/s. 
However, during the recent high-intensity operation, as described 
in Section~\ref{sec:t2kandmumon}, the radiation damage of Si sensors and signal saturation of 
ionization detectors started to become issues.

In this paper, we report the investigation of the 
electron-multiplier tube, or EMT, 
which is equivalent to a photomultiplier tube (PMT) 
without a photocathode, as a new muon monitor for the T2K experiment.
This is the first ever study of beam monitoring using EMTs at 
the intensity level of $10^{6}$ particles per hundred nanoseconds.
After a description of the T2K muon monitor and its current status
in Section~\ref{sec:t2kandmumon},
we describe the principle of muon monitoring using electron-multiplier tubes
in Section~\ref{sec:emtconcept} and our EMT prototype sensors in Section~\ref{sec:developemt}.
Then, we report the initial performance test results obtained with the muon beam at the 
T2K muon monitor location in Section~\ref{sec:performance}, followed by 
the conclusion in Section~\ref{sec:conclusion}.


%% file: t2kmuonmonitor.tex
\section{The muon monitor for the T2K experiment} 
\label{sec:t2kandmumon}

\subsection{The T2K experiment and J-PARC neutrino beamline} 

The T2K experiment measures neutrino oscillation parameters precisely using 
accelerator-produced (anti-)neutrinos~\cite{t2k}. 
A muon neutrino or anti-neutrino beam is produced at the Japan Proton Accelerator 
Research Complex (J-PARC) and then measured at the Super-Kamiokande (SK) detector 
\cite{superk}, which is located 295~km from J-PARC. 
By measuring the disappearance of muon (anti-)neutrinos and 
the appearance of electron (anti-)neutrinos, 
T2K determines the neutrino oscillation parameters and probes CP violation 
in the lepton sector \cite{oa2016short,oa2016long}.

A 30~GeV proton beam from the main ring (MR) synchrotron of 
J-PARC impinges on a graphite target to produce predominantly pions and some 
fraction of kaons.
The generated particles are parallel-focused by three magnetic horns \cite{ichikawa,sekiguchi}, 
currently operated at positive or negative 250~kA, in order to
focus positively or negatively charged charged outgoing particles, respectively. 
The hadrons then decay into neutrinos and muons 
in a 96-m-long decay volume, producing a neutrino beam or an anti-neutrino beam 
depending on the focusing horn polarity. 
Both the direction and intensity of the neutrino beam must be known precisely
in order to precisely predict the neutrino flux at SK.
The neutrino beam profile spreads as wide as a few km at SK, but a 1~mrad change 
in the direction of the beam axis causes a significant change in the neutrino event rate and 
a distortion of the energy spectrum, which leads to large systematic errors 
on the neutrino flux. 
Near detectors sit 280~m downstream of the target to monitor the beam direction and 
intensity, as well as to study neutrino-nucleus interactions.
One of the near detectors, INGRID, is composed of 14 modules arranged in a cross
spanning 10~m $\times$ 10~m around the beam center axis, and monitors 
the neutrino beam direction 
and intensity by measuring the event rate at each module \cite{ingrid}.
The INGRID detector gives direct information about the neutrino beam properties, however, it requires more than 
one day to accumulate sufficient data for a neutrino beam profile measurement
because of the 
small interaction cross section of neutrinos.
On the other hand, as explained below, the T2K muon monitor can reconstruct the beam 
profile for each beam bunch. 

\begin{figure}[htbp]
  \begin{center}
   \includegraphics[clip,width=11.0cm]{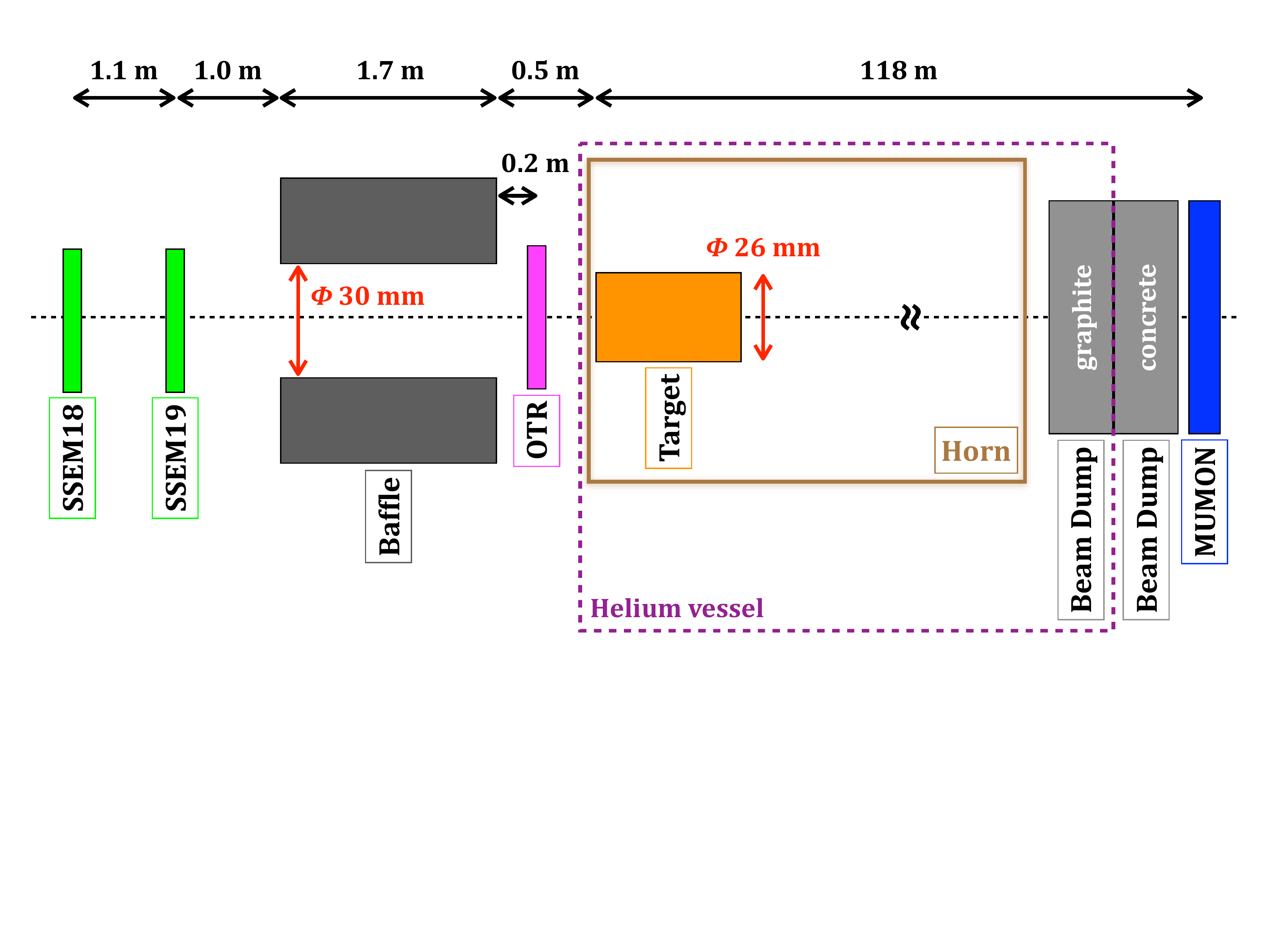}
  \end{center}
  \vspace{-85truept}
  \caption{Schematic drawing of the neutrino beamline equipment from the most downstream part 
           of the primary proton beamline to the muon monitor. 
	   Though an optical transition radiation monitor (OTR) \cite{bhadra} is placed just before the target, 
	   its information is not used in this analysis.} 
  \label{fig:secondaryline} 
 \end{figure}

The neutrino beam direction is affected by various sources, i.e. the position of 
the proton beam at the target, misalignment of the target or horns, etc.
In each 2.48~s beam cycle, a proton beam spill composed of 
eight bunches, which are separated by 581~ns and typically have an 80~ns {bunch width (3$\sigma$)}, 
is extracted to the neutrino beamline.
Bunch-by-bunch beam monitoring is indispensable for primary proton beam tuning
and diagnostics of the proton beam and beamline equipment.
The proton beam intensity is monitored by five current transformers (CT),  
and the position and profile of the proton beam are measured 
by nineteen segmented secondary emission monitors (SSEM).
In this paper, we use the values measured by the most downstream monitors, CT5, SSEM18, and SSEM19. 
The fluctuation of the CT relative intensity
measurement is as small as 0.5\%.
%
In addition to the proton beam monitors, T2K has a muon
monitor, which monitors 
the profile and intensity of the muons produced along with the neutrinos.
The muon monitor sits 118~m away from the target, downstream of a beam dump, at the end of 
the T2K decay volume.
The target, horns, and first part of the beam 
dump, which is made of graphite blocks, are in a vessel 
filled with helium gas. 
The second part of the beam dump is downstream of the He vessel and is composed of concrete blocks. 
This is followed by the muon monitor, which also sits outside the helium vessel. 
The beam dump thickness is designed such that only muons with momentum higher than 5~GeV
penetrate it, while hadrons and low-momentum muons are absorbed.
These monitors give bunch-by-bunch information and have been used for proton 
beam tuning and alignment checks of the target, horns, and baffle located upstream of the target. 
Figure~\ref{fig:secondaryline} shows a schematic drawing of the beamline equipment 
from the most downstream part of the primary beamline to the muon monitor.  

\subsection{The T2K muon monitor (MUMON)} 

\begin{figure}[htbp]
  \begin{center}
   \includegraphics[clip,width=9.0cm]{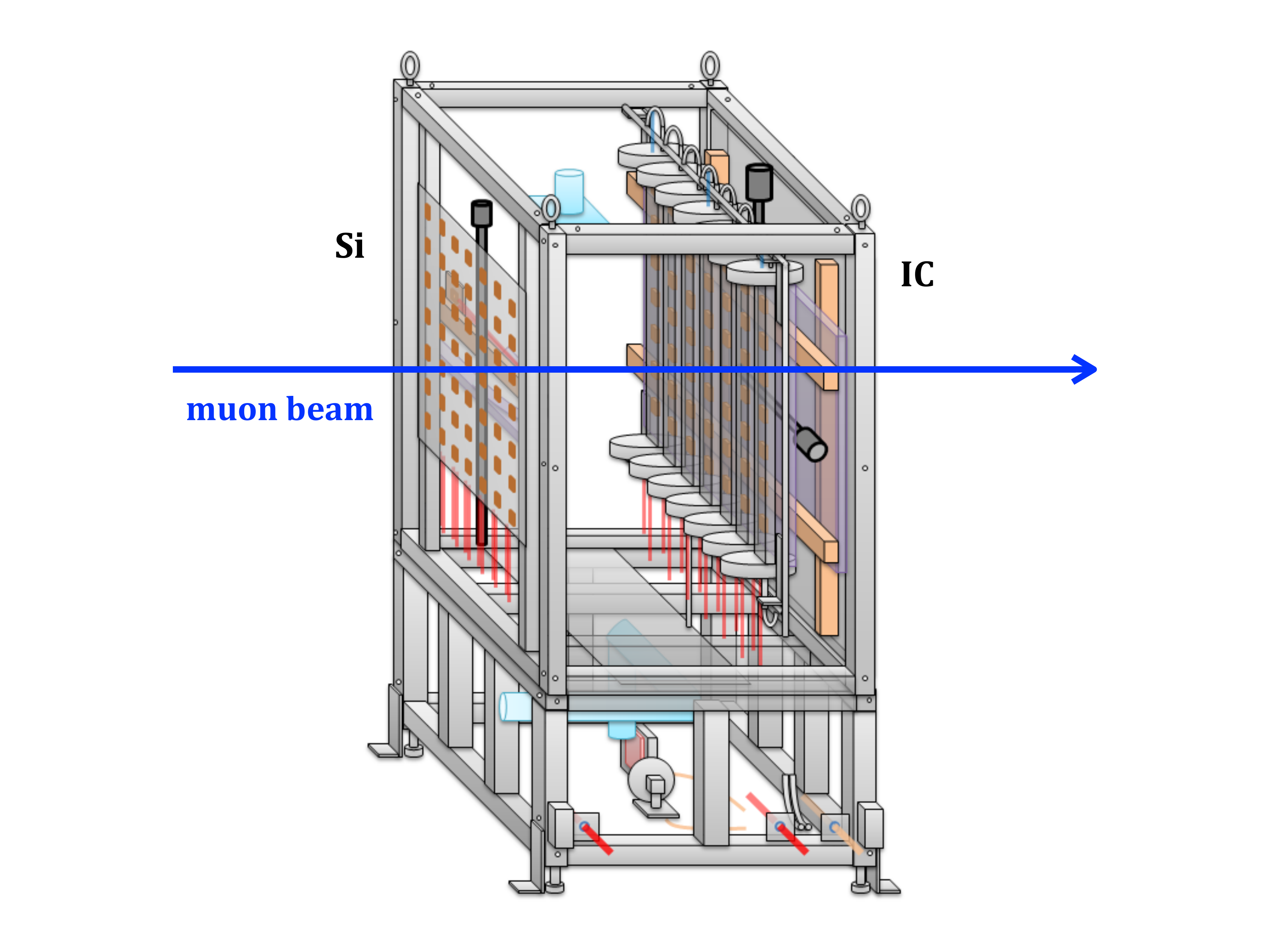}
  \end{center}
  \vspace{-15truept}
  \caption{Schematic view of the T2K muon monitor \cite{matsuoka}.
           Each array of Si and IC sensors covers a $150 \times 150 \ {\rm cm^2}$
	   region with $7 \times 7$ channels.}
  \label{fig:mumonschematic} 
 \end{figure}

The current muon monitor (MUMON) system is composed of two arrays of sensors,  
one of silicon PIN photodiodes (Si; ${\rm HAMAMATSU^{\textregistered}}$ S3590-08) 
and another of ionization chambers (IC; Ar$+{\rm N_2}$ or He$+{\rm N_2}$) \cite{matsuoka},
where the Si array sits upstream of the IC array.
Each array covers a $150 \times 150 \ {\rm cm^2}$ region that
has $7 \times 7$ channels 
installed with a 25~cm pitch. 
A schematic view of the MUMON is shown in Figure~\ref{fig:mumonschematic}.
Seventy-meter-long polyimide cables for signal and high voltage are drawn from the underground muon monitor 
location to an electronics hut on the ground.
The sensor waveforms are stored in a 65~MHz flash analog-to-digital converter (ADC) in the COmmon Pipelined Platform
for Electronics Readout (COPPER) system \cite{higuchi}.
At high-intensity beam operation, additional attenuation modules are used to attenuate the signal size 
to be within the flash ADC dynamic range. 
The precision (systematic uncertainty) of the MUMON beam direction measurement is 0.28 mrad 
and the resolution is better than 0.03 mrad \cite{suzuki}.
During 
beam operation, the MUMON continuously monitors the muon beam profile and intensity, 
and if any major deviation from the expected muon beam profile is measured 
an alarm is issued and the proton beam is tuned.
The system has been working well since the beginning of the T2K experiment.
When the horn current is set to $+250$~kA (neutrino beam mode), the muon flux at the MUMON is 
$1.09 \times 10^5$~${\rm /cm^2/10^{12}}$~POT (protons on target).
At 485~kW proton beam power (the present maximum power), 
the muon flux is $3.5 \times 10^{6}$~${\rm /cm^2}$ per 80~ns beam bunch.
For $-250$~kA (anti-neutrino beam mode), the muon flux becomes about two-thirds of that at $+250$~kA,
due to the difference in the focused pion flux. 

\subsection{Concerns for the current muon monitors} 

There is a plan to increase the J-PARC beam power up to 1.3~MW from 485~kW 
and the horn current to 320~kA from 250~kA \cite{t2k2}.
At the 1.3~MW operation phase, a 1.16~s beam cycle is planned, 
where the number of beam bunches becomes nearly doubled.
In this situation, the averaged muon flux is expected to become 
about three times larger than the current flux, and the flux for the beam bunch is 
$4.9 \times 10^{6}$~${\rm /cm^2}$ per 80~ns beam bunch.
Under such a large muon flux irradiation, 
we expect certain issues with the current muon monitors, 
some of which have already been observed even at the present muon flux.

\begin{figure}[htbp]
  \begin{center}
   \includegraphics[clip,width=10.5cm]{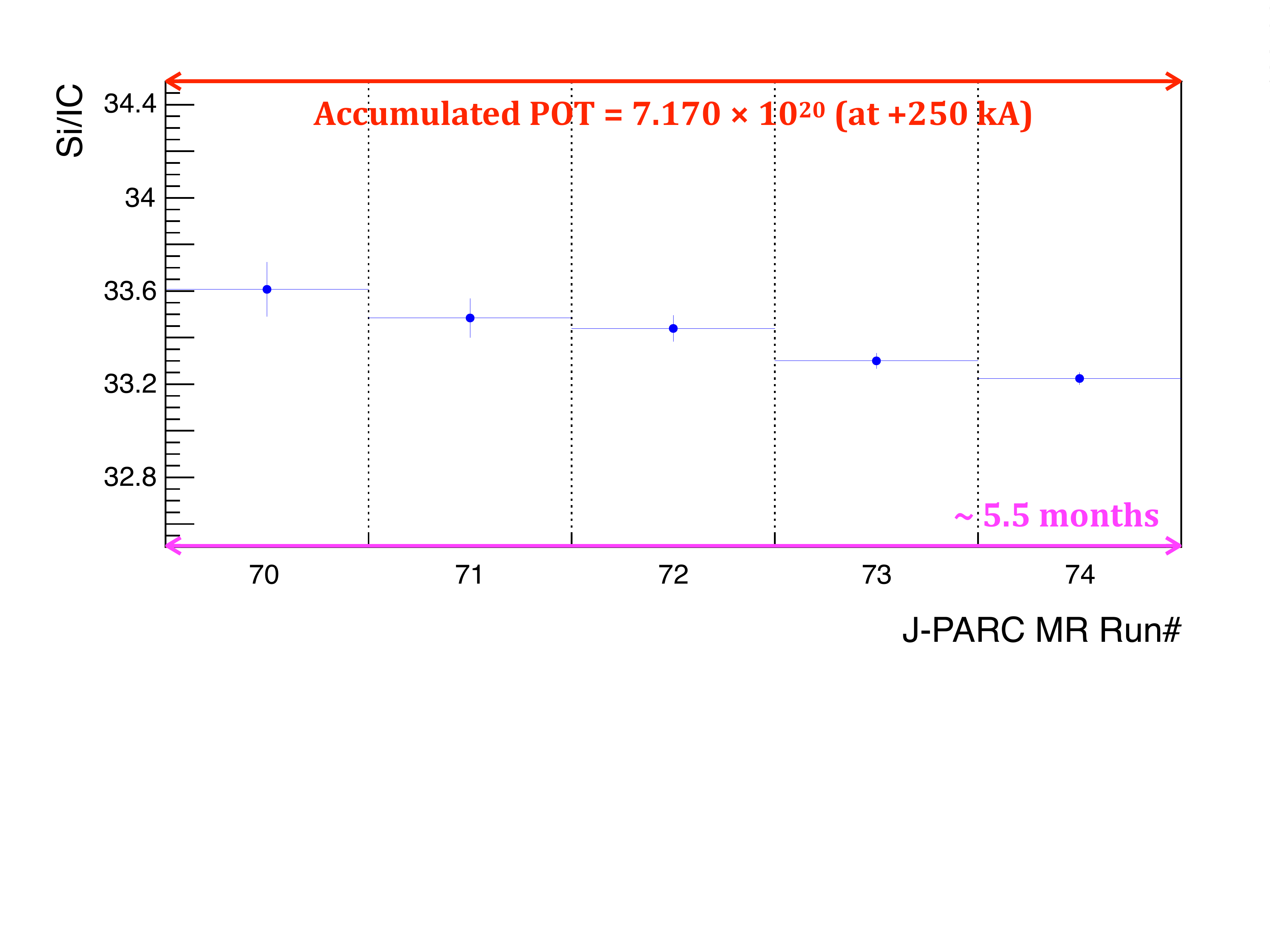}
  \end{center}
  \vspace{-75truept}
   \caption{Ratio of the average yield of Si over IC (Ar) over 
            the period of several MR runs 
            with $+$250~kA horn current.
	    A 1.2\% decrease is seen after an accumulated POT of $7 \times 10^{20}$.} 
  \label{fig:sioveric}
 \end{figure}

The Si sensors will suffer radiation damage, and monthly regular replacements 
may be necessary.
Figure~\ref{fig:sioveric} shows the evolution of the ratio of the total yield 
of the Si and IC (Ar gas) sensors over several J-PARC operation periods. 
By taking the ratio of the two, fluctuations from other sources, such as the proton beam condition 
or horn current, are canceled. 
The ratio 
decreased by 1.2\% after irradiation of about $7 \times 10^{20}$ POT 
at $+$250~kA horn current ($7.82\times10^{13}/$cm$^2$ integrated protons).
The fact that the IC yield was stable within $\pm$0.2\% during that period indicates 
that the Si sensors themselves show signal degradation due to radiation damage.  

\begin{figure}[htbp]
 \begin{minipage}{0.5\hsize}
  \begin{center}
   \includegraphics[clip,width=7.5cm]{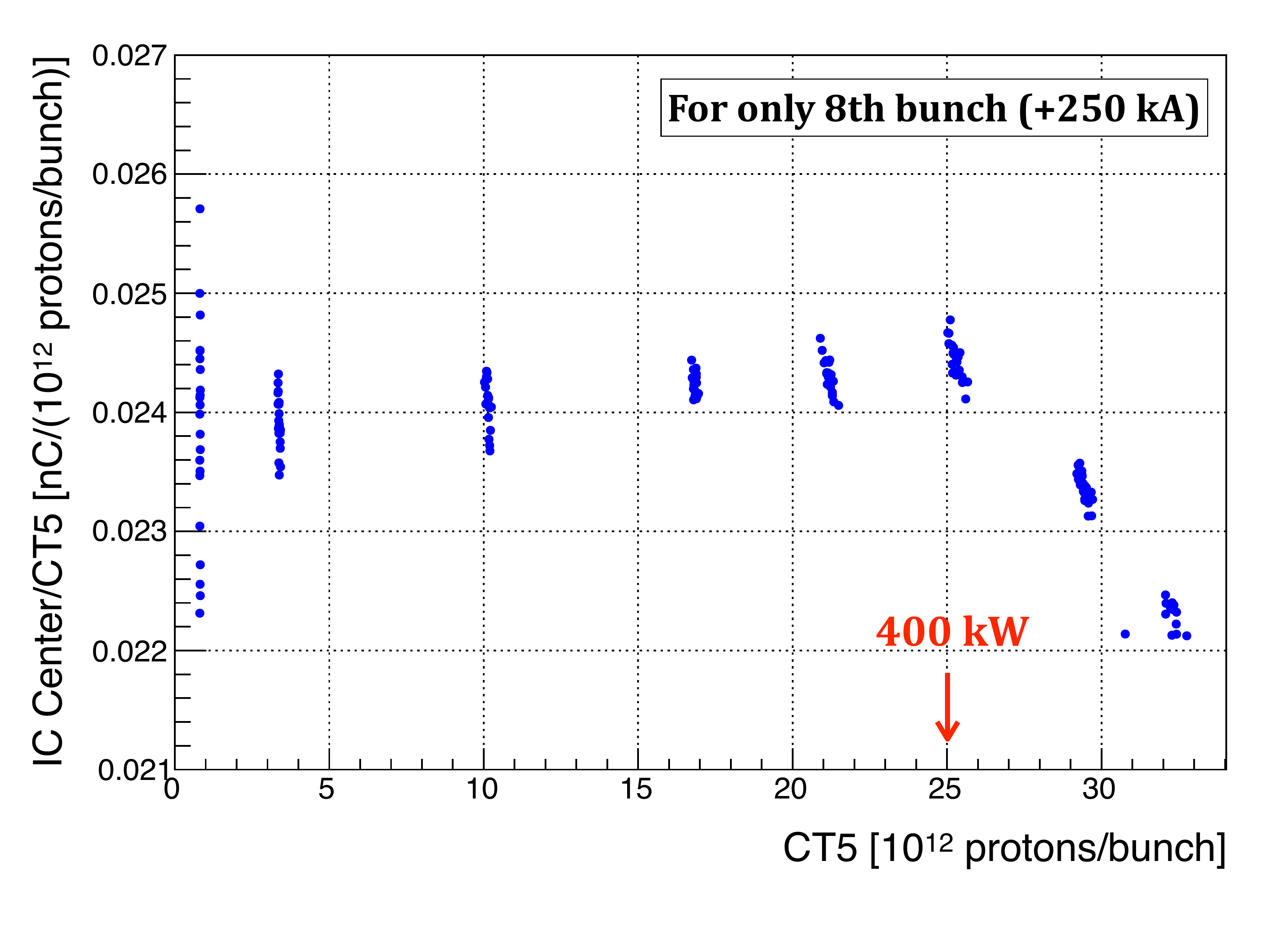}
  \end{center}
 \end{minipage}
 \begin{minipage}{0.5\hsize}
  \begin{center}
   \includegraphics[clip,width=7.5cm]{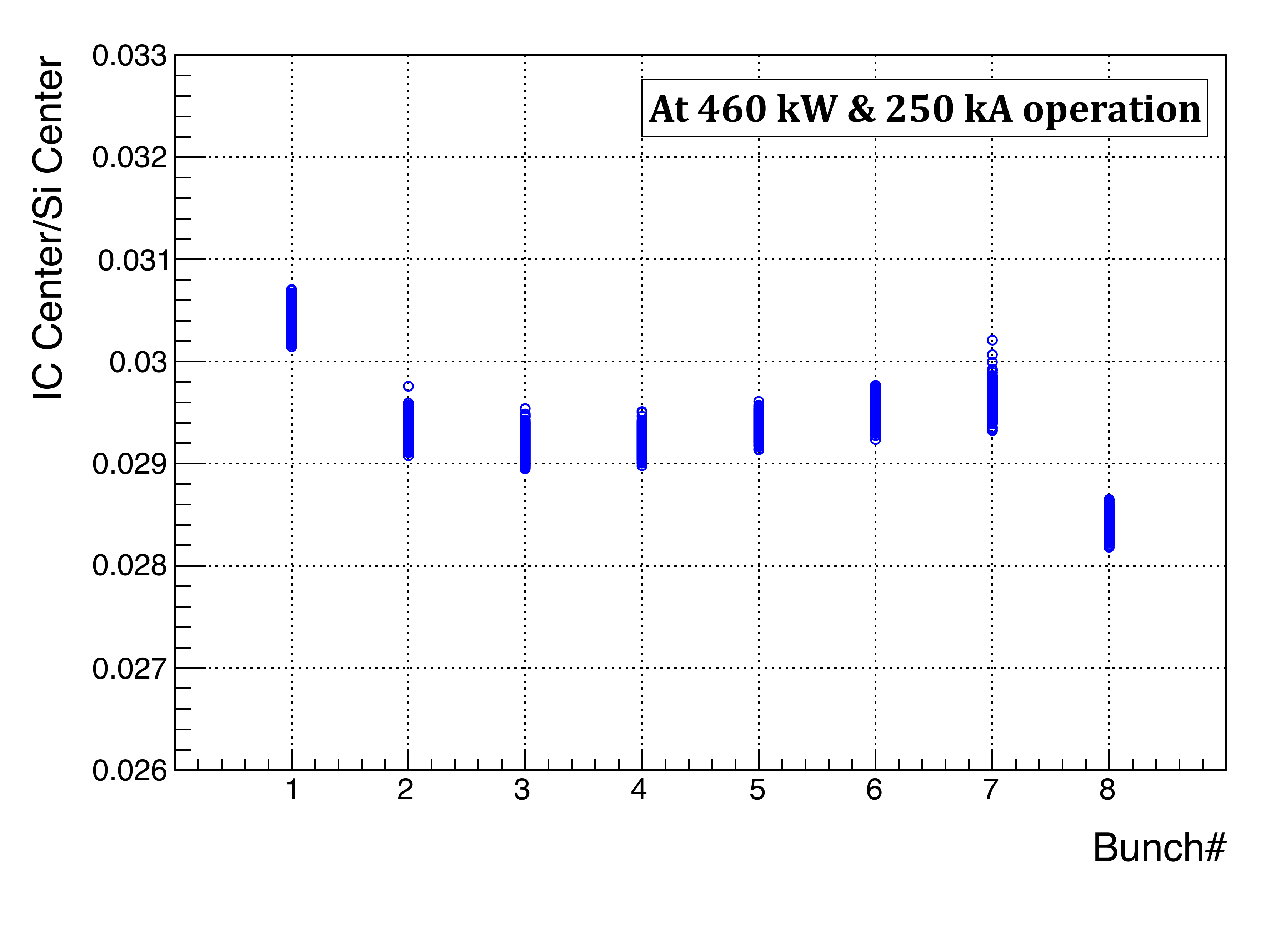}
  \end{center}
 \end{minipage}
 \vspace{-15truept}
 \caption{IC yield for the eighth bunch at various beam powers at $+$250~kA horn current (left) 
          and the ratio of the IC yield to Si yield for each bunch at  
          460~kW and $+$250~kA horn current 
          (right).
          The features seen in bunches 1$\sim$7 can be understood 
          by the signal tail properties: the Si sensors have 
          larger signals but faster decay time tails
            than the IC, as shown in Figure~\ref{fig:mumontimeresponse}. 
          The decrease in the eighth bunch is 
          due to space charge effects in the IC.}
 \label{fig:icnonlinearity}
 \end{figure}

\begin{figure}[htbp]
 \begin{minipage}{0.5\hsize}
  \begin{center}
   \includegraphics[clip,width=7.8cm]{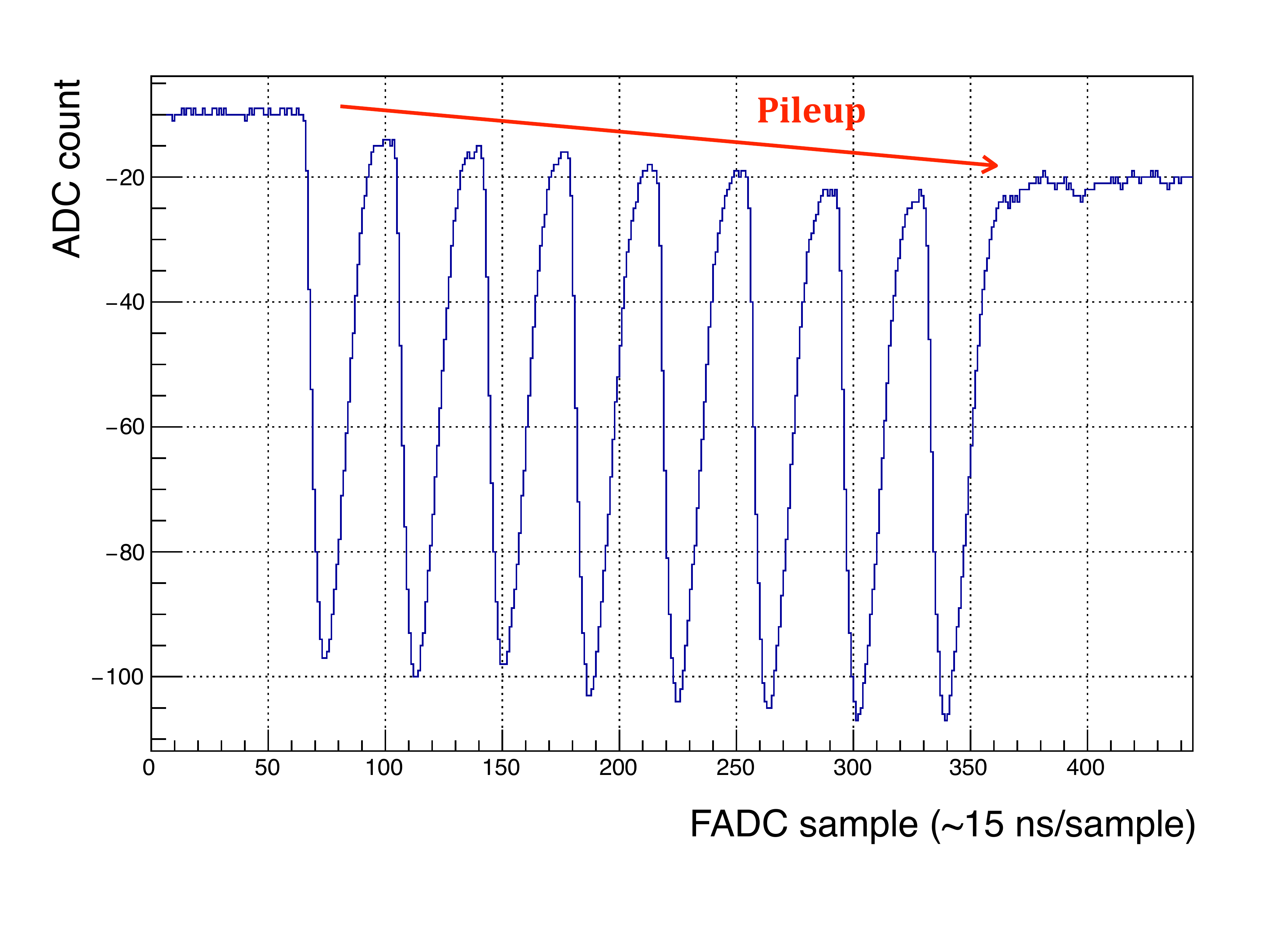}
  \end{center}
 \end{minipage}
 \begin{minipage}{0.5\hsize}
  \begin{center}
   \includegraphics[clip,width=7.8cm]{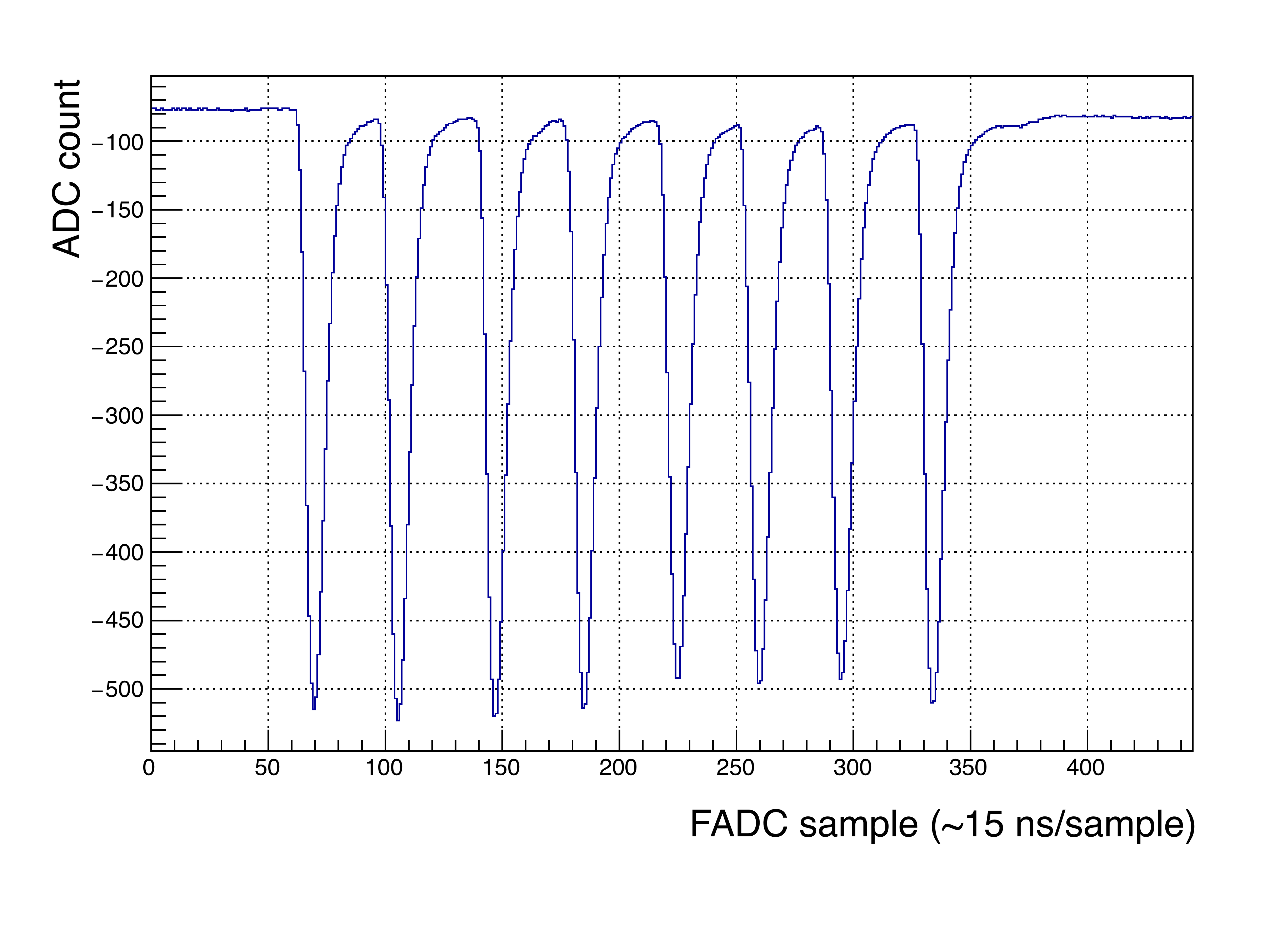}
  \end{center}
 \end{minipage}
 \vspace{-15truept}
   \caption{An example of the IC (He) signal waveform with beam power 340~kW and horn current $+$250~kA 
            ($2.5\times10^6/$cm$^2$ per 80~ns beam bunch, left).
            For comparison, an IC (Ar) waveform with beam power 480~kW and horn current $-$250~kA 
	    ($2.3\times10^6/$cm$^2$ per 80~ns beam bunch, right) is shown.}
 \label{fig:ichelium}
 \end{figure}

For the ionization chambers, non-linear signal response due to space charge effects 
becomes an issue at high intensity.
When many electrons and ions are generated, the electric field is distorted
by accumulated ions, which affects the signal yield.
This effect appears more severely at the latter bunches because more ions are accumulated. 
The left panel of Figure~\ref{fig:icnonlinearity} shows the IC (Ar) yield for the eighth bunch 
at various beam powers (all at $+$250~kA horn current), and indicates that non-linearities appear 
at 400 kW (with a muon flux of $2.9 \times 10^{6}$~${\rm /cm^2}$ per 80~ns beam bunch). 
The right panel shows the IC yield divided by the Si yield for each bunch, and  
a drop in the ratio is clearly seen for the final bunch. 
Here, the beam power and horn current are 460~kW and $+$250~kA, respectively.
%
There are three potential solutions to this problem. One would be to increase the applied 
high voltage so that generated ions are swept out faster; however, 
there is a risk of reaching the chamber breakdown voltage.
Another solution would be to use a lighter gas, such as He, 
in which the number of generated electron/ion pairs is much smaller than Ar, 
and therefore less charge accumulation would occur. 
However, lighter gases can suffer from signal pileup caused by the fast drift speed 
of the lighter ions, as shown in Figure~\ref{fig:ichelium} for He, which can affect 
bunch-by-bunch monitoring.
Another lighter noble gas under consideration is Ne; however, Ne gas is currently prohibitively expensive. 
Usage of a thinner gap between the electrodes could be another solution, 
but this would require a full replacement of the current T2K IC system.

For these reasons, a new T2K muon monitoring system is desired for future operation. 
electron-multiplier tubes (EMTs) are one candidate, as described below. 


%% file: emt.tex
\section{Beam monitoring by electron-multiplier tubes} 
\label{sec:emtconcept}

\begin{figure}[htbp]
  \begin{center}
   \includegraphics[clip,width=10.0cm]{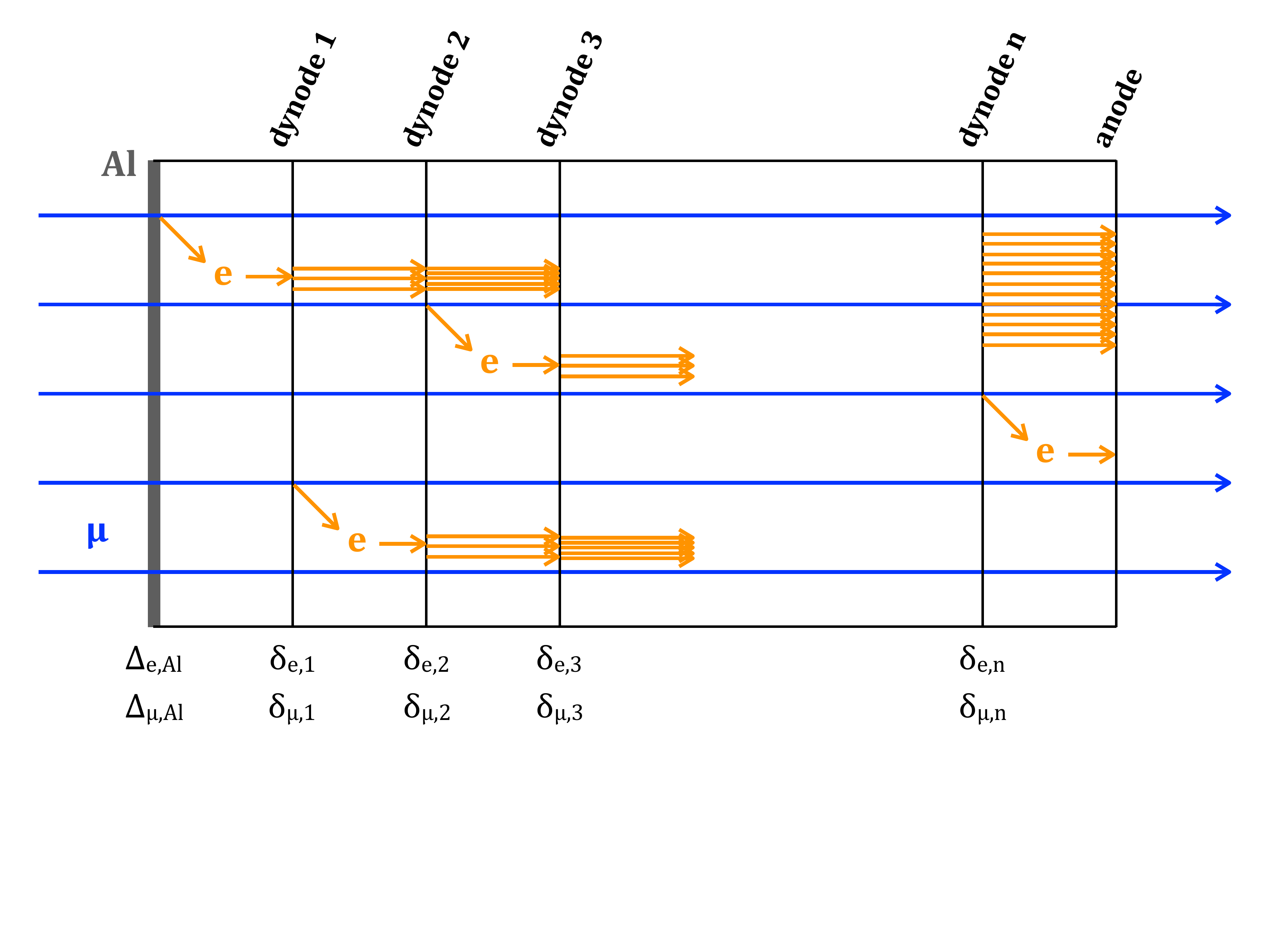}
  \end{center}
  \vspace{-55truept}
   \caption{Schematic diagram of the signal multiplication by an electron-multiplier tube
           as a muon monitor. Electron- and muon-induced secondary
           emission occurs at the Al surface and at each dynode.}
  \label{fig:emtschematic}
 \end{figure}

Several requirements must be satisfied when monitoring 
the T2K high-intensity muon beam. 
The signal size must be large enough to be read out without 
in situ electronic amplification. 
Fast signal response is also required for 
bunch-by-bunch monitoring.  The device must also be radiation tolerant. 
Secondary electron emission monitors (SEMs) are
proven to be radiation tolerant and have a fast response. 
However, the SEM expected signal size is too small to be used as 
a T2K muon monitor.
%
Photomultiplier tubes (PMTs) use the secondary emission process to multiply 
electrons from the PMT photocathode. 
Secondary emission electrons are also produced by the passage of muons
through a PMT, and the muon-induced secondary electron signal can 
be multiplied as usual by the PMT.
For muon monitoring, however, the PMT photocathode is not necessary.
Hence an electron-multiplier tube (EMT), a PMT without a photocathode,
is considered as a new MUMON candidate sensor.
In our study, EMTs were produced by depositing aluminum on the cathodes of PMTs. 
Figure~\ref{fig:emtschematic} shows a schematic diagram of the signal multiplication by an EMT. 

When a muon passes through the EMT, secondary electrons are produced either at 
the surrounding aluminum cathode or at the dynodes. 
The emitted electrons are accelerated, bombard the downstream dynodes, and produce more electrons.
The secondary emission efficiency is defined as the number of emitted electrons divided 
by the number of incident particles \cite{hamamatsu}, and
is a characteristic of the material type, the incident particle type, and the particle energy. 
In Figure~\ref{fig:emtschematic}, $\Delta$ represents the secondary emission efficiency for aluminum 
and $\delta$ is that for dynodes.
To specify the incident particle, an ``$e$" subscript is used for electrons and 
a ``$\mu$" is used for muons. 
The dynode number is given as a subscript of $\delta$.
The secondary emission efficiency $\delta$ is dependent on the particle energy; therefore,
it is possible that $\delta_{e,i}$ and $\delta_{e,j}$ ($i \neq j$) are not equal.
The product of the secondary emission efficiencies of all dynodes determines the gain of the PMT:

 \begin{eqnarray}
  G = \delta_{e,1} \times \delta_{e,2} \times \cdots \times \delta_{e,n} 
    = \prod_{i=1}^{n} \delta_{e,i}. 
 \label{eq:emtgain}
 \end{eqnarray}
 \vspace{1truept}

\noindent
In the case of the muon monitor, the signal is generated 
primarily by muons and electrons (delta-rays) knocked out by muons.
The energy of the delta-rays reaches several hundred MeV \cite{suzuki}. 
These muons and delta-rays penetrate the EMT, 
while most of the secondary emission electrons stop once they hit the dynodes. 
Therefore, it is possible that secondary electrons that contribute to the signal 
are produced at every dynode by muons or delta-rays. 
The final output signal $Q$ can be written as follows: 

 \begin{eqnarray}
  && Q = Q_\mu + Q_e,  
  \label{eq:signalyield1} \\
  && Q_\mu = e \cdot \phi_\mu \cdot
          \left\{ A_{\rm sur} \cdot \Delta_{\mu,{\rm Al}} \cdot \prod_{i=1}^{n} \delta_{e,i} + 
	          \sum_{i=1}^{n-1} \left( A_i \cdot \delta_{\mu,i} \cdot \prod_{j=i+1}^{n} \delta_{e,j} \right)
	  \right\},   
  \label{eq:signalyield2} \\
  && Q_e = e \cdot \phi_e \cdot
          \left\{ A_{\rm sur} \cdot \Delta_{e,{\rm Al}} \cdot \prod_{i=1}^{n} \delta_{e,i} + 
	          \sum_{i=1}^{n-1} \left( A_i \cdot \delta_{e,i} \cdot \prod_{j=i+1}^{n} \delta_{e,j} \right)
	  \right\}, 
  \label{eq:signalyield3}
 \end{eqnarray}
 \vspace{1truept}

\noindent
where $e$ is the elementary electric charge ($1.6\times10^{-19}$ C), 
$A_{\rm sur}$ and $A_i$ are the area of the aluminum cathode surface and each dynode 
surface [${\rm cm^2}$], respectively, and 
$\phi_{\mu}$ ($\phi_e$) is the muon (delta-ray) flux [$/{\rm cm^2}$]. 
The first term corresponds to the secondary electron production
at the aluminum cathode, and the second term to that at the i-th dynode.  

As a simple case for a rough estimation, the contribution from delta-rays 
is ignored here and the area
of each Al and dynode surface is assumed to be $A$.
Here we assume that particles are incident on the EMT perpendicular to the surface 
and any change of the dynode effective surface area is not considered.
In addition, the electron secondary emission efficiencies of all dynodes ($\delta_{e,i}$) 
are assumed to be the same (referred to as $\delta_{e}$, hereafter).
The typical gain of ${\rm HAMAMATSU^{\textregistered}}$ R9880, which was used for the EMT
studies shown here, 
is $5 \times 10^3$ when $-$500 V is applied,
and the number of dynodes is $n = 10$. Therefore, the gain per dynode can be calculated
as $\delta_{e} \sim 2.35$.
The secondary emission efficiencies for muons with several GeV energy,
$\Delta_{\mu,{\rm Al}}$ and $\delta_{\mu,i}$, are assumed to be 0.08 \cite{winn}.
The surface area is $A = 2.01 \ {\rm cm^2}$ (radius is 8 mm),
and the normalized muon flux at horn current $+$250~kA is
$\phi_\mu^{\rm normalized} = 1.09\times10^5 \ {\rm /cm^2/10^{12} \ POT}$.
Using the number of protons at beam power 460~kW, $N = 3.0\times10^{13} \ {\rm POT/bunch}$, 
the muon flux is calculated as $\phi_\mu = \phi_\mu^{\rm normalized} \cdot N$.
With these assumptions, the expected charge is calculated as 730~pC/bunch.


%% file: setup.tex
\section{Prototype detectors}  
\label{sec:developemt}

\begin{figure}[htbp]
  \begin{center}
   \includegraphics[clip,width=8.0cm]{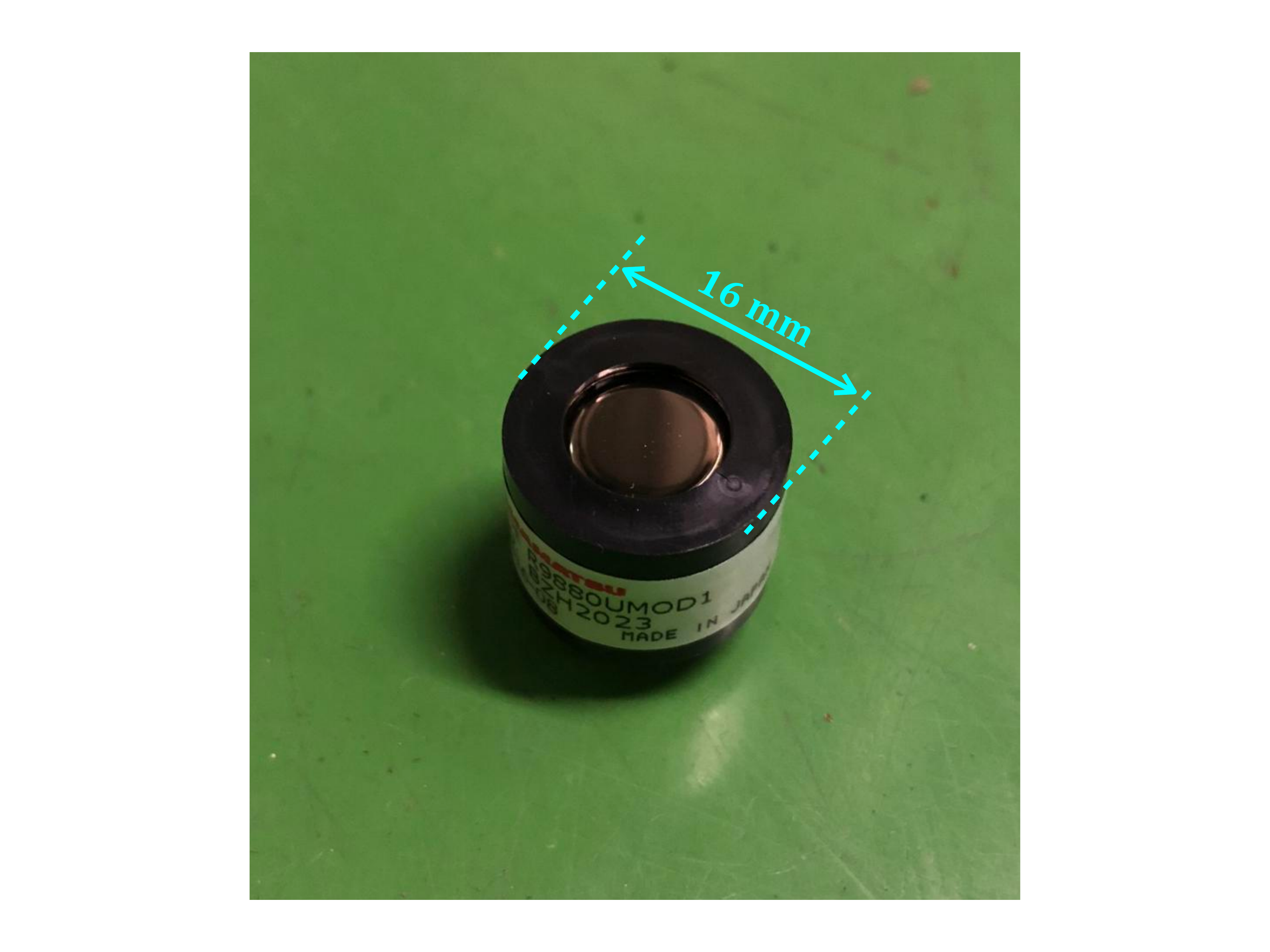}
  \end{center}
  \vspace{-15truept}
   \caption{A photograph of EMTC3. The monitor radius 
   is 8~mm.}
  \label{fig:emtpicture}
 \end{figure}
\begin{figure}[htbp]
  \begin{center}
   \includegraphics[clip,width=10.0cm]{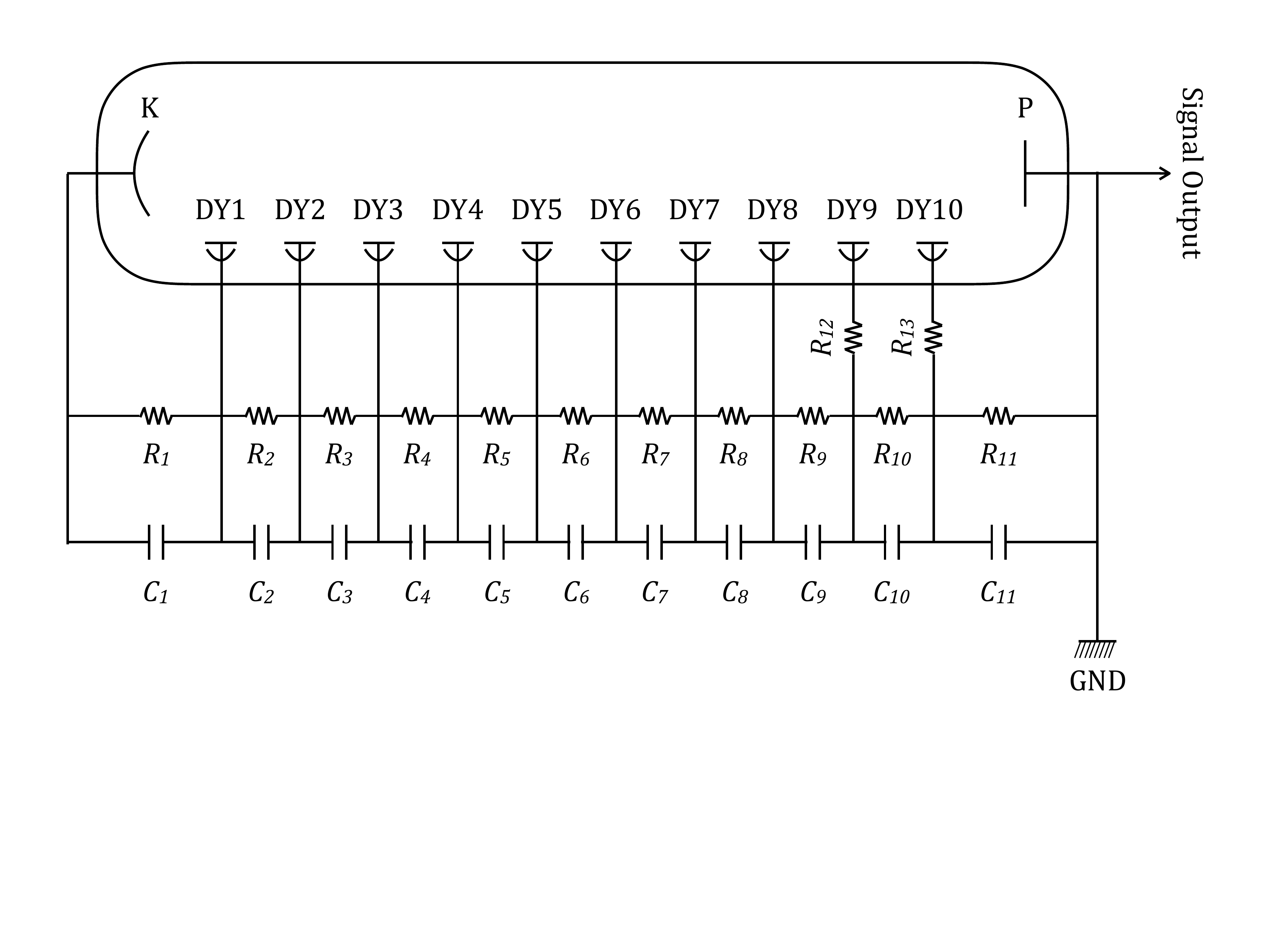}
  \end{center}
  \vspace{-65truept}
  \caption{Schematic diagram of the divider circuit of the prototype EMT.
           ``K", ``P'', ``DY'' and ``GND'' represent the cathode, anode, 
	   dynode, and ground, respectively.
           The resistances $R_1 \sim R_{10}$ are 330~k$\Omega$, and $R_{11}$ 
	   is 160$ \ {\rm k\Omega}$. 
	   The capacitances are given in Table~\ref{tab:capacitancevalue}.}
  \label{fig:emtcircuit}
 \end{figure}

Two EMTs were custom-made based on the ${\rm HAMAMATSU^{\textregistered}}$ 
metal-package PMT R9880 by depositing aluminum on the cathode.
R9880 was selected due to its small footprint and short length: it is 
short enough to be easily installed and aligned at the muon monitor site.
Figure~\ref{fig:emtpicture} shows a photograph of the EMTC3 sensor and
Figure~\ref{fig:emtcircuit} shows a schematic diagram of the divider circuit 
used to operate these EMTs. 
In Figure~\ref{fig:emtcircuit}, the resistances $R_1 \sim R_{10}$ are 330~k$\Omega$, and $R_{11}$ is  
160~${\rm k\Omega}$, giving uniform voltage differences between the dynodes.
The maximum negative high voltage applied to the cathode was limited to $-$500~V during 
our tests due to 
the currently installed connectors in the T2K muon monitor system.
The capacitors, $C_1 \sim C_{11}$, compensate for the charge consumed on the 
dynodes in the process of multiplication.
To maintain linearity in the EMT response, sufficient charge, usually $100 \sim 1000$ 
times 
the consumed charge, must be stored in these capacitors \cite{hamamatsu}.
The 51~${\Omega}$ damping resistances, $R_{12}$ and $R_{13}$, are usually inserted 
to reduce waveform ringing, but we removed them so that the charge consumed in 
earlier bunches can be quickly compensated for from the capacitors.
 
We made four sets of divider circuits.
The capacitances of the later versions were optimized based on 
initial measurement results 
using the earlier versions.
In this paper, results of measurements with two later versions, 
referred to as EMTC3 and EMTC4, are reported.
Table~\ref{tab:capacitancevalue} summarizes the capacitances and stored charges
of each capacitor used in EMTC3 and EMTC4 when a negative bias of $-500$~V is applied. 
Only $C_8$ to $C_{11}$ for EMTC3 and $C_6$ to $C_{11}$ for EMTC4
were used 
and the other capacitors were removed (represented as ``-" in Table~\ref{tab:capacitancevalue}). 

 \begin{table}[htbp]
  \begin{center}
    \caption{Capacitances and stored charges for the EMTs when $-$500~V is applied.}
  \label{tab:capacitancevalue}
  \vspace{2truept}
   \begin{tabular}{l c c c c} \hline \hline
     &  \multicolumn{2}{c}{EMTC3} & \multicolumn{2}{c}{EMTC4} \\ \hline
     & capacitance (nF) & charge ($\mu$C) & capacitance (nF) & charge ($\mu$C) \\
     \hline
     K-DY1 ($C_1$) & \multicolumn{2}{c}{-} & \multicolumn{2}{c}{-}     \\
    DY1-2 ($C_2$)  & \multicolumn{2}{c}{-} & \multicolumn{2}{c}{-}     \\
    DY2-3 ($C_3$)  & \multicolumn{2}{c}{-} & \multicolumn{2}{c}{-}     \\
    DY3-4 ($C_4$)  & \multicolumn{2}{c}{-} & \multicolumn{2}{c}{-}     \\
    DY4-5 ($C_5$)  & \multicolumn{2}{c}{-} & \multicolumn{2}{c}{-}     \\
    DY5-6 ($C_6$)  & \multicolumn{2}{c}{-}     & 100 & 4.8  \\
    DY6-7 ($C_7$)  & \multicolumn{2}{c}{-}     & 100 & 4.8  \\
    DY7-8 ($C_8$)  & 10 & 0.48   & 100 & 4.8  \\ 
    DY8-9 ($C_9$)  & 10 & 0.48   & 330 & 15.7  \\ 
    DY9-10 ($C_{10}$)   & 10 & 0.48   & 330 & 15.7\\ 
    DY10-GND ($C_{11}$) & 15 & 0.35   & 330 & 7.6  \\ \hline \hline 
   \end{tabular}
  \end{center}
 \end{table}
 
\begin{figure}[htbp]
  \begin{minipage}{0.4\hsize}
   \begin{center}
    \includegraphics[clip,width=7.8cm]{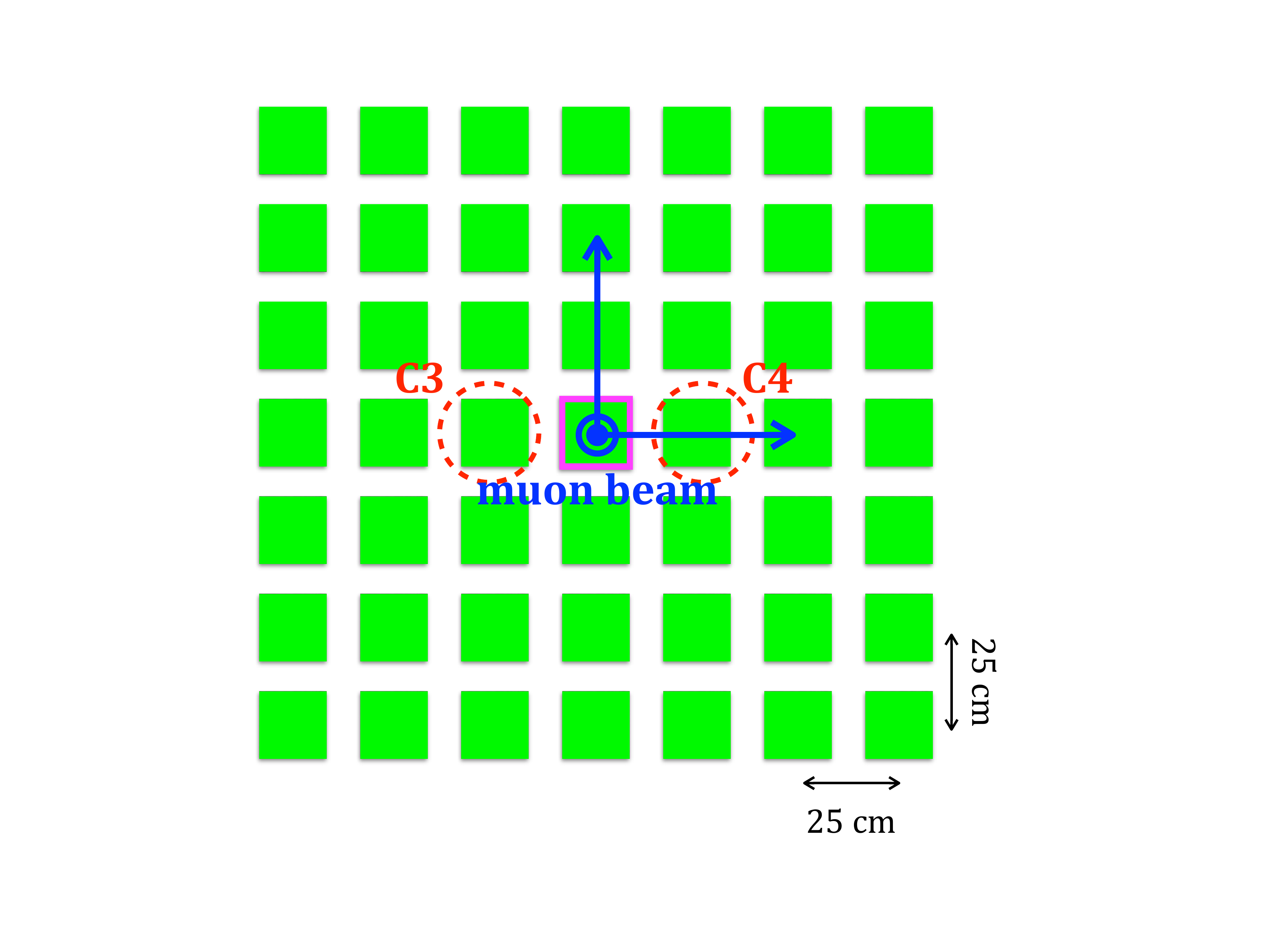}
   \end{center}
  \end{minipage}
  \begin{minipage}{0.6\hsize}
   \begin{center}
    \includegraphics[clip,width=9.2cm]{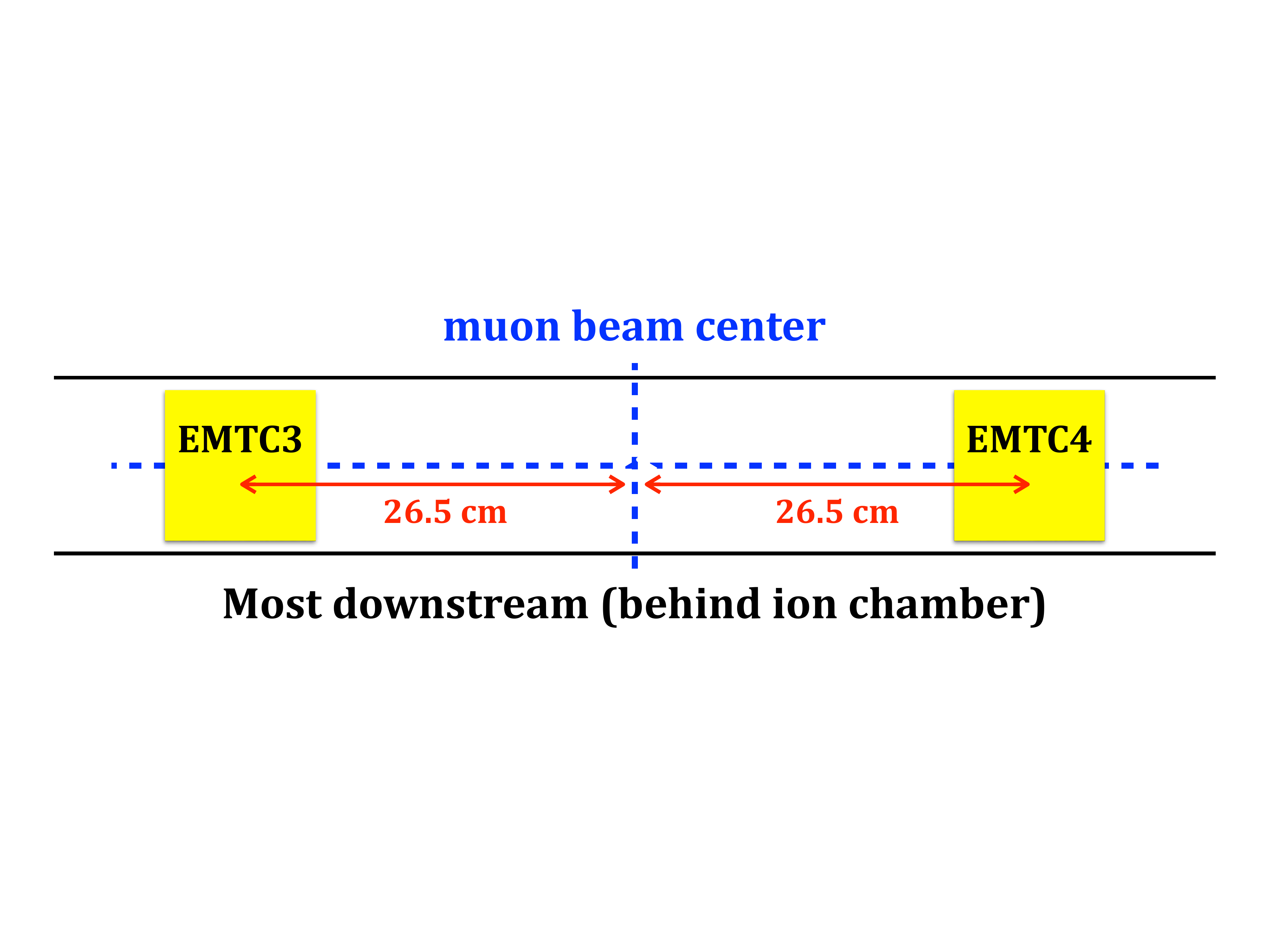}
   \end{center}
  \end{minipage}
  \vspace{-20truept}
  \caption{Prototype EMTs' horizontal and vertical installation positions 
           overlaid with the IC sensor positions (left) and
           a zoom-out focusing on the geometry around the EMTs (right).  
	   Green squares in the left panel indicate the 49 channels of the ionization chambers;
	   the one whose edge is colored magenta is the center channel.}
  \label{fig:emtinstallplace}
 \end{figure}

\newpage
EMTC3 and EMTC4 were installed directly downstream of the T2K MUMON ionization chambers.
Figure~\ref{fig:emtinstallplace} shows the EMT installation positions 
compared with the IC sensor positions.
Relative to the MUMON, the EMT sensors are vertically centered and 
horizontally offset from center by 26.5~cm.
The signal and high-voltage cables are the same as those used for the Si and IC sensors.
The read-out system is also the same; however, no attenuator module is used for the EMTs.

%% file: performance.tex
\section{Muon beam test}  
\label{sec:performance}

The performance of the EMTs (C3 and C4), including checks of the signal size, time response, 
intensity resolution, linearity, and stability, was studied 
in situ with the T2K muon beam.

\subsection{Output charge} 

The expected output charge of an EMT was calculated in Section~\ref{sec:emtconcept} 
as 730~pC/bunch at 460~kW beam power and $+$250~kA horn current 
(corresponding to a muon flux of $3.3 \times 10^{6}$~${\rm /cm^2}$ per 80 ns beam bunch).
The measured charge output per beam bunch for EMTC3 and C4 when $-$500~V was applied  
is summarized in Table~\ref{tab:emtcharge}. 
Reasonable agreement with our expectation is seen.
The difference in signal size between the two sensors is compatible with 
the expected difference between individual PMTs.

 \begin{table}[htbp]
  \begin{center}
  \caption{EMT output signal size per bunch at 460~kW beam intensity and $+$250~kA horn current 
           (muon flux of $3.3 \times 10^{6}$~${\rm /cm^2}$ per 80 ns beam bunch) with applied voltage of $-$500~V. 
	   The expected charge is 730~pC/bunch.} 
  \label{tab:emtcharge}
  \vspace{2truept}
   \begin{tabular}{c c c} \hline \hline
    Bunch\# & EMTC3 charge [pC] & EMTC4 charge [pC] \\ \hline
    1       & 873.1 & 784.9 \\
    2       & 870.7 & 787.3 \\ 
    3       & 866.9 & 788.6 \\ 
    4       & 855.9 & 787.9 \\
    5       & 860.4 & 797.0 \\
    6       & 850.6 & 795.5 \\ 
    7       & 847.8 & 798.5 \\ 
    8       & 854.1 & 797.7 \\ \hline 
    Average & 859.9 & 792.2 \\ \hline \hline
   \end{tabular}
  \end{center}
 \end{table}

\subsection{Signal waveform} 

\begin{figure}[h]
  \begin{minipage}{0.5\hsize}
   \begin{center}
    \includegraphics[clip,width=7.5cm]{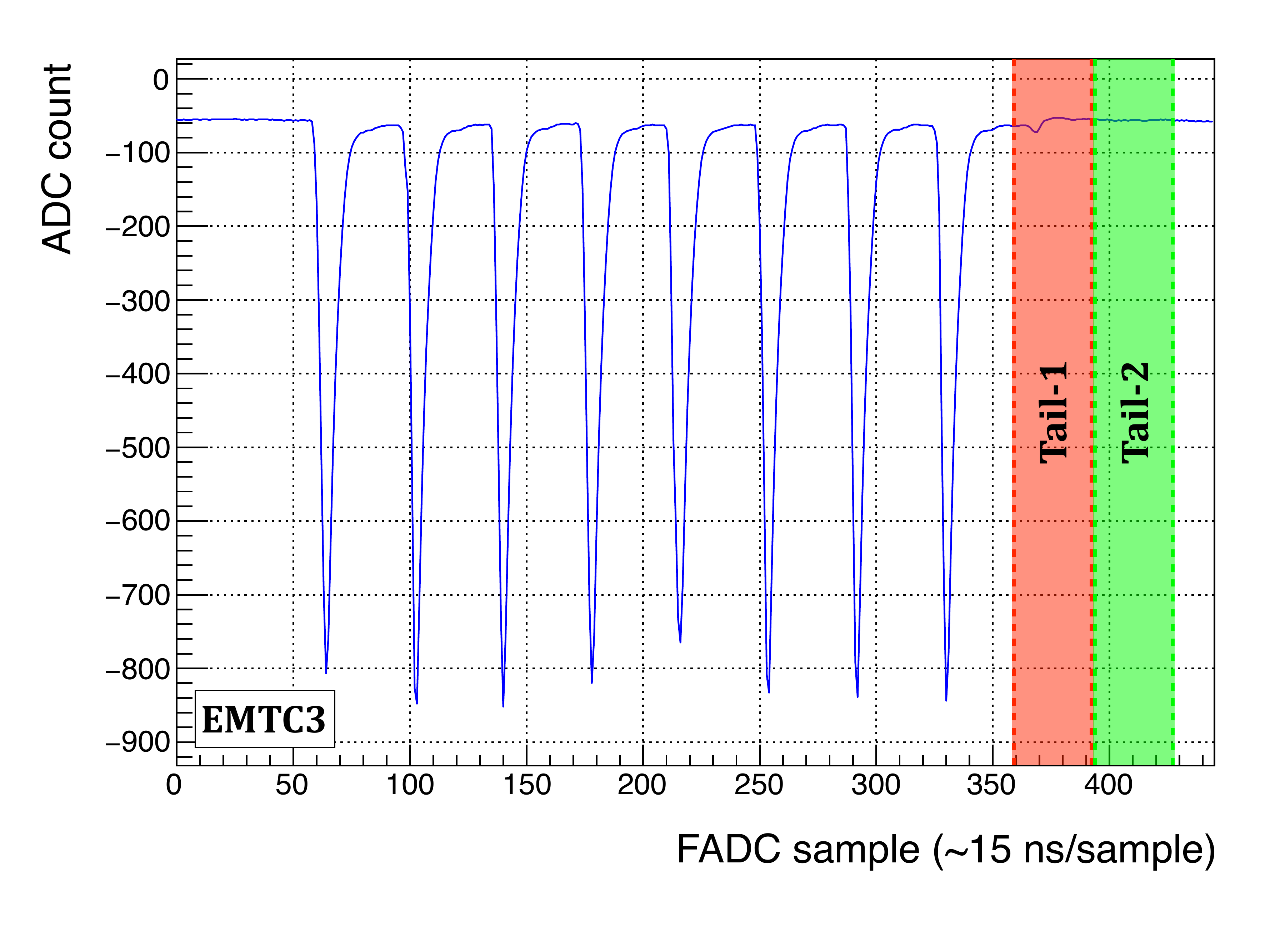}
   \end{center}
  \end{minipage}
  \begin{minipage}{0.5\hsize}
   \begin{center}
    \includegraphics[clip,width=7.5cm]{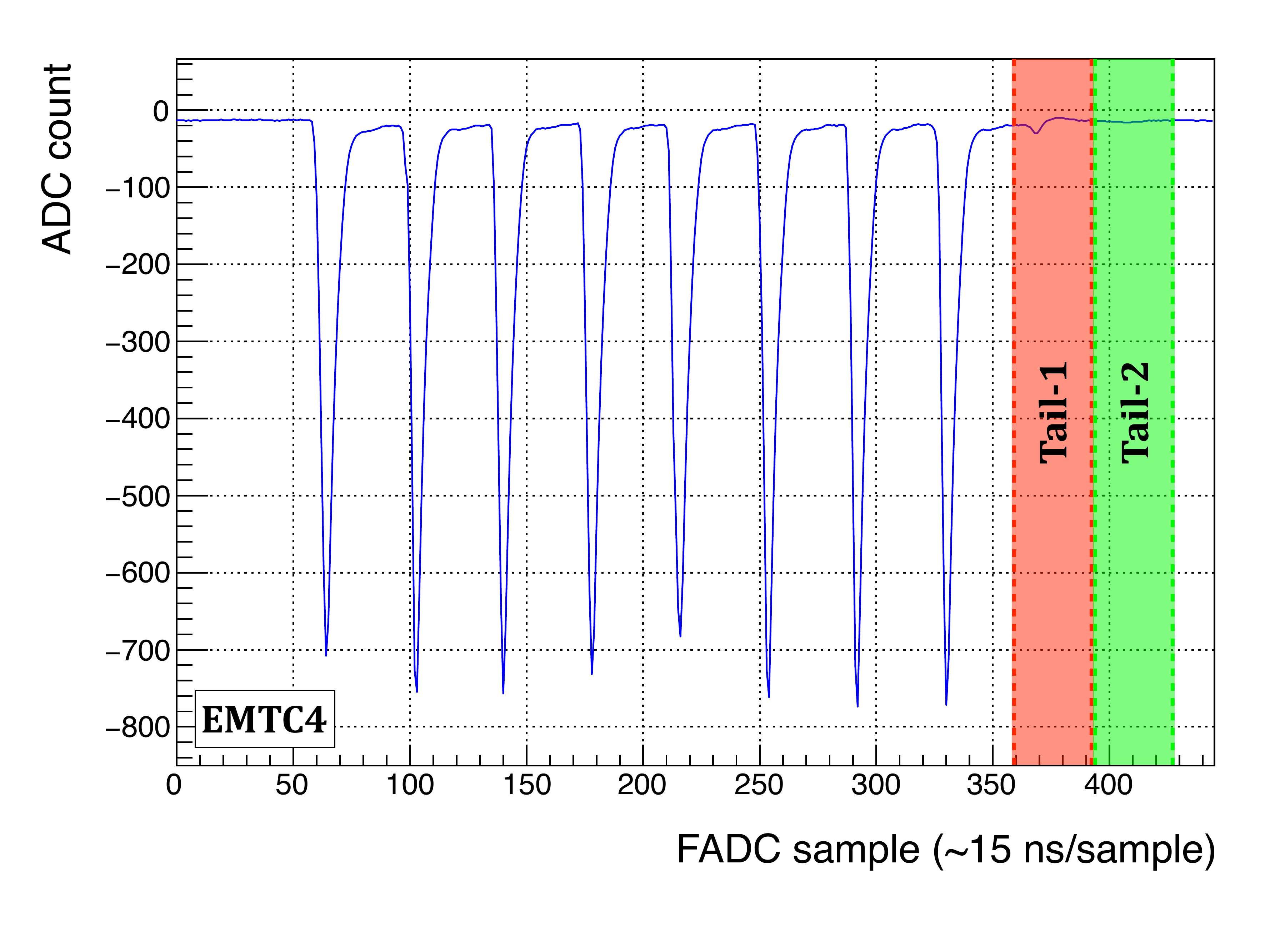}
   \end{center}
  \end{minipage}
  \vspace{-15truept}
  \caption{EMTC3 (left) and C4 (right) waveform examples. Two tail regions, 
           Tail-1 and Tail-2, are selected 
           to estimate the tail component contribution.}
  \label{fig:emtwaveform}
 \end{figure}

\begin{figure}[h]
  \begin{minipage}{0.5\hsize}
   \begin{center}
    \includegraphics[clip,width=7.5cm]{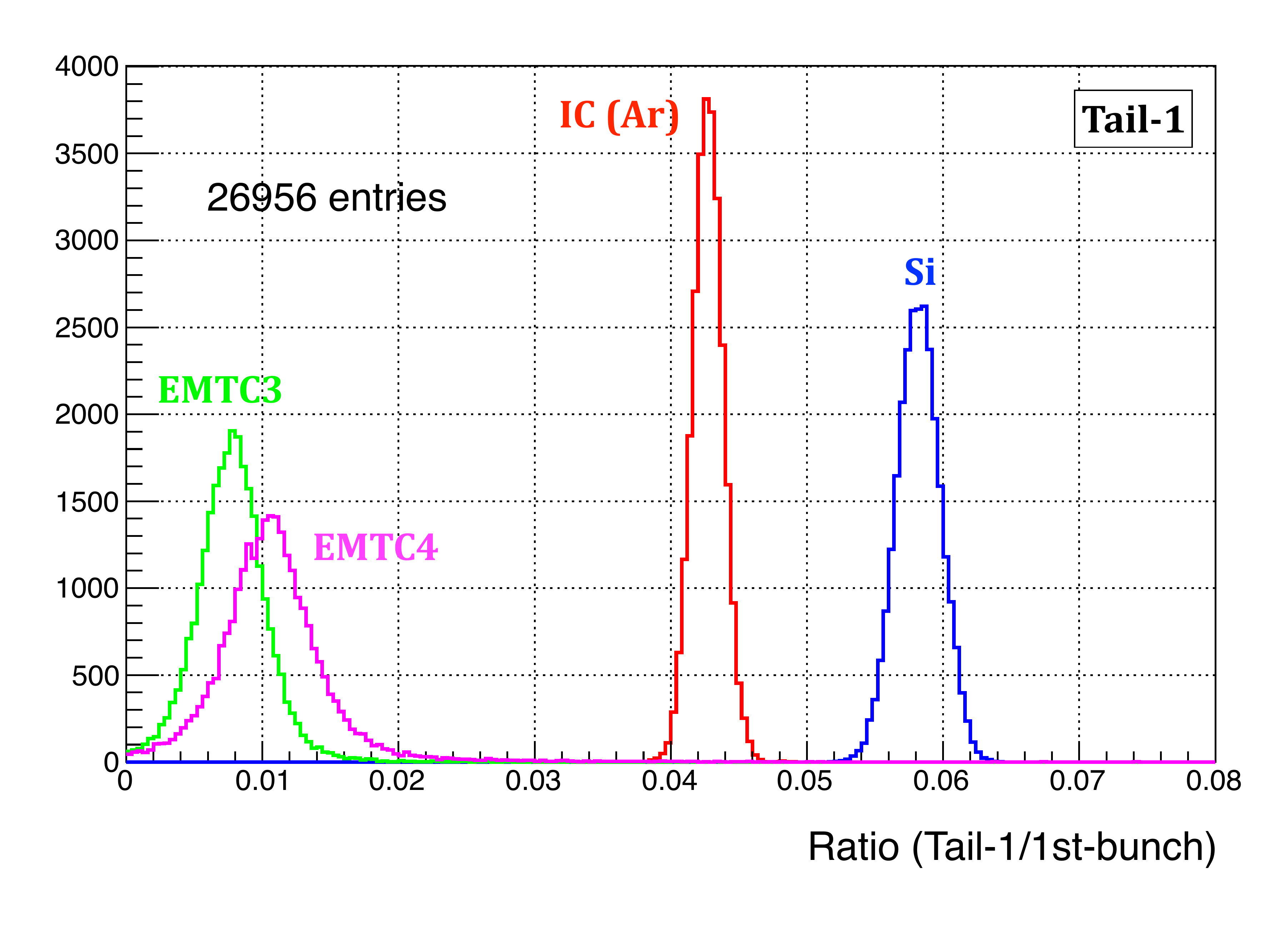}
   \end{center}
  \end{minipage}
  \begin{minipage}{0.5\hsize}
   \begin{center}
    \includegraphics[clip,width=7.5cm]{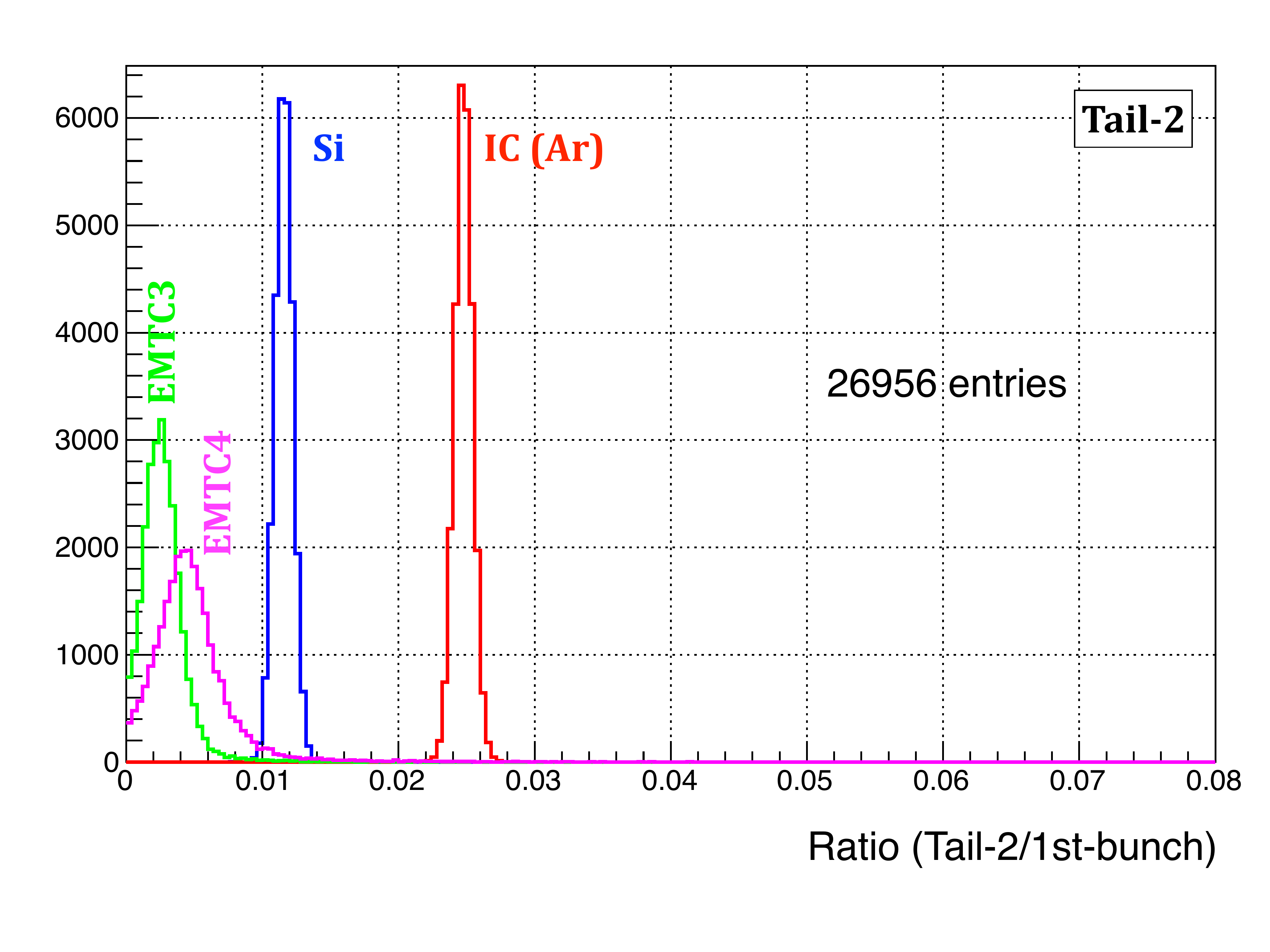}
   \end{center}
  \end{minipage}
  \vspace{-15truept}
  \caption{Tail sizes for Tail-1 (left) and Tail-2 (right).
	   The tail integration regions are shown in Figure~\ref{fig:emtwaveform}.
           The proton beam intensity was 450 kW and the horn current
           setting was $+$250~kA (corresponding to a 
           muon flux of $3.2 \times 10^{6}$~${\rm /cm^2}$ per 80 ns beam bunch).
	   The EMTs show a smaller tail component than the Si and IC (Ar) sensors. 
           The Si tail component has a larger signal but faster decay time than the IC (Ar) tail.}
  \label{fig:mumontimeresponse}
 \end{figure}

Figure~\ref{fig:emtwaveform} shows examples of the EMTC3 (left panel) and 
C4 (right panel) signal waveform. 
The signal has a tail component caused by both detector intrinsic 
properties and cable and electronics reflections. 
Such tails could cause the performance of bunch-by-bunch monitoring to deteriorate.
To quantify the tail contribution, we calculate the ratio of the integral of 
the tail region to that of the first bunch signal for the EMTs, 
the Si center channel (Si channel at the center position of the 49 channels), 
and the IC center channel (IC channel at the center position of the 49 channels). 
%
The tail size is calculated by integrating the regions 
Tail-1 and Tail-2 shown in Figure~\ref{fig:emtwaveform},
and normalizing by the signal size of the first bunch. 
The integration 
periods are the same for all three regions.
The tail sizes are shown in Figure~\ref{fig:mumontimeresponse}
when
the beam power and horn current settings were
450~kW and $+$250~kA, respectively. 
It can be seen that the EMTs have a smaller tail component (about 1\%) 
than the Si or IC sensors (a few \%).
Note that the tail of the eighth bunch of the IC is affected by space charge 
effects, as discussed in Section~\ref{sec:t2kandmumon},
which cause a larger tail.
In the future, with higher beam intensities, this effect is expected to become more severe, 
since space charge effects will begin to affect earlier bunches as well.
%
These measurements indicate that EMTs perform 
better than Si and IC sensors for bunch-by-bunch beam monitoring.

\newpage
\subsection{Intensity resolution} 

\begin{figure}[h]
  \begin{minipage}{0.5\hsize}
   \begin{center}
    \includegraphics[clip,width=7.2cm]{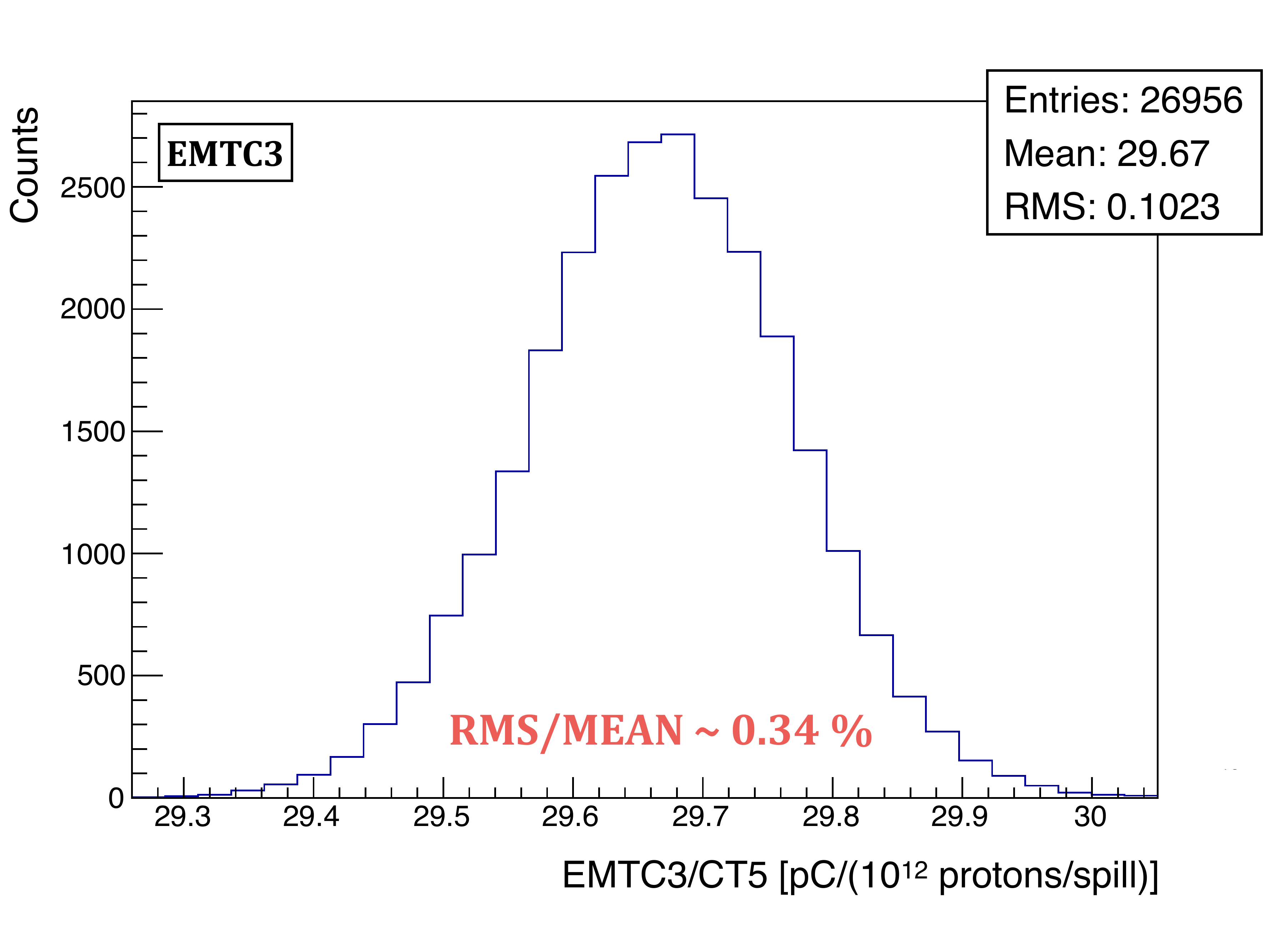}
   \end{center}
  \end{minipage}
  \begin{minipage}{0.5\hsize}
   \begin{center}
    \includegraphics[clip,width=7.2cm]{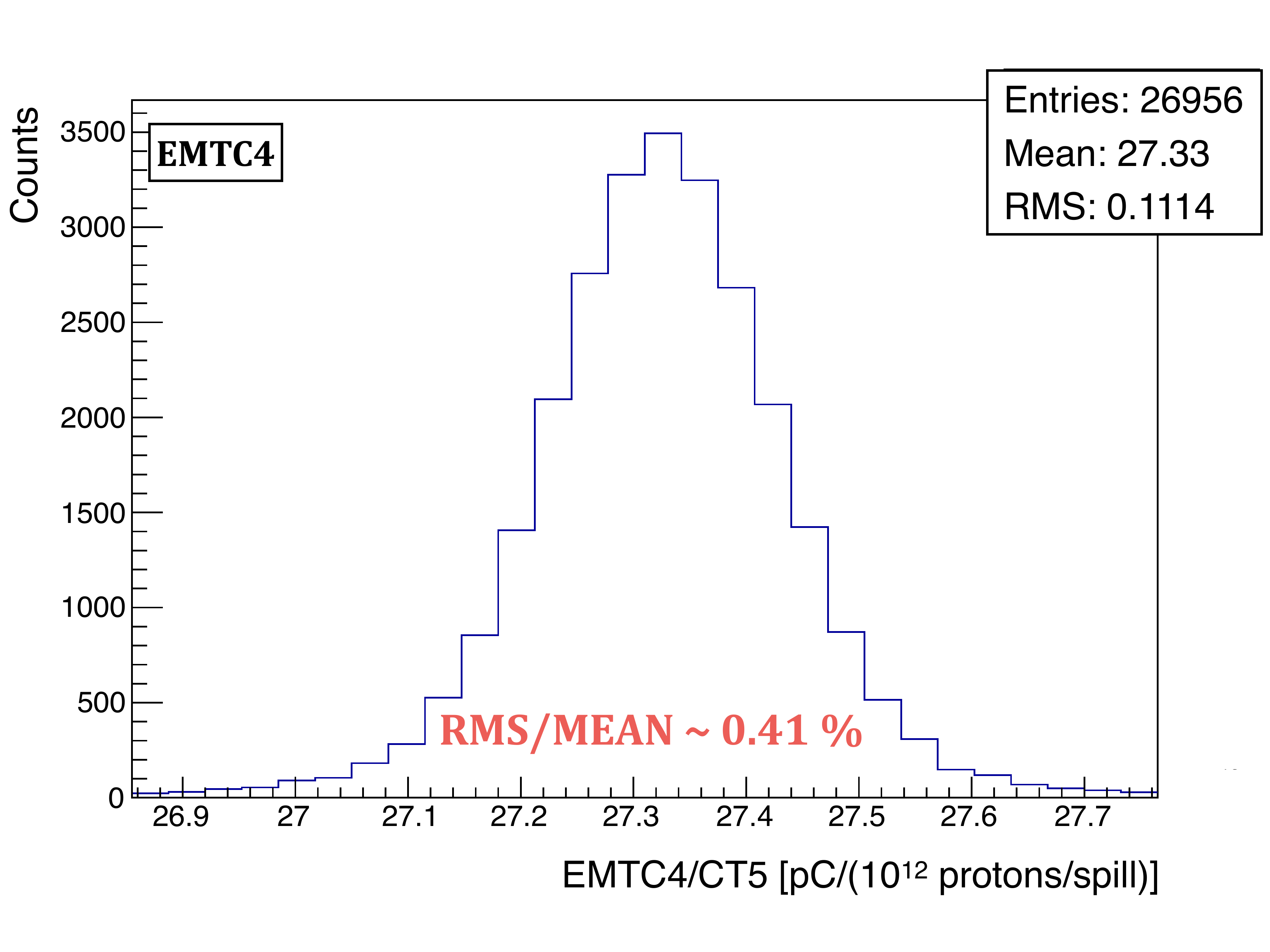}
   \end{center}
  \end{minipage}
  \begin{minipage}{0.5\hsize}
   \vspace{-5truept}
   \begin{center}
    \includegraphics[clip,width=7.2cm]{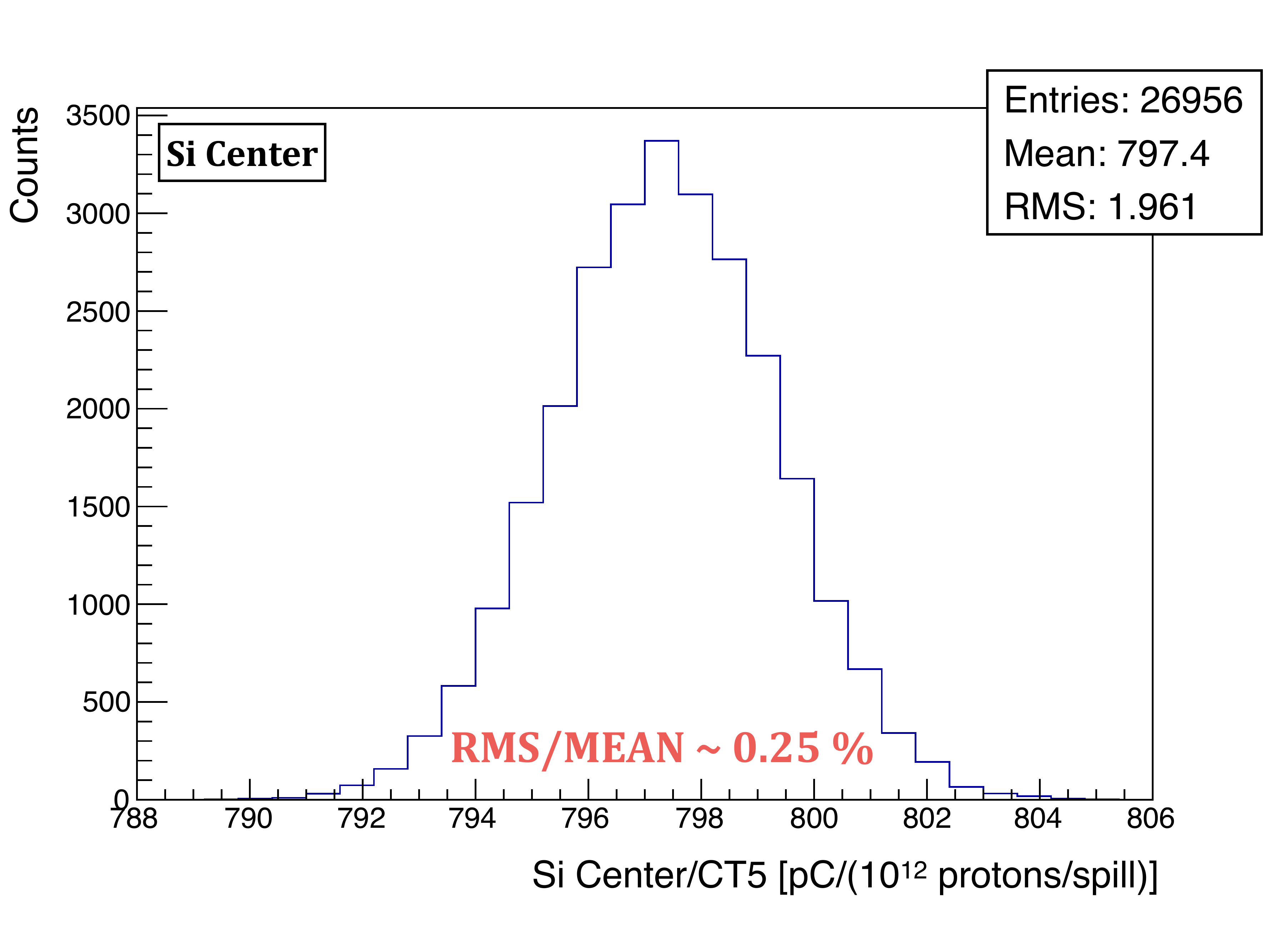}
   \end{center}
  \end{minipage}
  \begin{minipage}{0.5\hsize}
   \vspace{-5truept}
   \begin{center}
    \includegraphics[clip,width=7.2cm]{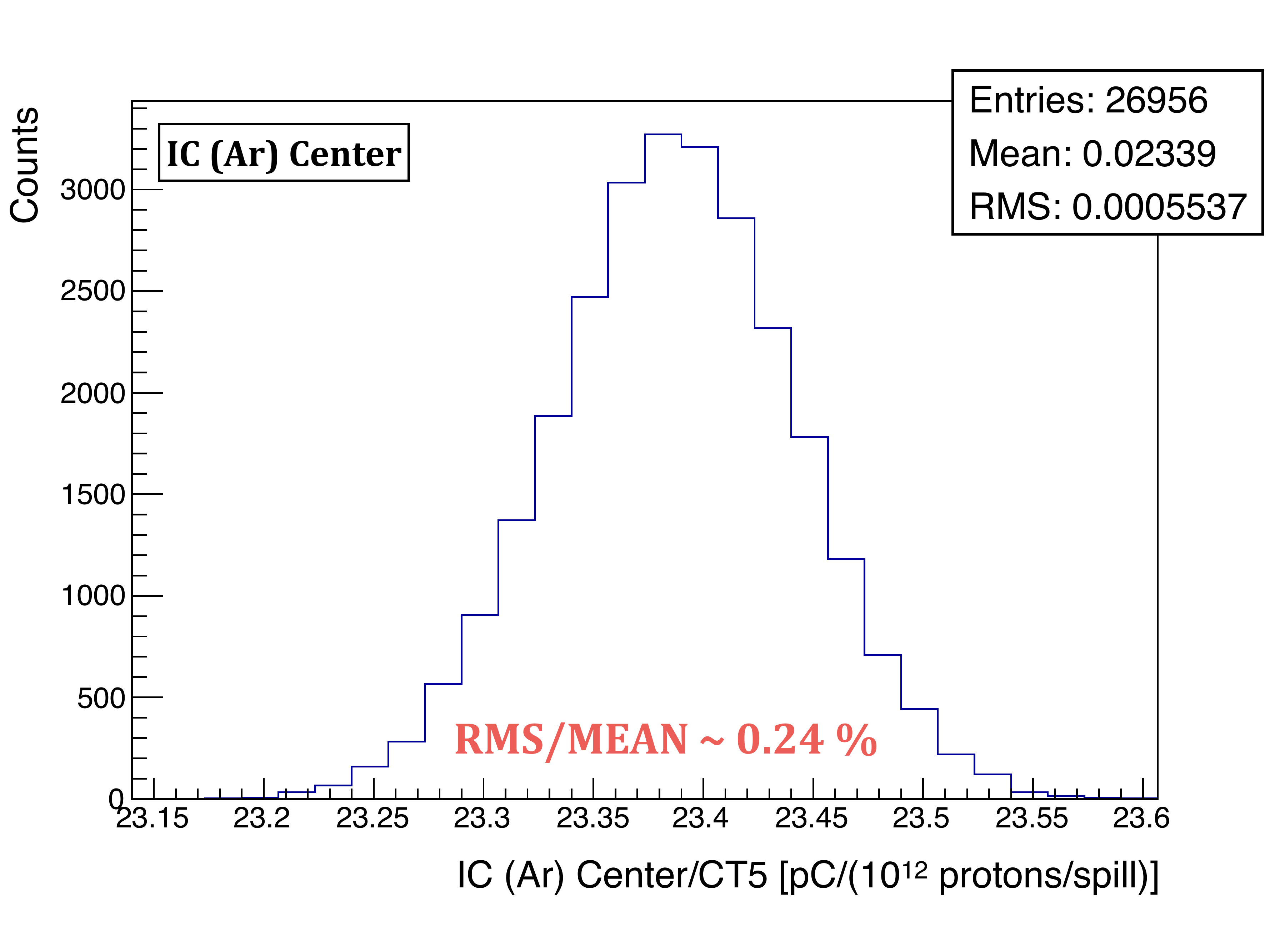}
   \end{center}
  \end{minipage}
  \vspace{-10truept}
  \caption{The normalized signal size of EMTC3 (top left), C4 (top right), 
           Si center (bottom left), and IC (Ar) center (bottom right) 
           at 450 kW beam power and $+$250~kA horn current (corresponding to a
           muon flux of $3.2 \times 10^{6}$~${\rm /cm^2}$ per 80 ns beam bunch).
           The intensity resolutions for each sensor type are
           summarized in Table~\ref{tab:emtresolutionsummary}.} 
  \label{fig:emtresolution}
 \end{figure}

Figure~\ref{fig:emtresolution} shows the spill-by-spill detector signal size 
normalized to the proton beam intensity for the EMTs, Si center channel, and IC (Ar) center channel 
at a proton beam power of 450~kW and horn current of $+$250~kA.
From these distributions, the EMT intensity resolution is calculated to be 0.34\% for C3 
and 0.41\% for C4, while the Si and IC (Ar gas) center channels show 0.25\% and 0.24\% resolutions, respectively. 
The resolutions of each detector are summarized in Table~\ref{tab:emtresolutionsummary},
including the intensity resolution for only the first bunch. 
The statistical error on each value is quite small (less than 0.01\%).
The contribution from noise due to the read-out system including the cable, attenuator, and flash ADC was 
estimated by integrating the baseline pedestal before the beam spill and was found to be less than 0.1\%.
Compared to the current monitors, EMTs show slightly poor resolution.
However, simulation studies show that
a 1\% uncertainty on the intensity measurement for each sensor corresponds to 
a 0.06~mrad uncertainty in the beam direction measurement, 
which is much smaller than the total precision of 0.28~mrad.
In fact, the total intensity monitoring precision is limited by the calibration of the read-out system, 
the precision of which is a few \%.
Therefore, 
the intensity resolution of the EMTs satisfies the requirements for 
muon beam direction and flux measurements. 

 \begin{table}[htbp]
  \begin{center}
  \caption{Summary of the intensity resolution of each detector for 
           spill-by-spill and first bunch measurements.
           The proton beam intensity is 450 kW and the horn current
     setting is $+$250~kA, corresponding to a
     muon flux of 
     $3.2 \times 10^{6}$~${\rm /cm^2}$ per 80 ns beam bunch. 
     The statistical uncertainty is less than 0.01\%.} 
  \label{tab:emtresolutionsummary}
  \vspace{2truept}
   \begin{tabular}{c c c} \hline \hline
    Detector     & Spill (8 bunch sum) & First bunch \\ \hline
    EMTC3        & 0.34\%              & 0.73\%    \\
    EMTC4        & 0.41\%              & 0.78\%    \\ 
    Si center    & 0.25\%              & 0.37\%    \\
    IC (Ar) center & 0.24\%            & 0.33\%    \\ \hline \hline
   \end{tabular}
  \end{center}
 \end{table}

\newpage
\subsection{Linearity} 

A linear sensor response is desired to correctly monitor the beam at low power 
during beam tuning up to high power during physics operation.
In order to test the linearity performance of the EMTs, 
beam intensity scans were carried out. 
During scans, the proton beam power was set to 13 kW, 50 kW, 150 kW, 260 kW, 
340 kW, 400 kW, 460 kW, and 500 kW, and the horn current setting was $+$250~kA.
Two different voltages were applied to the EMTs to study space charge effects
inside the EMT, 
which occur when the number of produced electrons is so large that 
the electric field is significantly distorted, causing a
decreased signal yield, similar to the IC case described in Section~\ref{sec:t2kandmumon}.
We applied $-$450 V to reduce the number of secondary electrons, and compared 
the linearity performance to the case with $-$500~V applied.
The scan conditions are summarized in Table~\ref{tab:intensityscaninfo}.

 \begin{table}[htbp]
  \begin{center}
    \caption{Beam powers and EMT applied voltages during beam intensity scans. 
           Note that the the number of protons per kW at 
     a repetition rate of 2.48~s is $5.3 \times 10^{11}$ protons/spill/kW.}
  \label{tab:intensityscaninfo}
  \vspace{2truept}
   \begin{tabular}{c c c} \hline \hline
    Scan & Beam power [kW] & Applied HV [V] \\ \hline
    I   & 150, 260, 340, 400, 460 & $-$500      \\
    II  & 260, 340, 400, 460      & $-$450      \\ 
    III & 13, 50              & $-$500, $-$450  \\
    IV  & 500                 & $-$500          \\ \hline \hline
   \end{tabular}
  \end{center}
 \end{table}

\begin{figure}[htbp]
  \begin{center}
   \includegraphics[clip,width=8.5cm]{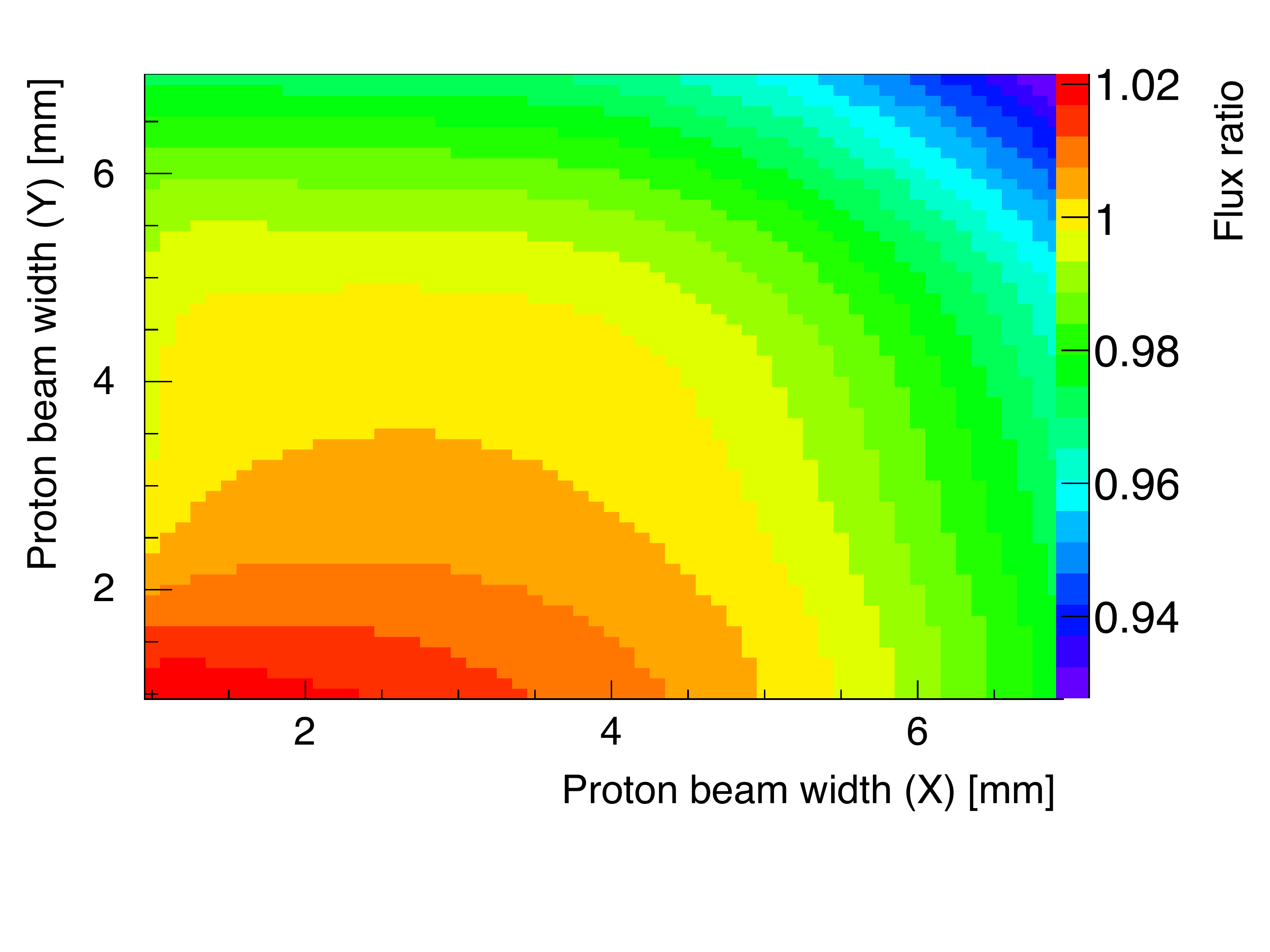}
  \end{center}
  \vspace{-30truept}
   \caption{Ratio of the simulated muon flux at the EMT position for various proton beam widths 
           to the simulated muon flux for beam width $= (4, 4)$~mm.} 
  \label{fig:relativeflux}
 \end{figure}

Since the muon yield is correlated with the horn current, 
a correction to the calculated yield is applied using horn current monitor information. 
The correction factor was determined based on horn current scans
(see Ref.~\cite{suzuki} for details about the correction).
During the intensity scans, the proton beam position was kept within $\pm$0.5~cm, 
monitored by SSEM19 and 18, and the effect on the muon flux by this fluctuation is less than 1\%. 
Therefore, no correction for the proton beam
position is needed. 
%
The proton beam width changes depending on the beam power, 
  from 2~mm (1.5~mm) at the lowest power to 4.8~mm (4~mm) at the highest 
  power in the horizontal (vertical) direction,
which does affect the muon flux.
A Monte Carlo (MC) simulation \cite{flux} was conducted to estimate the change of the muon flux at the 
EMT position for various values of the beam width, and the result is shown in 
Figure~\ref{fig:relativeflux}. 
The measured yields were corrected by factors determined from the simulation results to 
extract the expected yields at the nominal beam size (4~mm in both 
the horizontal and vertical directions).
During the beam intensity scan, the beam width values 
were taken from SSEM18, since SSEM19 had several unreliable points 
at very low beam powers, 
where the beam width was too narrow for the monitor sensitivity
and profile reconstruction was therefore not possible. 
The consistency between SSEM19 and 18 has been confirmed by other, reliable, points.
To check the width correction validity, two different beam widths, 
nominal 2.6~mm (2.2~mm) and 4.2~mm (3.2~mm) for the horizontal (vertical) direction, 
were used at the 150~kW point ($\sim 80 \times 10^{12}$ protons/spill) in Scan I. 

\begin{figure}[htbp]
  \begin{minipage}{0.5\hsize}
   \begin{center}
    \includegraphics[clip,width=7.8cm]{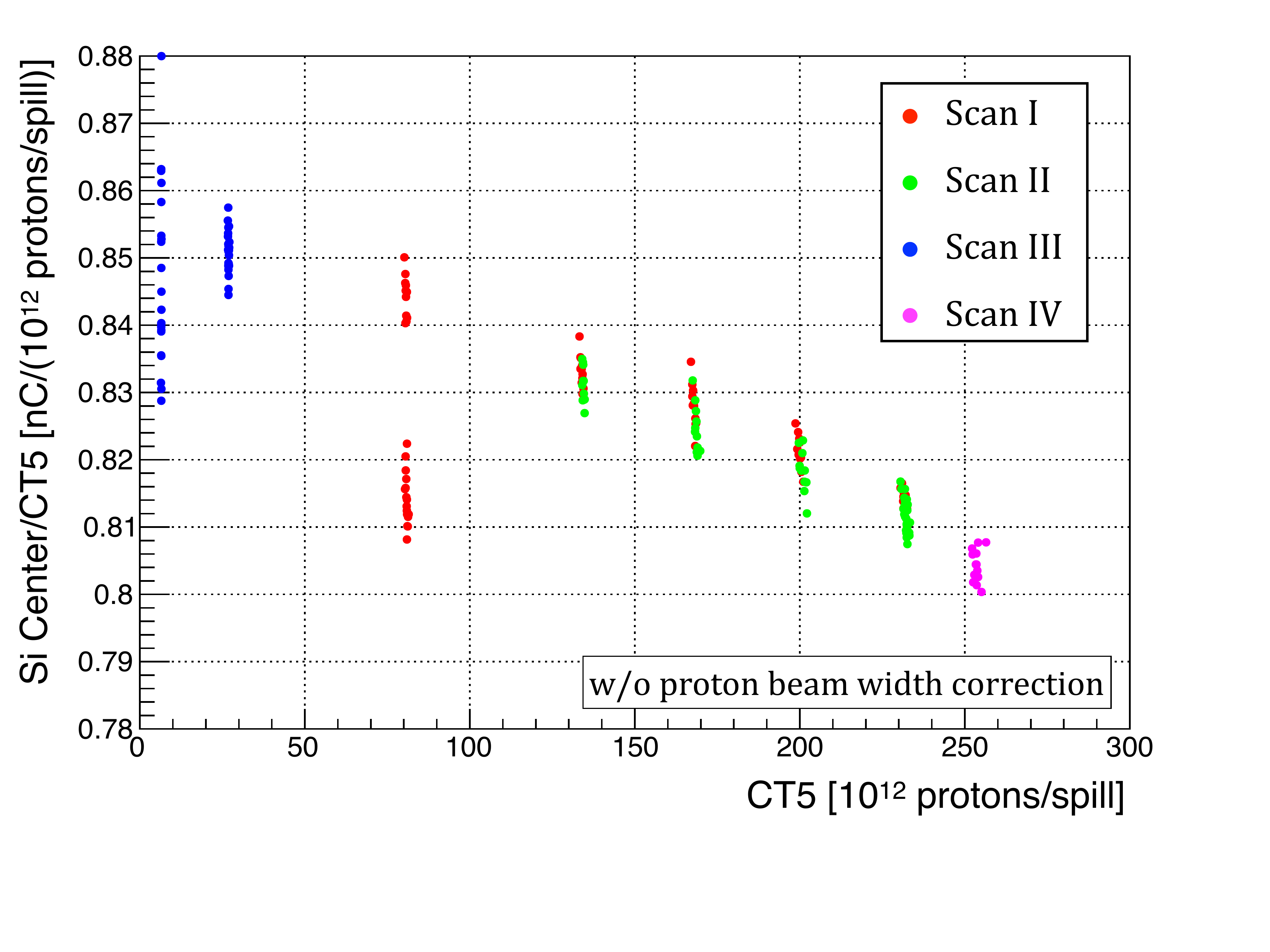}
   \end{center}
  \end{minipage}
  \begin{minipage}{0.5\hsize}
   \begin{center}
    \includegraphics[clip,width=7.8cm]{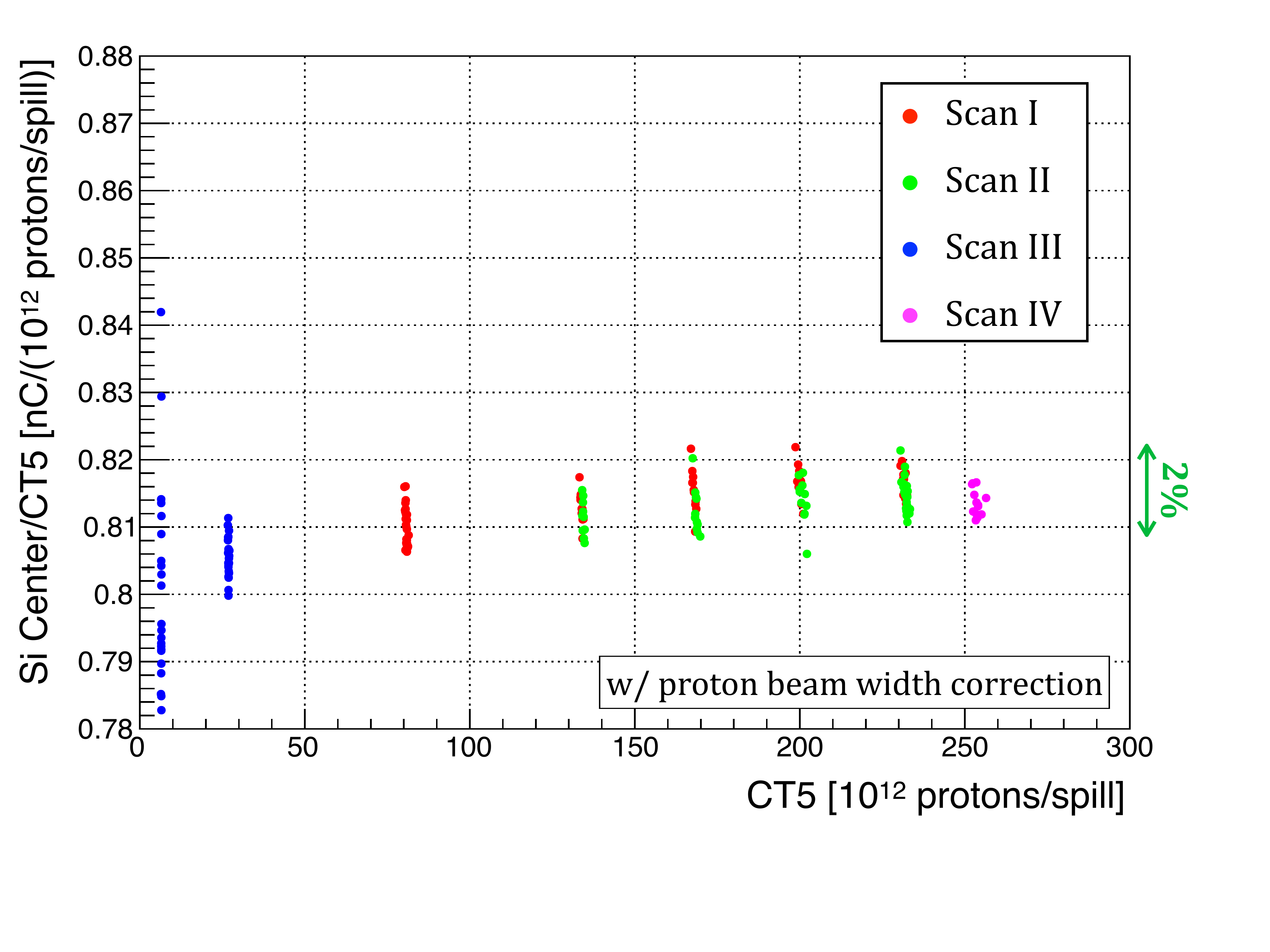}
   \end{center}
  \end{minipage}
  \vspace{-25truept}
  \caption{Results of the beam intensity scan for the Si center channel before (left) 
           and after (right) the proton beam width correction.
     Different colors correspond to different scans,
	   as shown in Table~\ref{tab:intensityscaninfo}.}
  \label{fig:silinearity}
 \end{figure}

Figure~\ref{fig:silinearity} shows the scan results for the Si center channel.
Here, Si is selected because IC shows non-linearity as described in Section~\ref{sec:t2kandmumon}.
The plot in the left panel shows the result before the beam width correction, 
and the plot in the right panel shows that after the correction.
The horn current correction is applied for both results.
At low intensity, points are more scattered due to a poor signal-to-noise ratio because of 
the small signal size.
Therefore, the linearity at scan points above 100~kW ($\sim 50 \times 10^{12}$ protons/spill)
is considered more reliable. 
It is found that the beam width correction works very well, especially for the 150~kW points 
($\sim 80 \times 10^{12}$ protons/spill) in Scan I, where, before the MC correction is applied,
the distributions are separated due to the different beam widths.
The linearity is stable 
within $\pm$1\% from 100~kW to 500~kW.

\begin{figure}[h]
  \begin{minipage}{0.5\hsize}
   \begin{center}
    \includegraphics[clip,width=7.8cm]{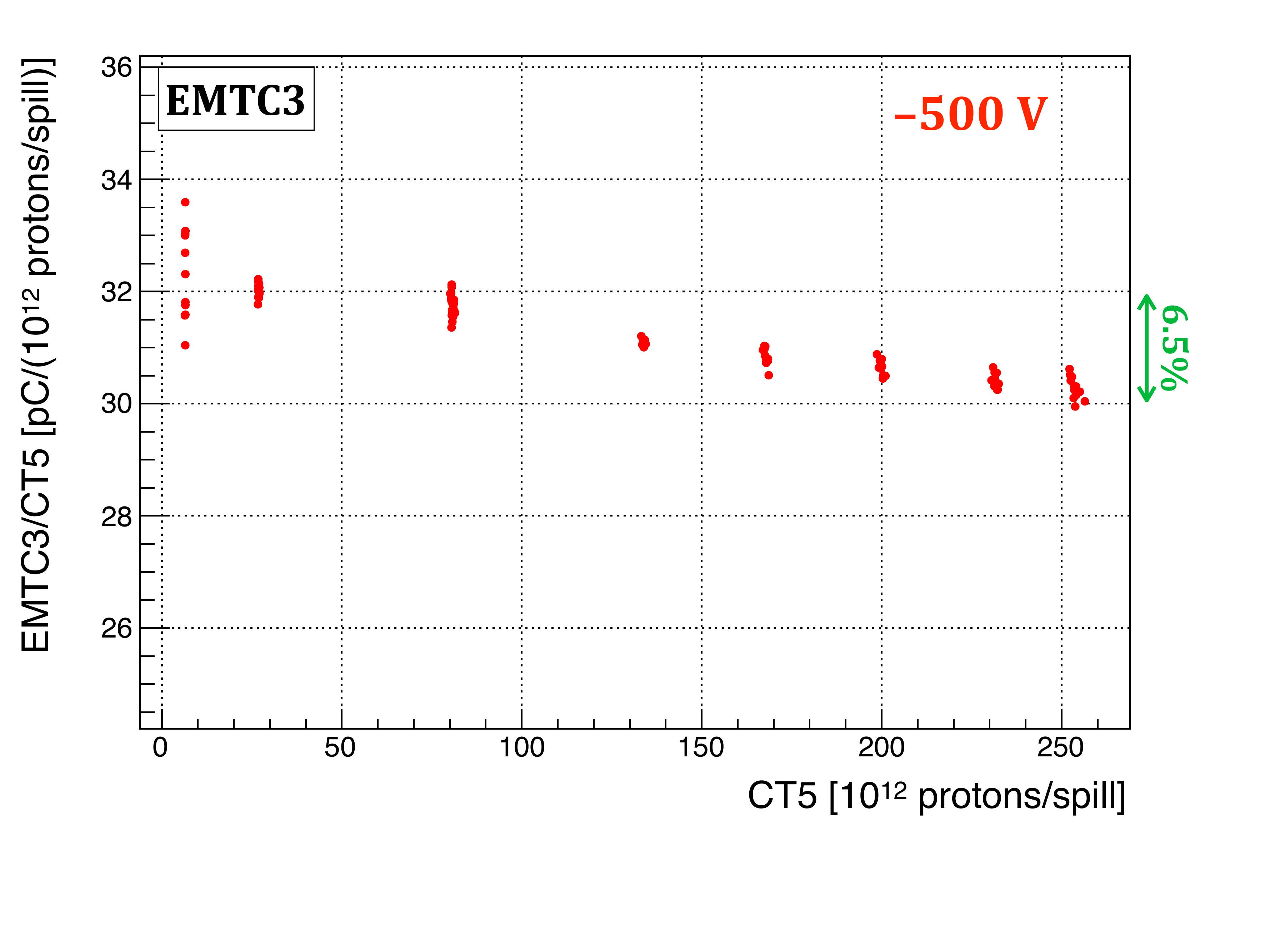}
   \end{center}
  \end{minipage}
  \begin{minipage}{0.5\hsize}
   \begin{center}
    \includegraphics[clip,width=7.8cm]{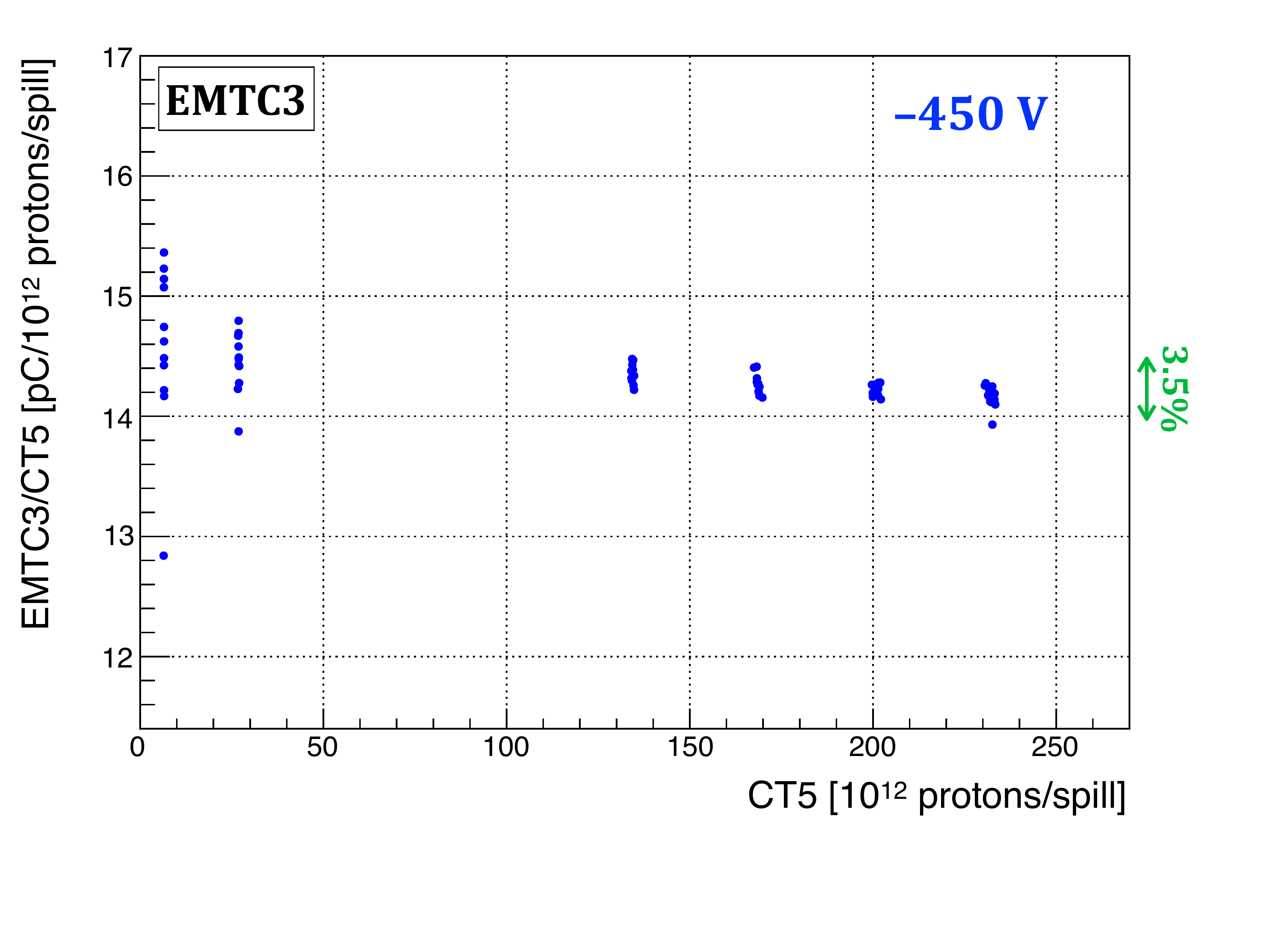}
   \end{center}
  \end{minipage}
  \vspace{-25truept}
  \caption{Results of the beam intensity scan for EMTC3. 
           The left panel shows the result with $-$500~V 
           applied, and 
           the right panel shows the result with $-$450~V applied.}
  \label{fig:emtc3linearity}
 \end{figure}

\begin{figure}[h]
  \begin{minipage}{0.5\hsize}
   \begin{center}
    \includegraphics[clip,width=7.8cm]{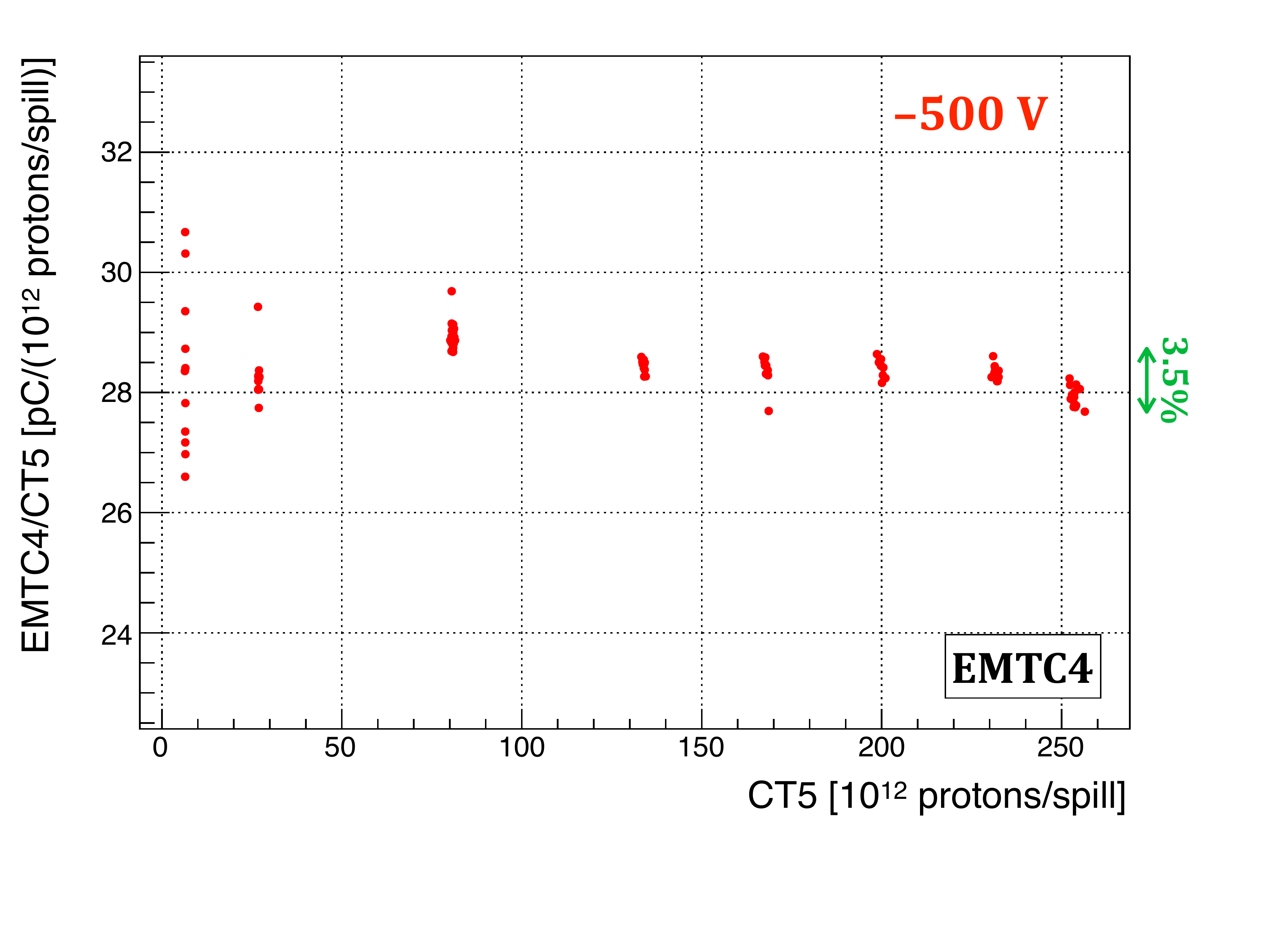}
   \end{center}
  \end{minipage}
  \begin{minipage}{0.5\hsize}
   \begin{center}
    \includegraphics[clip,width=7.8cm]{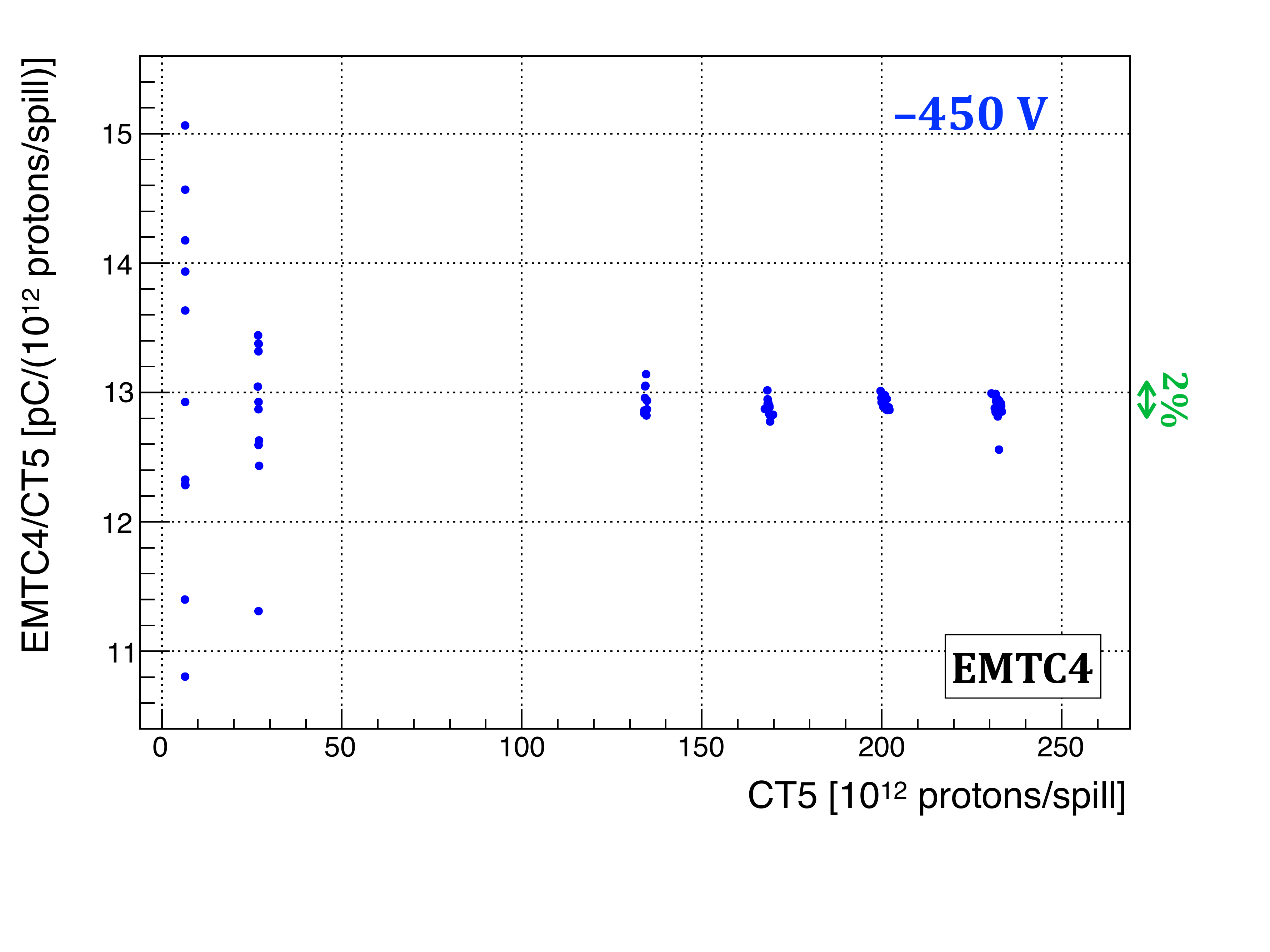}
   \end{center}
  \end{minipage}
  \vspace{-25truept}
  \caption{Results of the beam intensity scan for EMTC4. 
           The left panel shows the result with $-$500 V 
           applied, and 
           the right panel shows the result with $-$450~V applied.}
  \label{fig:emtc4linearity}
 \end{figure}

The linearity results for EMTC3 and C4 are shown in Figures~\ref{fig:emtc3linearity} and 
\ref{fig:emtc4linearity}, respectively.
The results use both horn current and proton beam width corrections.
The plots in the left panels show the results with $-$500 V high voltage (HV) applied, while 
those in the right panels show the results with $-$450 V
applied. 
EMTC4 shows better linearity performance than EMTC3,  
because the capacitance used in the circuit for C4 was improved as
described 
in Section~\ref{sec:developemt}.
Measurements with lower voltage applied show better linearity, and this indicates that 
space charge effects exist when a higher voltage is applied.
EMTC4 with $-$450 V applied shows as good linearity as the Si sensor up to 460 kW 
(muon flux of $3.3 \times 10^{6}$~${\rm /cm^2}$
per 80 ns beam spill).
At higher intensities, non-linearities may appear.
To further improve linearity, use of a different dynode divider 
ratio can be considered, with a possible compromise on signal size. 

\clearpage
\subsection{Stability} 

The muon monitor must work stably during
continuous running.
Yield fluctuations are required to be within 3\%. 
The EMTs were continuously exposed to the T2K muon beam over five data-taking periods. 
The beam conditions and applied HVs for each period are 
summarized in Table~\ref{tab:stabilityperiod}. 
Periods I and II are separated because EMTC4 was installed between them. 
EMTC3 was installed before Period I, when several tests, 
such as HV and attenuation level tuning, were performed. 
Therefore, the EMTC3 stability during the period before Period I is not shown here.  
There is an approximately half-year beam-off period between Periods II and III, 
during which the EMTs' HVs were turned off. 
Other than that, we turned off the HV of both EMTs for just one day in the middle of 
Period V in order to check the stability.
%

 \begin{table}[htbp]
  \begin{center}
    \caption{Beam conditions and HVs for EMTC3 and C4 during each
    beam exposure period.}
  \label{tab:stabilityperiod}
  \vspace{2truept}
   \begin{tabular}{l c c c} \hline \hline
    Period                                     & Horn current [kA] & HV [V] & POT [$\times 10^{18}$] \\ \hline
    I \ \ (23, Feb., 2017 $\sim$ 30, Mar., 2017) & $+$250            & $-$500         & 224.1                  \\
    II \ (31, Mar., 2017  $\sim$ 12, Apr., 2017) & $+$250            & $-$500         & 76.8                   \\
    III (16, Oct., 2017   $\sim$ 22, Oct., 2017) & $+$250            & $-$505         & 20.5                   \\ 
    IV  (22, Oct., 2017   $\sim$ 2, Nov., 2017)  & $-$250            & $-$505         & 59.4                   \\
    V \  (2, Nov., 2017   $\sim$ 22, Dec., 2017) & $-$250            & $-$450         & 305.8                  \\ \hline \hline
   \end{tabular}
  \end{center}
 \end{table}

\begin{figure}[h]
  \hspace{-50truept}
  \begin{minipage}{0.5\hsize}
   \begin{center}
    \includegraphics[clip,width=9.0cm]{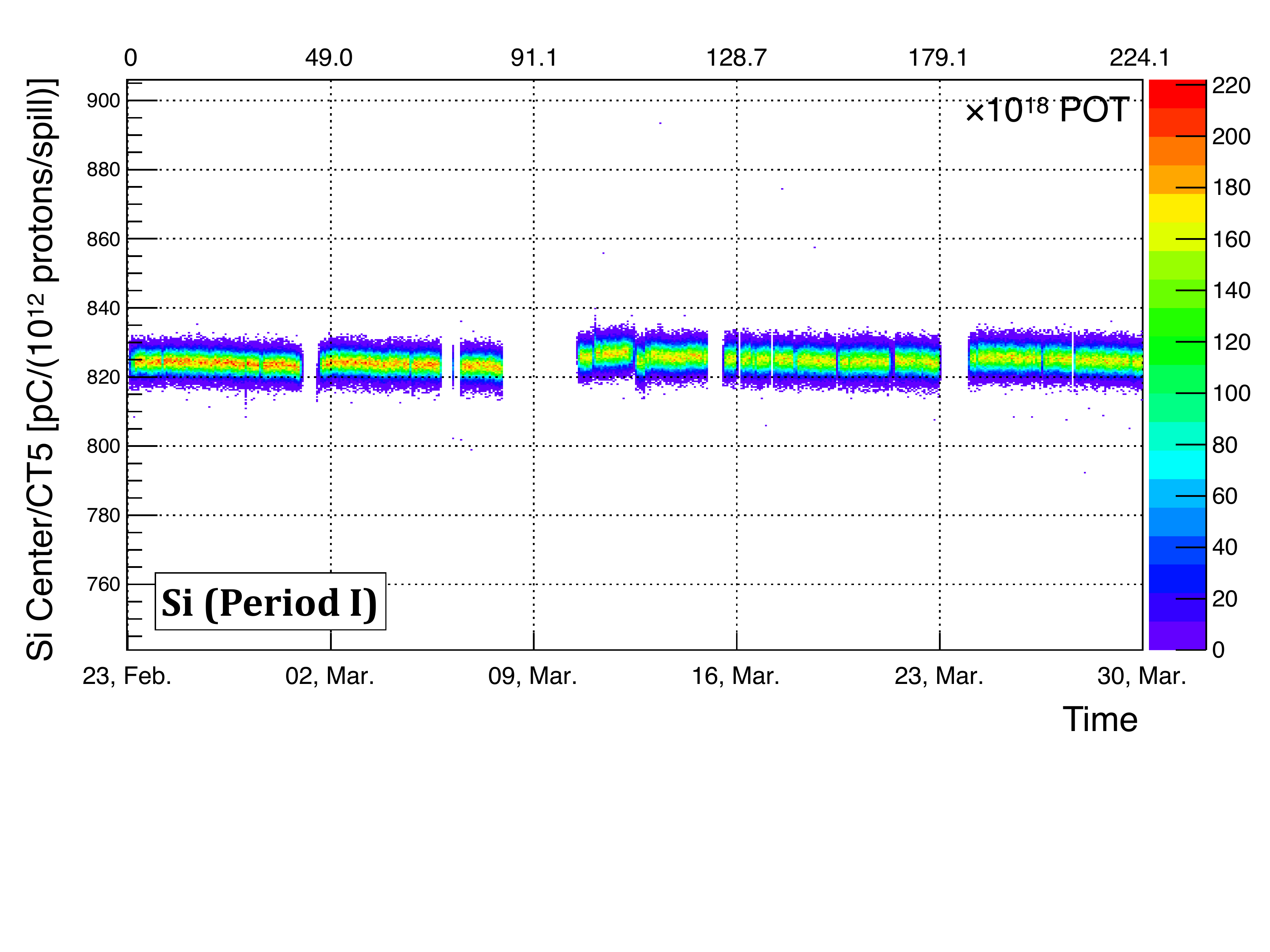}
   \end{center}
  \end{minipage}
  \hspace{45truept}
  \begin{minipage}{0.5\hsize}
   \begin{center}
    \includegraphics[clip,width=9.0cm]{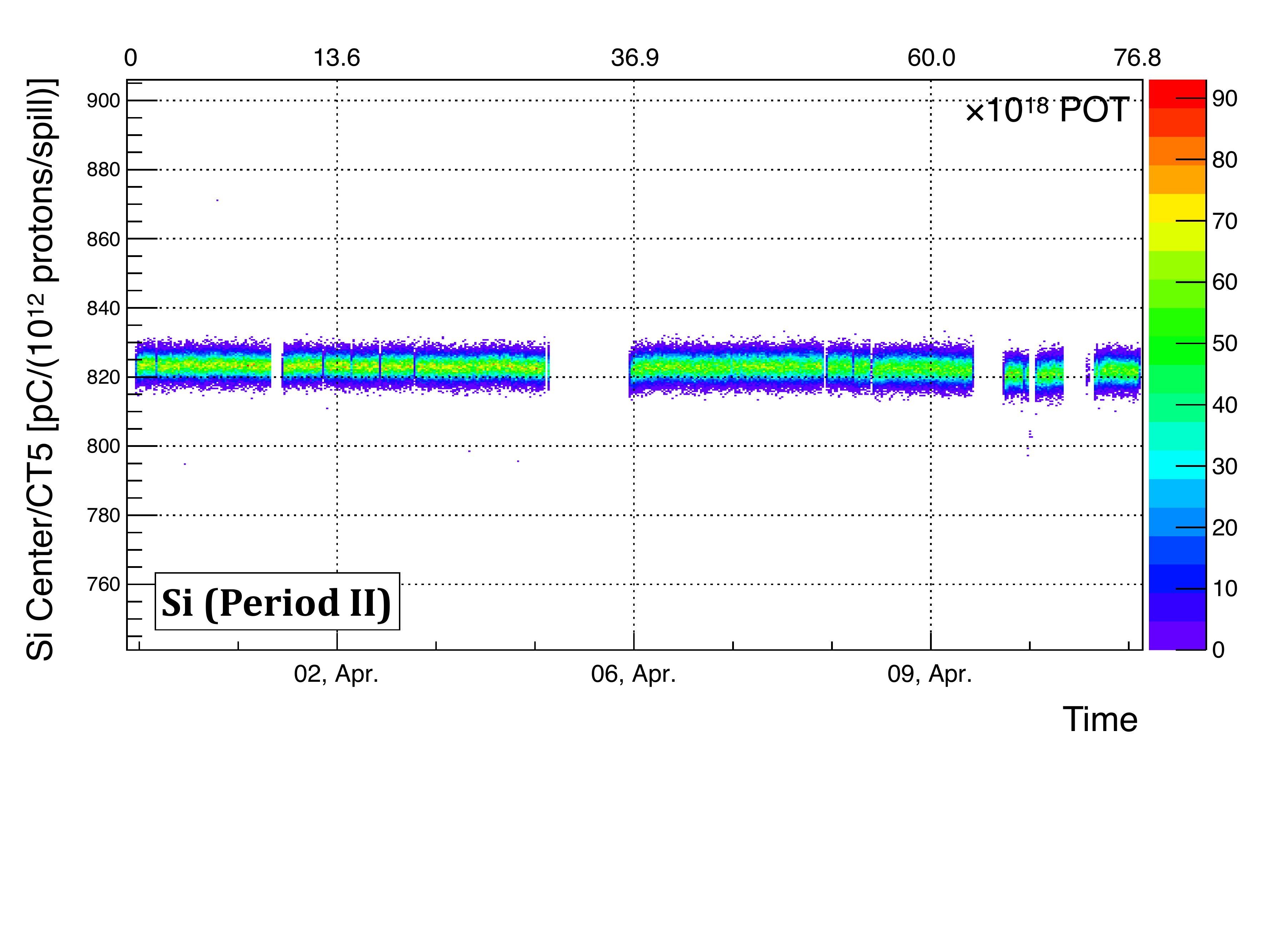}
   \end{center}
  \end{minipage}
  \\
  \noindent
  \vspace{20truept}
  \hspace{-50truept}
  \begin{minipage}{0.5\hsize}
   \vspace{-10truept}
   \begin{center}
    \includegraphics[clip,width=9.0cm]{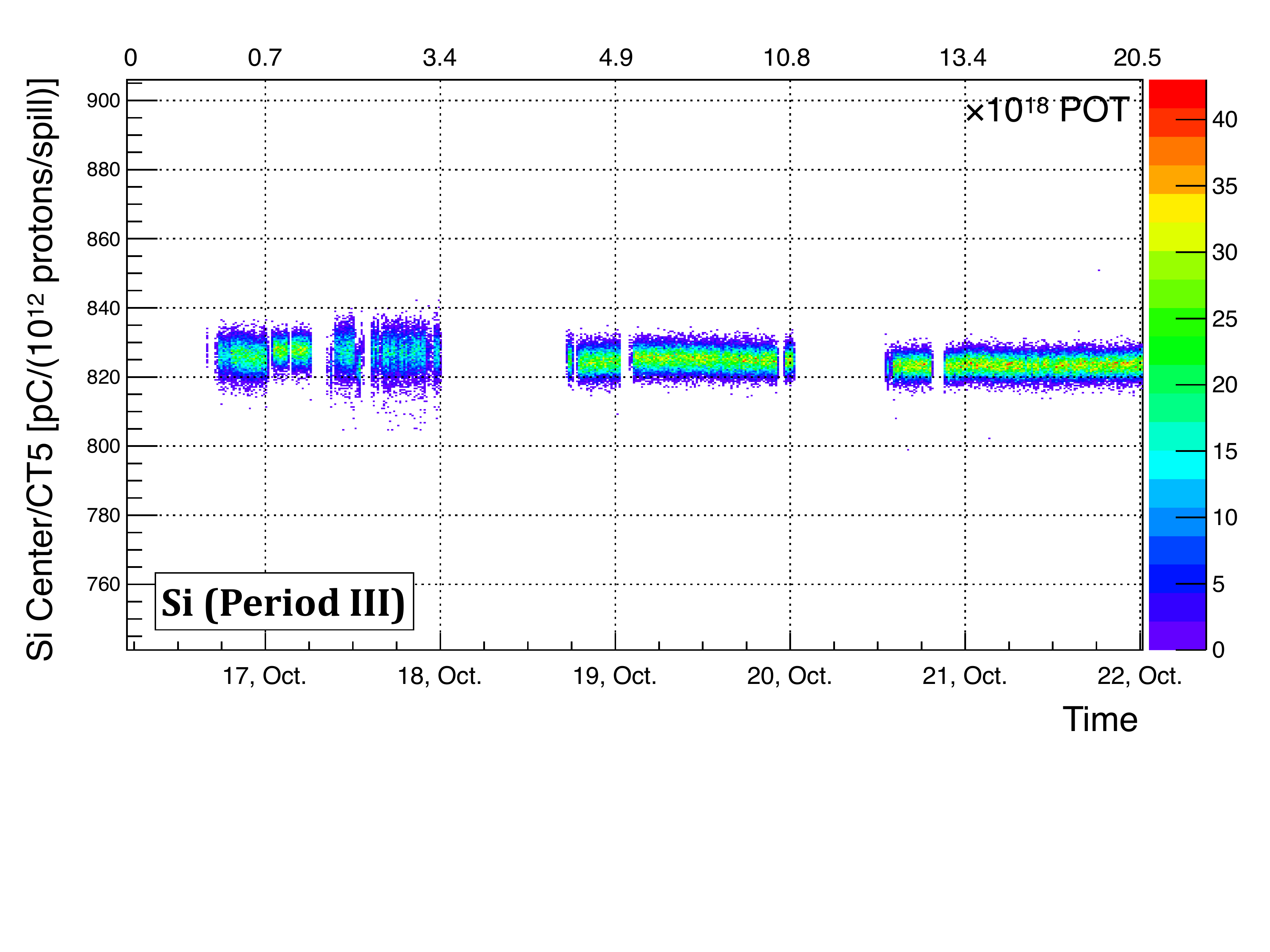}
   \end{center}
  \end{minipage}
  \hspace{45truept}
  \begin{minipage}{0.5\hsize}
   \vspace{-10truept}
   \begin{center}
    \includegraphics[clip,width=9.0cm]{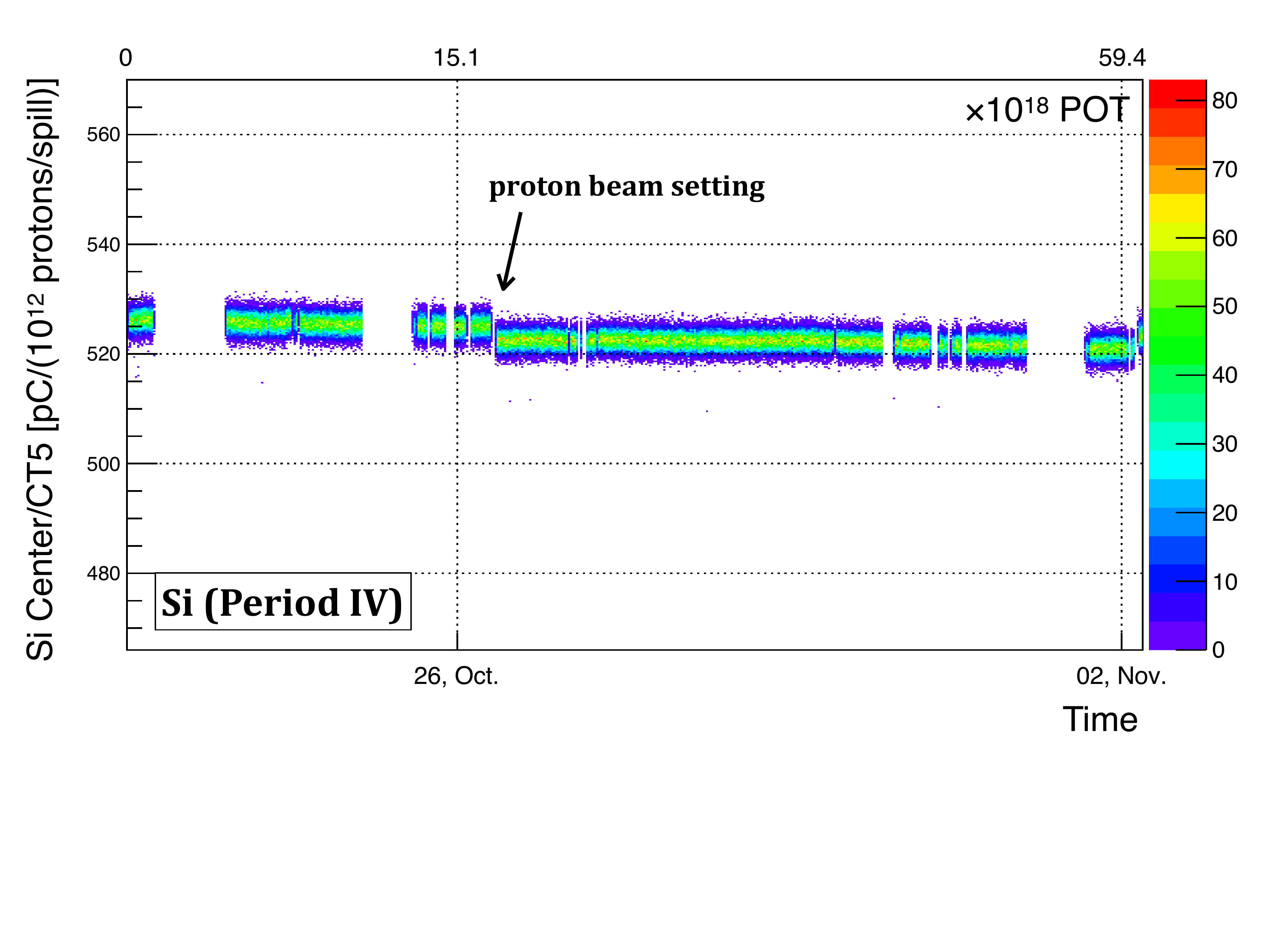}
   \end{center}
  \end{minipage}
  \\
  \noindent
  \hspace{-50truept}
  \vspace{-10truept}
  \begin{center}
   \includegraphics[clip,width=9.0cm]{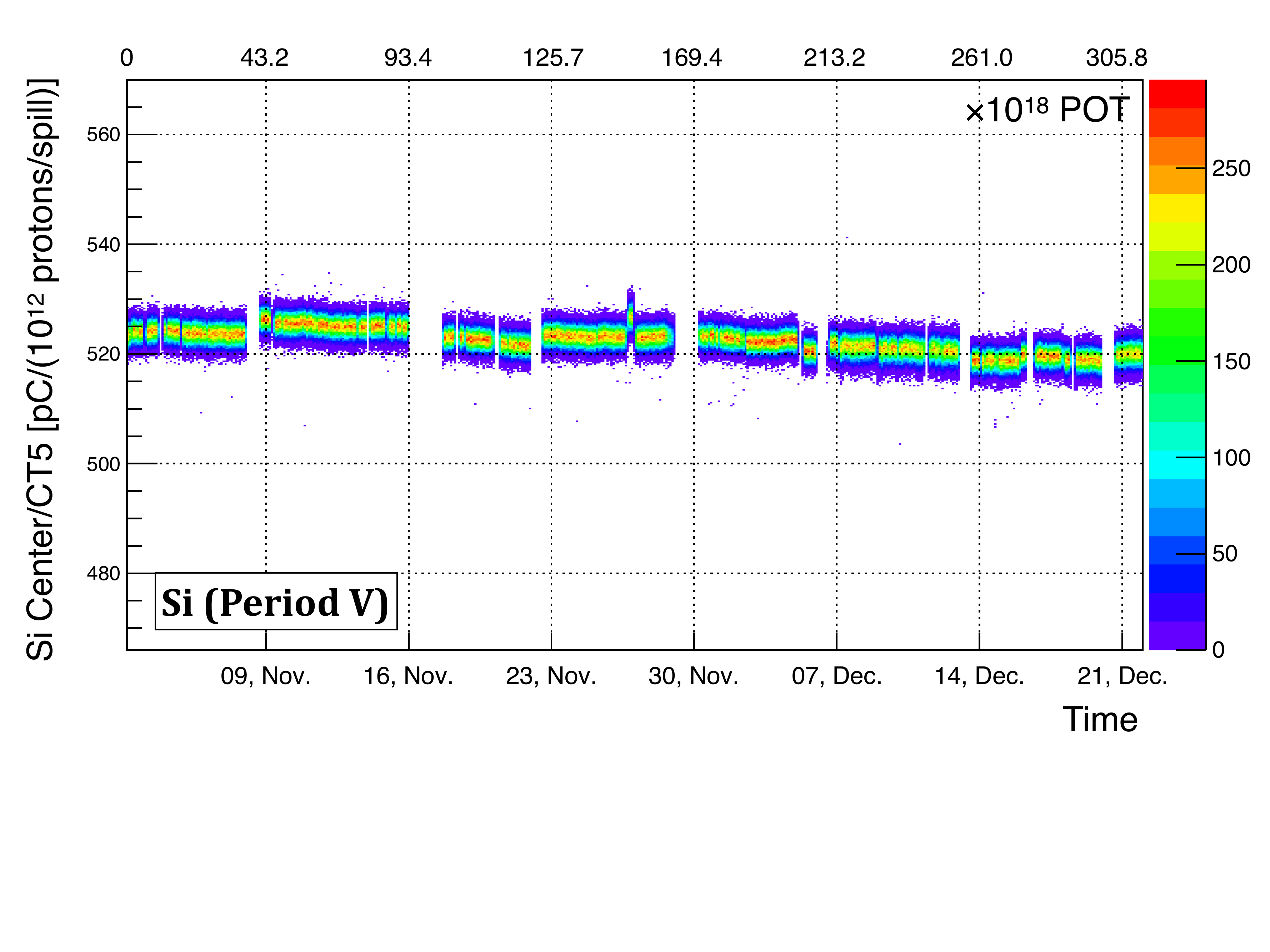}
  \end{center}
  \vspace{-20truept}
  \caption{Signal yield of Si center channel as a function of time.}
  \label{fig:sistability}
 \end{figure}

\begin{figure}[h]
  \hspace{-50truept}
  \begin{minipage}{0.5\hsize}
   \begin{center}
    \includegraphics[clip,width=9.0cm]{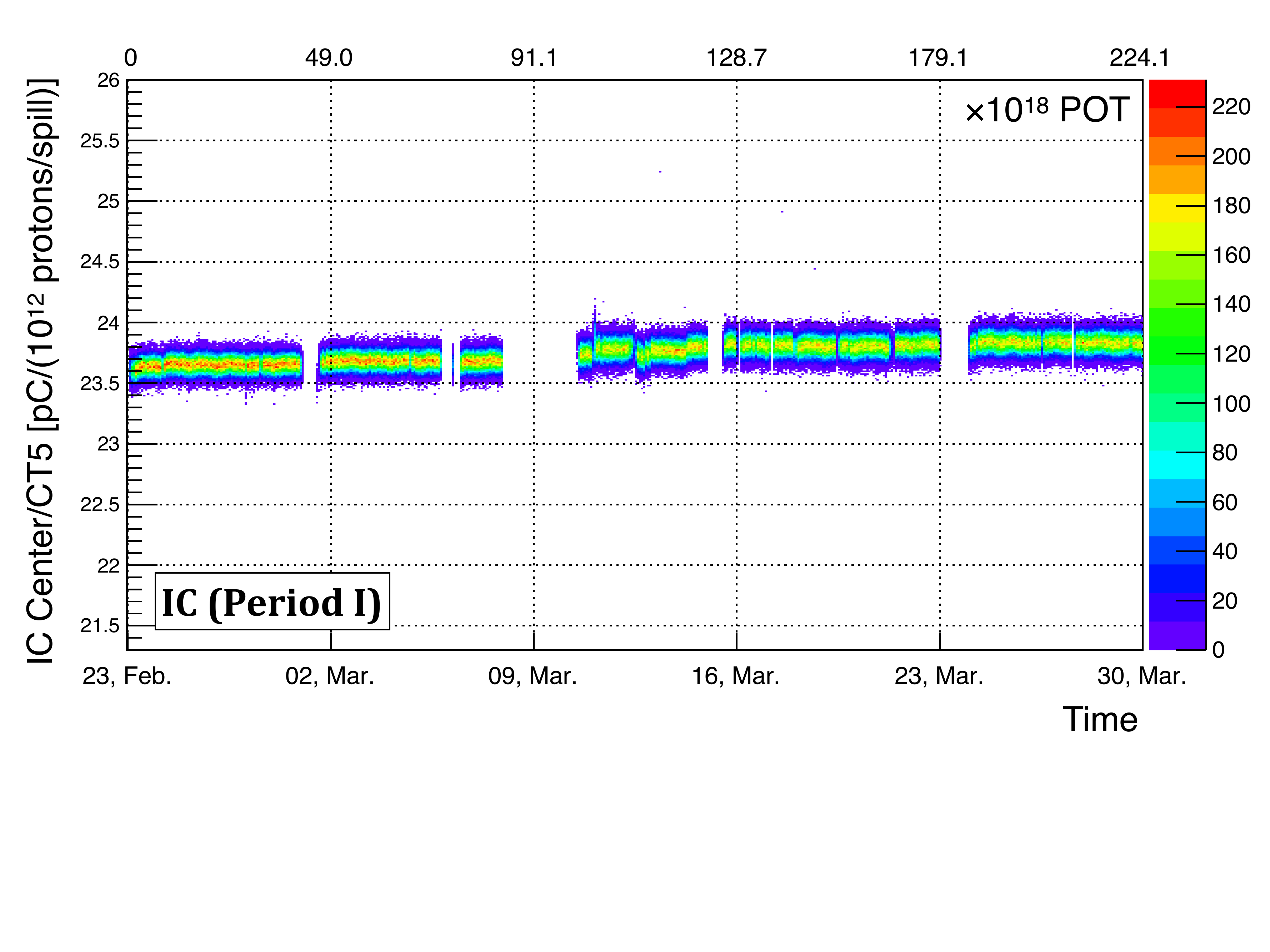}
   \end{center}
  \end{minipage}
  \hspace{45truept}
  \begin{minipage}{0.5\hsize}
   \begin{center}
    \includegraphics[clip,width=9.0cm]{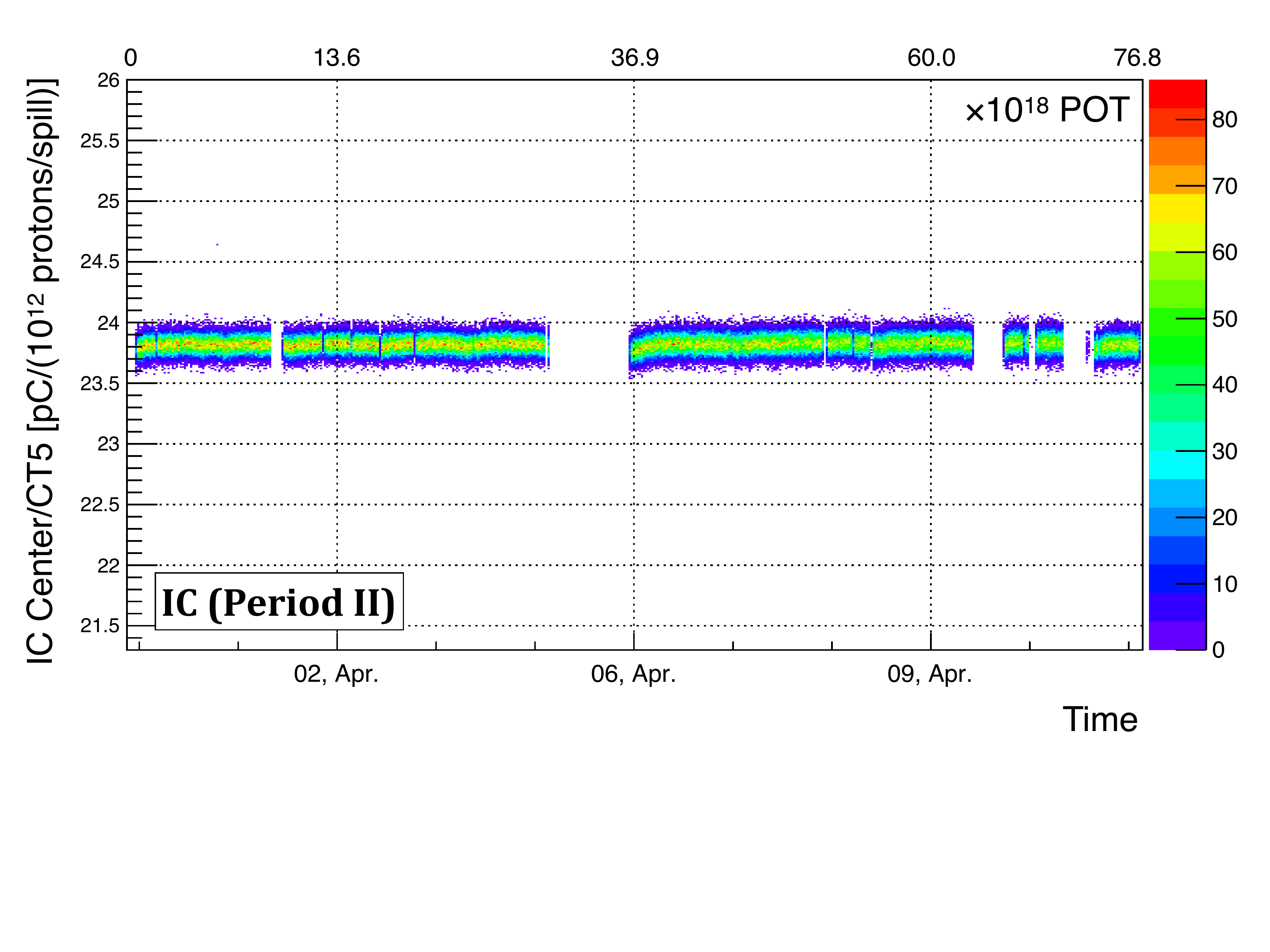}
   \end{center}
  \end{minipage}
  \\
  \noindent
  \vspace{20truept}
  \hspace{-50truept}
  \begin{minipage}{0.5\hsize}
   \vspace{-10truept}
   \begin{center}
    \includegraphics[clip,width=9.0cm]{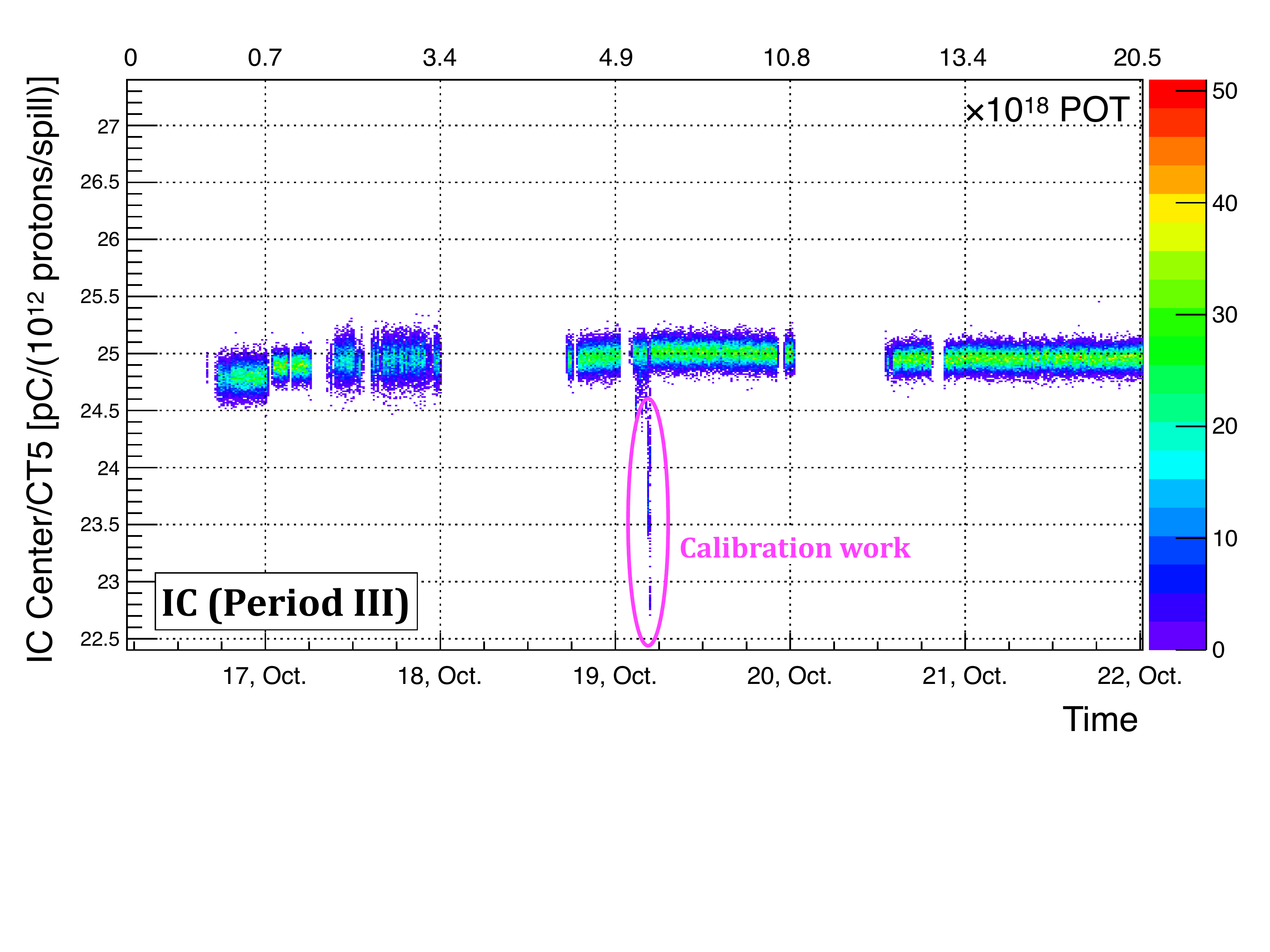}
   \end{center}
  \end{minipage}
  \hspace{45truept}
  \begin{minipage}{0.5\hsize}
   \vspace{-10truept}
   \begin{center}
    \includegraphics[clip,width=9.0cm]{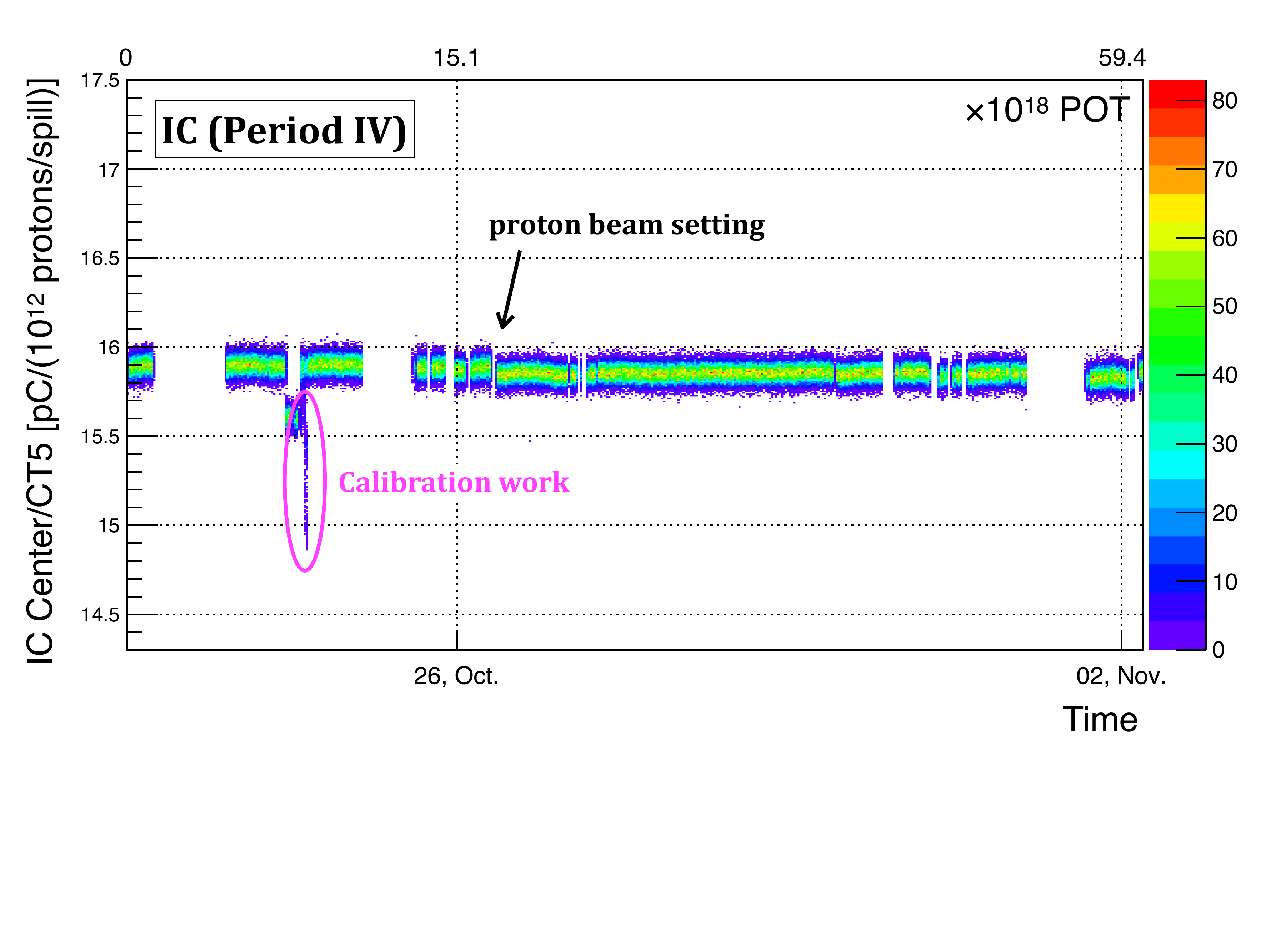}
   \end{center}
  \end{minipage}
  \\
  \noindent
  \hspace{-50truept}
  \vspace{-10truept}
  \begin{center}
   \includegraphics[clip,width=9.0cm]{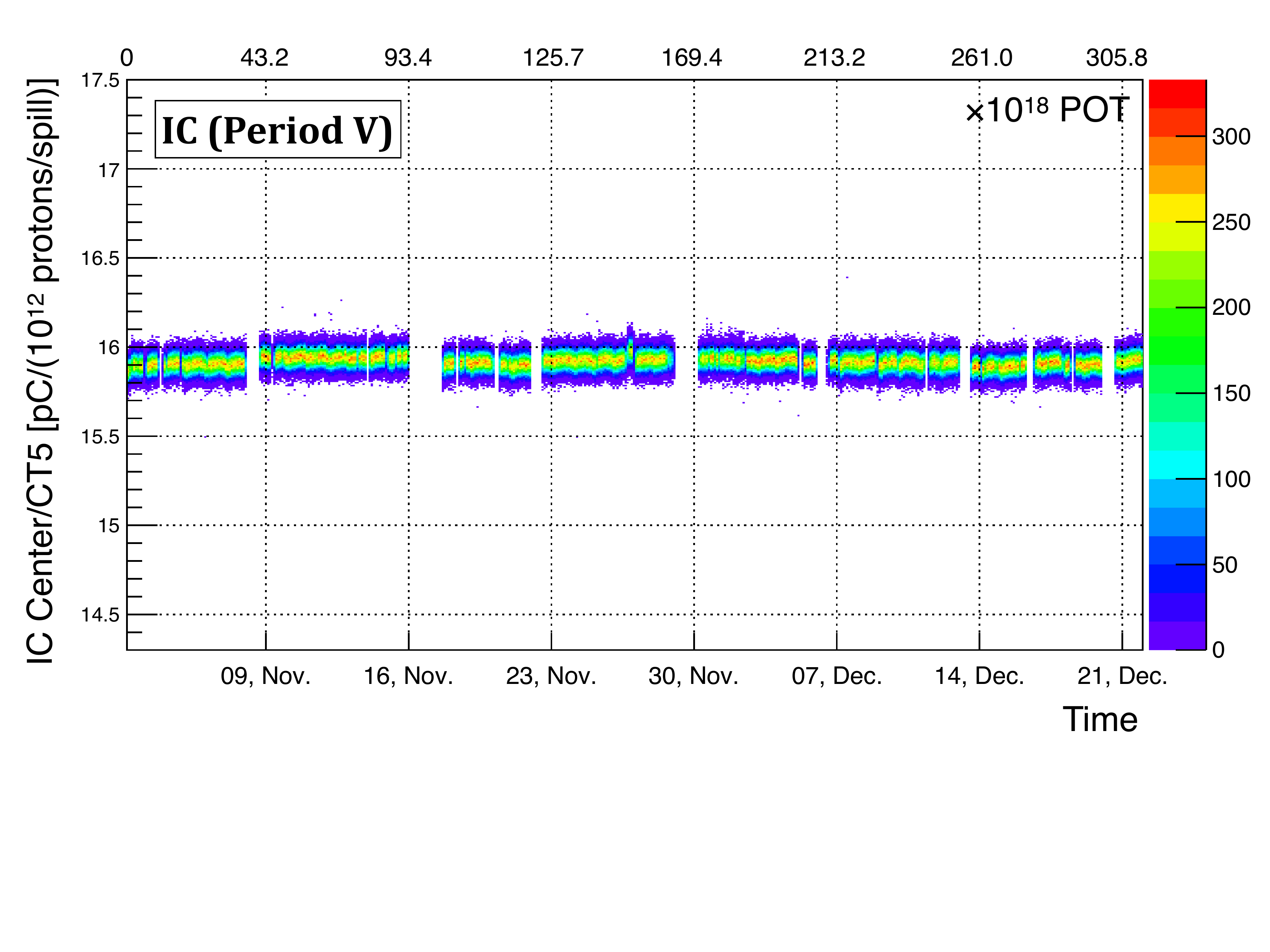}
  \end{center}
  \vspace{-20truept}
  \caption{Signal yield of IC (Ar) center channel as a function of time.
	   The yield jumps seen in Periods III and IV, marked with magenta circles, 
	   are due to calibration work where the entire IC system was moved 
	   (see Ref.~\cite{suzuki} for the calibration method).}
  \label{fig:icstability}
 \end{figure}

\begin{figure}[h]
  \hspace{-50truept}
  \begin{minipage}{0.5\hsize}
   \begin{center}
    \includegraphics[clip,width=9.0cm]{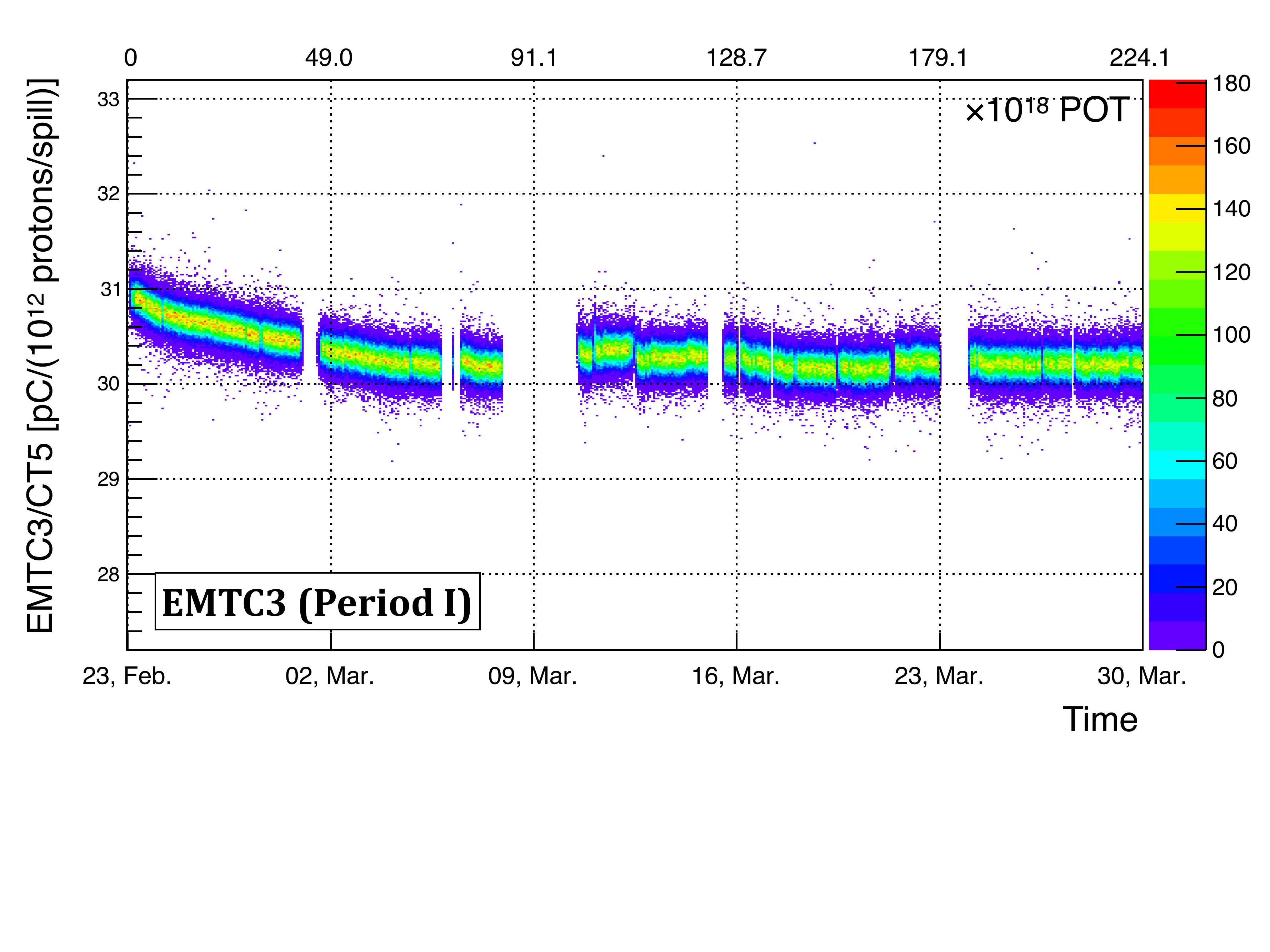}
   \end{center}
  \end{minipage}
  \hspace{45truept}
  \begin{minipage}{0.5\hsize}
   \begin{center}
    \includegraphics[clip,width=9.0cm]{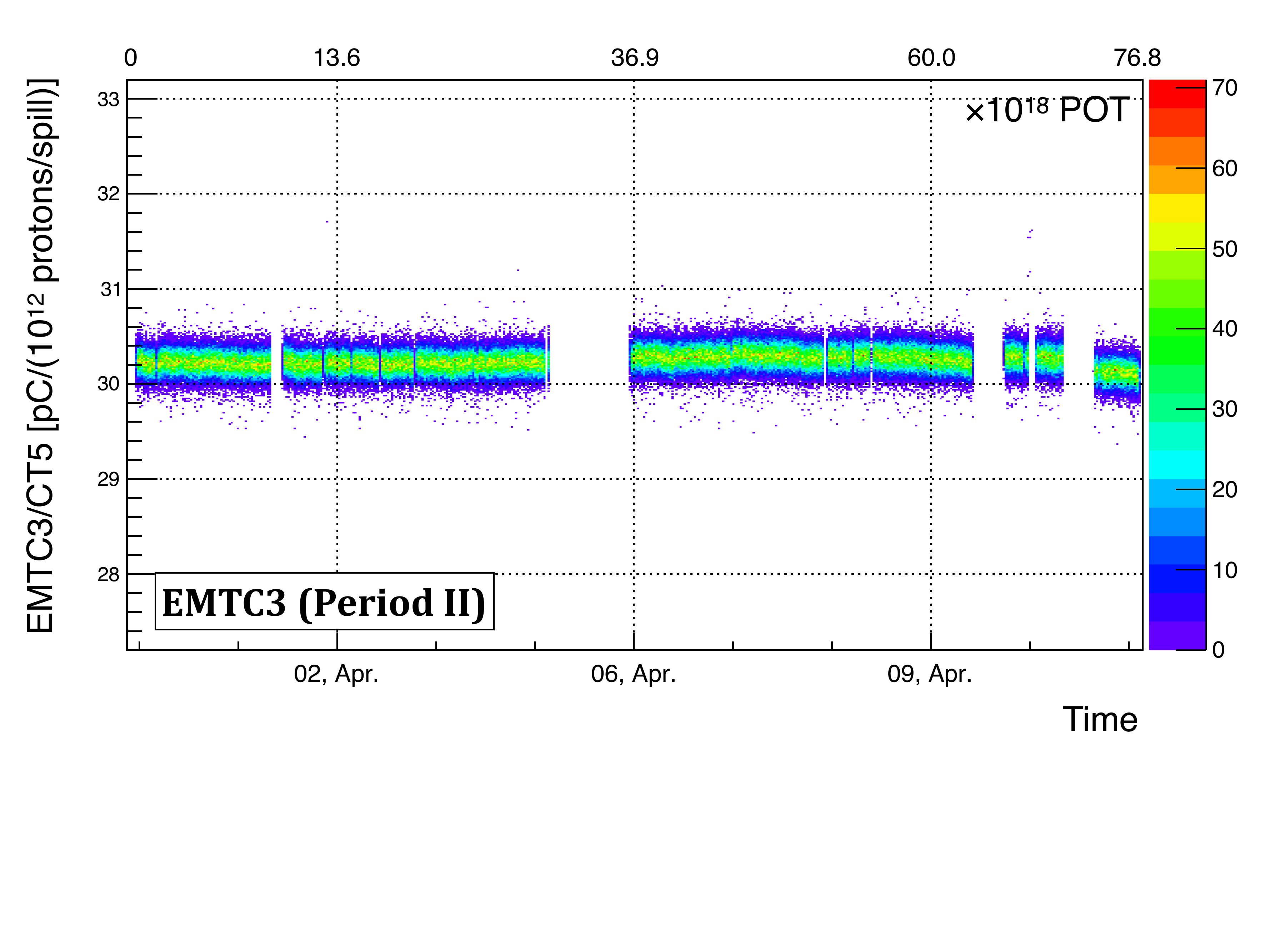}
   \end{center}
  \end{minipage}
  \\
  \noindent
  \vspace{20truept}
  \hspace{-50truept}
  \begin{minipage}{0.5\hsize}
   \vspace{-10truept}
   \begin{center}
    \includegraphics[clip,width=9.0cm]{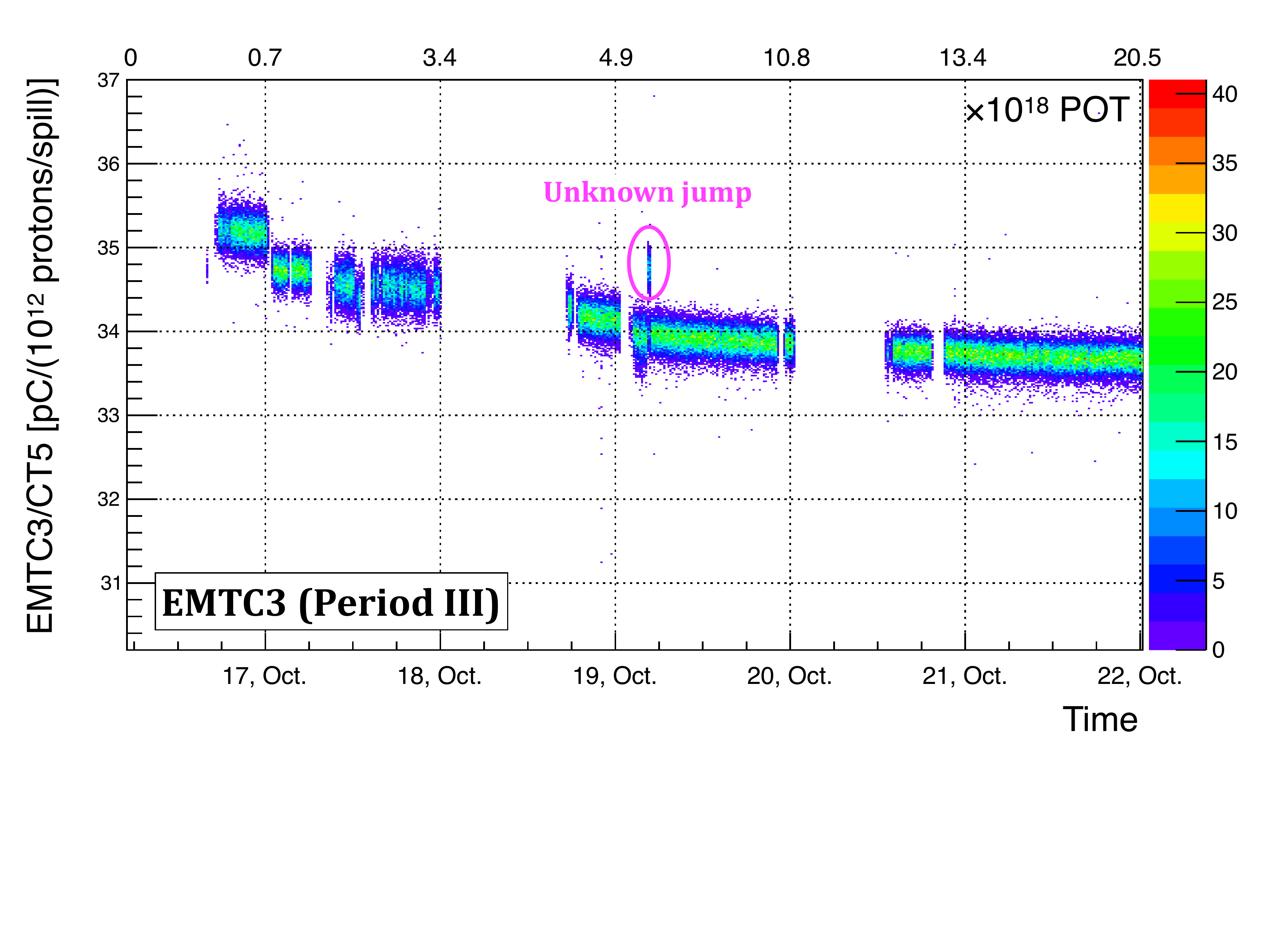}
   \end{center}
  \end{minipage}
  \hspace{45truept}
  \begin{minipage}{0.5\hsize}
   \vspace{-10truept}
   \begin{center}
    \includegraphics[clip,width=9.0cm]{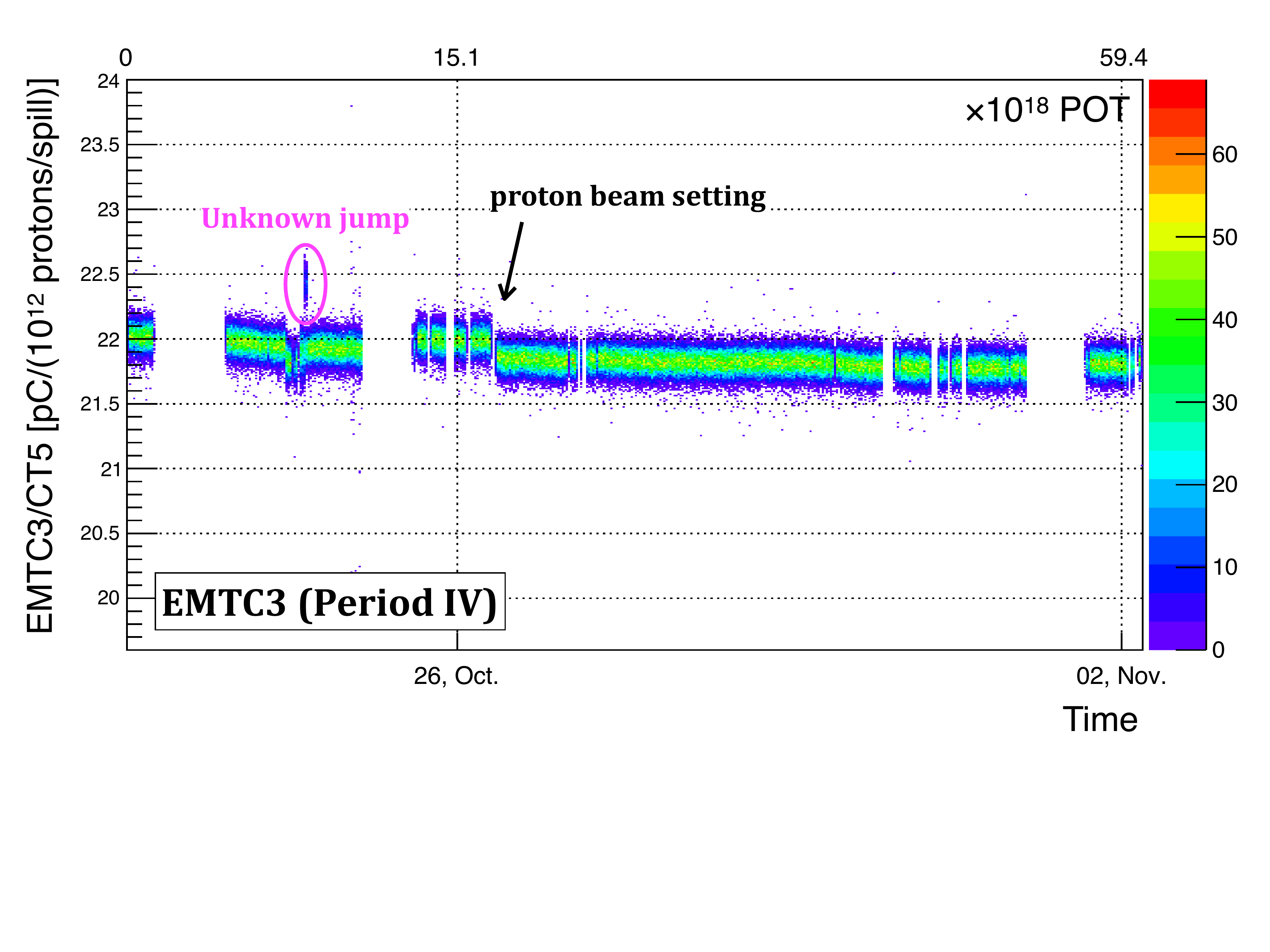}
   \end{center}
  \end{minipage}
  \\
  \noindent
  \hspace{-50truept}
  \vspace{-10truept}
  \begin{center}
   \includegraphics[clip,width=9.0cm]{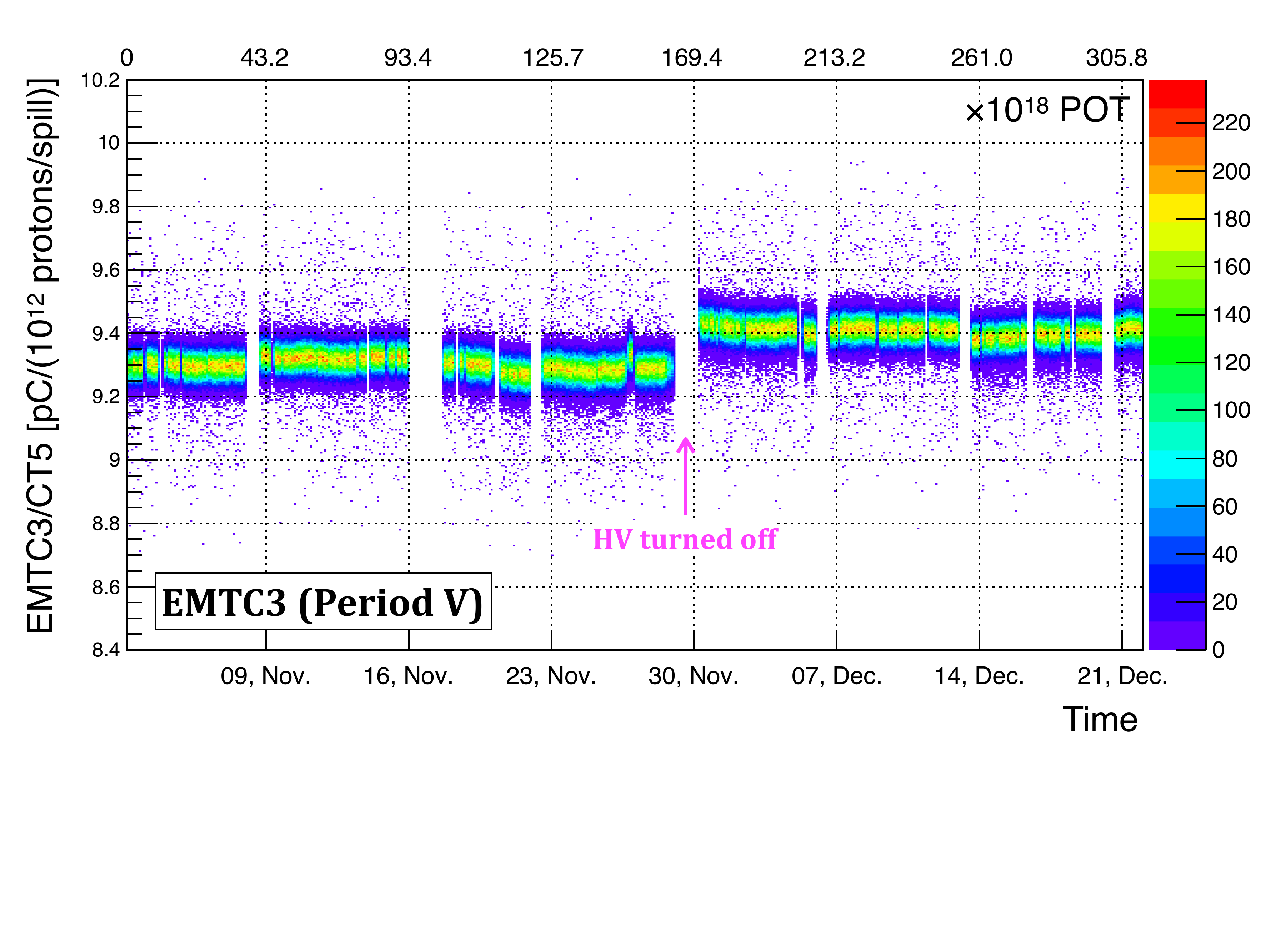}
  \end{center}
  \vspace{-20truept}
  \caption{Signal yield of EMTC3 as a function of time.
           Two yield jumps are seen and seem to be synchronized with 
	   IC calibration work, although the cause is not fully understood.
     After a short HV-off period during Period V, the yield changed.}
  \label{fig:emtc3stability}
 \end{figure}

\begin{figure}[h]
  \hspace{-50truept}
  \begin{minipage}{0.5\hsize}
   \begin{center}
    \includegraphics[clip,width=9.0cm]{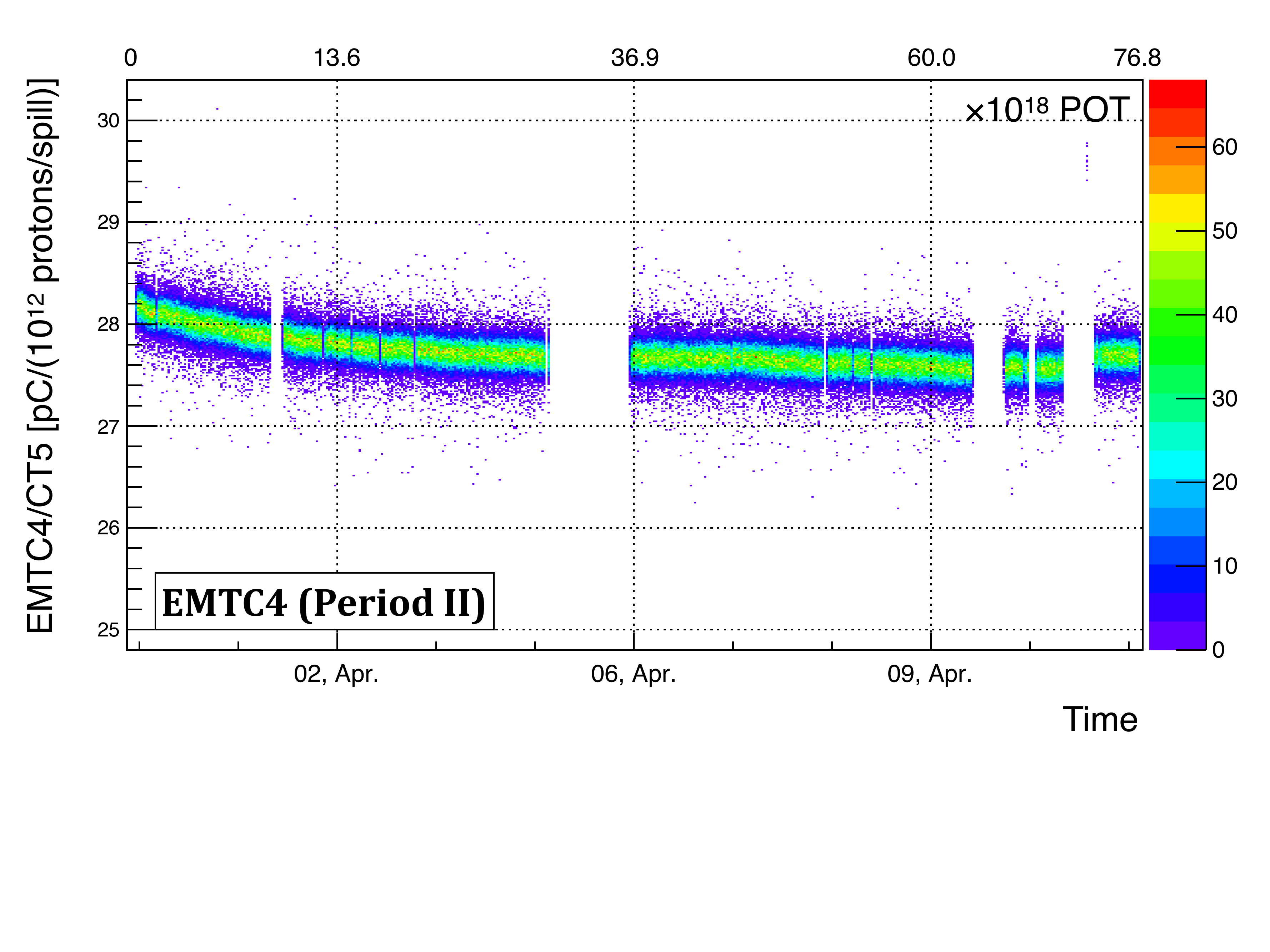}
   \end{center}
  \end{minipage}
  \hspace{45truept}
  \begin{minipage}{0.5\hsize}
   \begin{center}
    \includegraphics[clip,width=9.0cm]{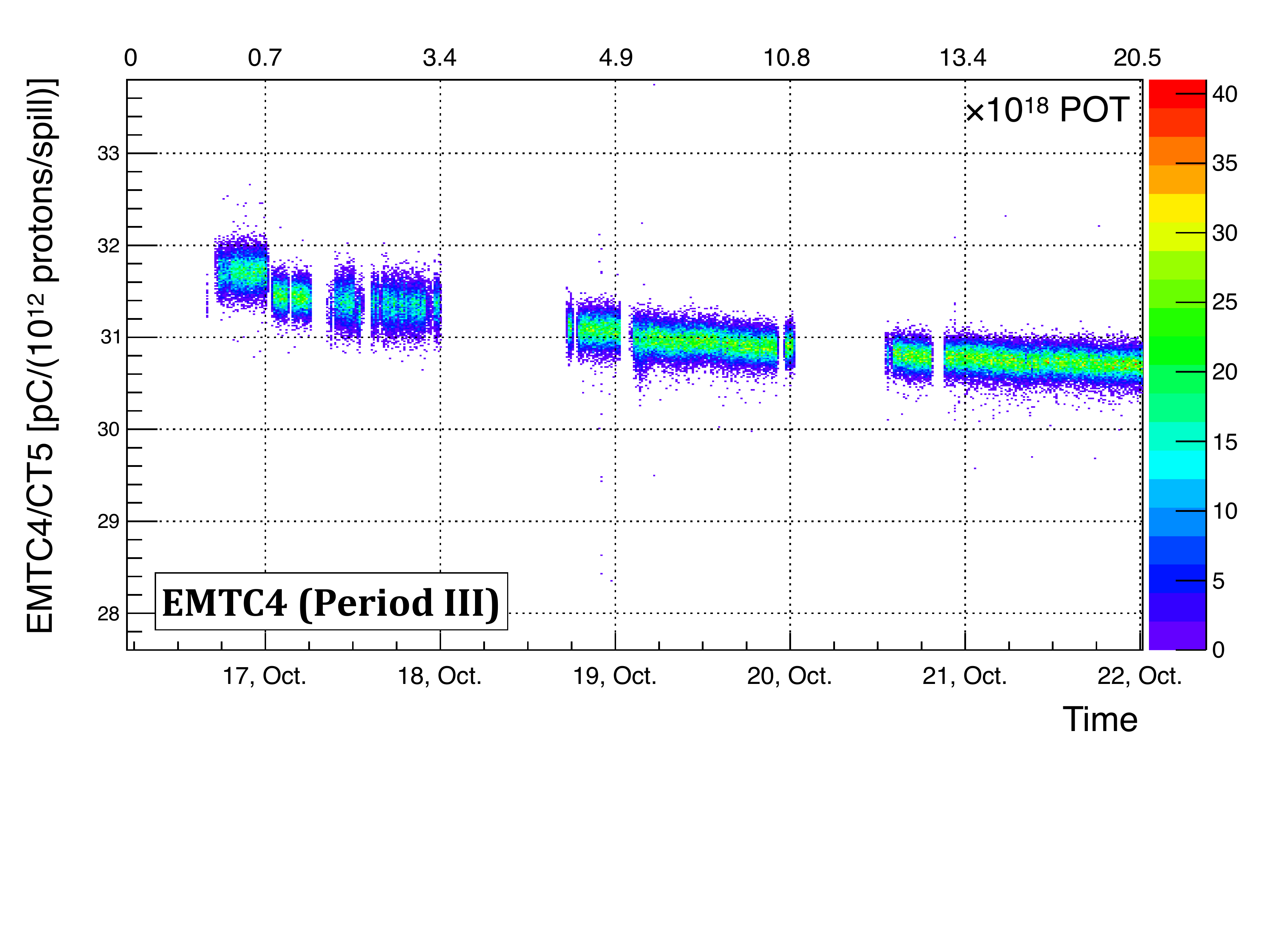}
   \end{center}
  \end{minipage}
  \\
  \noindent
  \vspace{20truept}
  \hspace{-50truept}
  \begin{minipage}{0.5\hsize}
   \vspace{-10truept}
   \begin{center}
    \includegraphics[clip,width=9.0cm]{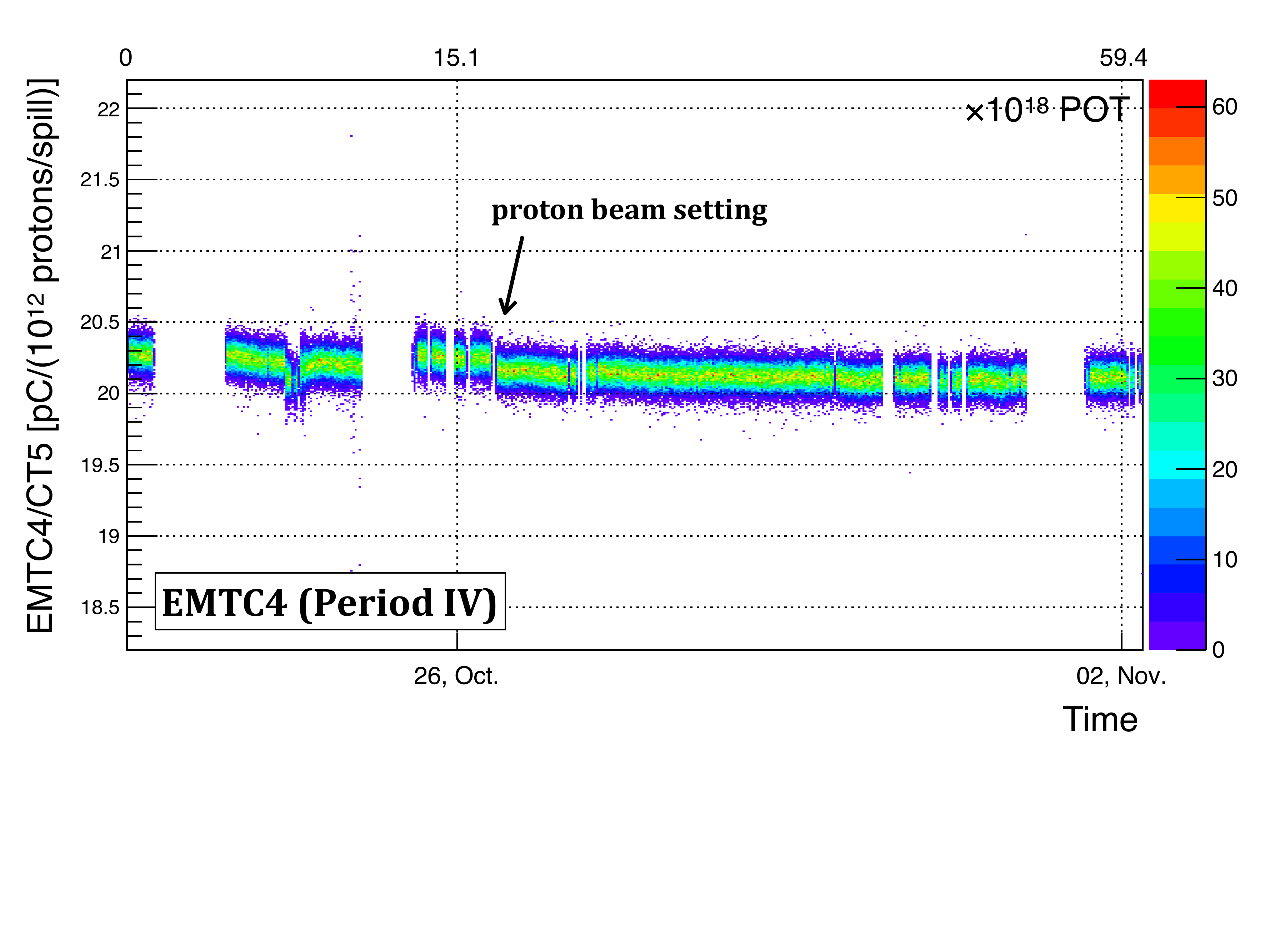}
   \end{center}
  \end{minipage}
  \hspace{45truept}
  \begin{minipage}{0.5\hsize}
   \vspace{-10truept}
   \begin{center}
    \includegraphics[clip,width=9.0cm]{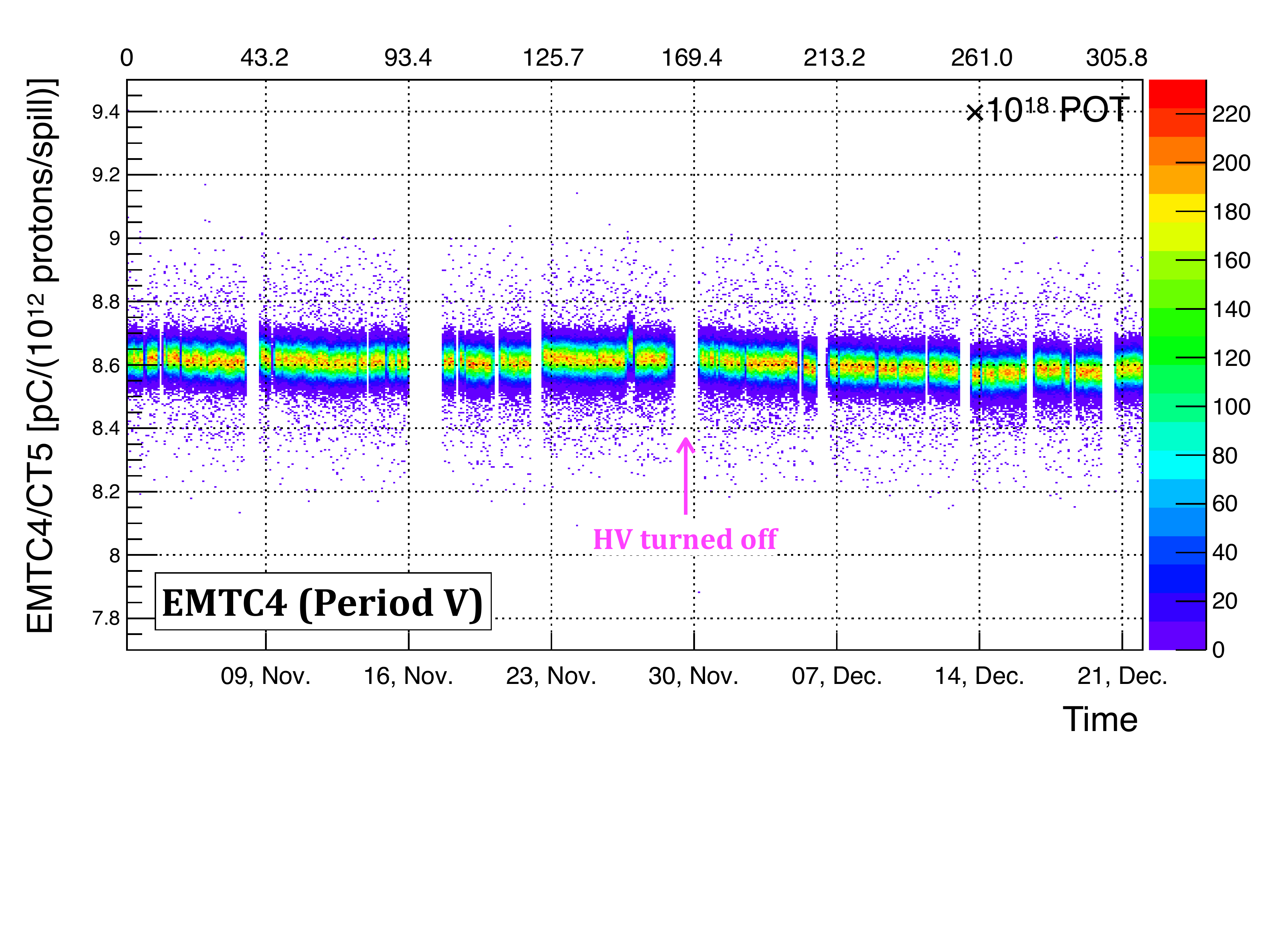}
   \end{center}
  \end{minipage}
  \vspace{-20truept}
  \caption{Signal yield of EMTC4 as a function of time.
           Other than the short periods just after the HV is turned on, C4 shows stable performance.}
  \label{fig:emtc4stability}
 \end{figure}

Figures~\ref{fig:sistability}, \ref{fig:icstability}, \ref{fig:emtc3stability}, and \ref{fig:emtc4stability} 
show the yield versus time for the Si center channel, IC center channel, EMTC3, and EMTC4, respectively.  
The effect of the horn current on the muon flux is 
corrected as described in Section~5.4. 
There are two jumps observed in the EMTC3 signal, the cause 
of which is still unknown; however,  
these appear to be synchronized with unrelated IC calibration work. 
Both EMTC3 and C4 show a drift in the yield following the initial application of 
HV,
in Periods I and III for C3 and Periods II and III for C4. 
This drift is assumed to be due to the stabilization of the dynode materials,  
which are generally alkali metals and antimony (Sb).  
Usually, PMTs require ``warming up" by irradiation with light for initial stabilization \cite{hamamatsu}; 
however, EMTs cannot be warmed up because they do not have a photocathode.
The required warm-up output charge is several ${\rm \mu A}$ for 
several minutes, which is equivalent to several mC. 
Table~\ref{tab:driftcharge} shows the integrated charge output by the time the EMT signal yield became stable 
during our tests.
The values are reasonable compared to the expected charge for PMT warm-up. 
For later runs, both EMTC3 and C4 
stabilized after fewer incident POT.
This is thought to be due to stabilization of the dynode materials at some level by previous irradiation. 
After the initial drift period, the yield is basically stable within $\pm$1\%
excluding some periods where the yield fluctuated 
due to manually changing the EMT HV or changes in the proton beam conditions.
This demonstrates that EMTs fulfill the requirement of $<$3\% signal fluctuation.  

 \clearpage

 \begin{table}[htbp]
  \begin{center}
    \caption{Total integrated charge output 
    prior to EMT signal yield stabilization.
           The calculations for Period III assume that the charge
           per POT is increased by a factor of 1.1,
	   since the applied HV is higher ($-$505~V).
	   The recommended value for stabilization is several~mC.}
  \label{tab:driftcharge}
  \vspace{2truept}
   \begin{tabular}{c c c c} \hline \hline
    EMT number & Period & POT amount [$\times 10^{18}$] & Integrated charge [mC] \\ \hline
    C3         & I      & $\sim$70 & $\sim$2.0 \\ \cline{2-4} 
               & III    & $\sim$13 & $\sim$0.4 \\ \hline
    C4         & II     & $\sim$50 & $\sim$1.3 \\ \cline{2-4}
               & III    & $\sim$13 & $\sim$0.4 \\ \hline \hline
   \end{tabular}
  \end{center}
 \end{table}


%% file: conclusion.tex
\section{Conclusion}
\label{sec:conclusion}

We are developing an electron-multiplier tube as a possible new muon monitor sensor
for the T2K experiment. 
Two prototype detectors were prepared by modifying the resistances and capacitances 
of the original divider circuits of Hamamatsu-made EMTs, 
and these sensors were installed in the muon monitor location. 
Various properties of EMTs were measured using the T2K muon beam.
The EMTs' time response was found to be faster than that of the T2K Si or IC sensors, 
which is of high importance for bunch-by-bunch monitoring. 
The EMT spill-by-spill intensity resolution at the current T2K operation conditions 
is better than the 1\% requirement. 
Linearity tests show that the EMT keeps its linearity within $\pm$1\% up to 460~kW 
at $+$250~kA horn current (corresponding to a 
muon flux of $3.3 \times 10^{6}$~${\rm /cm^2}$ per 80~ns beam bunch).
The linearity can be further improved by improving the HV divider circuit configuration.
The EMT signal yield gradually decreased during initial EMT operation, and then stabilized. 
One possible reason for this is the conditioning of the EMT dynode materials. 
This work is the first demonstration of the usage of EMTs for muon beam monitoring, 
and indicates a strong possibility of practical use in the future.    
Furthermore, this study suggests that EMTs could 
have additional applications as
general beam detectors for charged particles, 
with the merit that the gain is easily tunable for compatibility with 
various beam intensities.
To test the EMT performance 
at T2K future high beam intensities, 
an electron beam test of the EMTs is 
currently under consideration.
%


%% file: mumonemt_arxiv.bbl
\begin{thebibliography}{10}
\expandafter\ifx\csname url\endcsname\relax
  \def\url#1{\texttt{#1}}\fi
\expandafter\ifx\csname urlprefix\endcsname\relax\def\urlprefix{URL }\fi
\expandafter\ifx\csname href\endcsname\relax
  \def\href#1#2{#2} \def\path#1{#1}\fi

\bibitem{Pattison:1969sr}
J.~B.~M. Pattison, C.~A. Ramm, W.~A. Venus (Eds.),
  \href{http://www.slac.stanford.edu/spires/find/books/www?cl=QCD175:I5:1969}{{Neutrino
  Meeting, CERN, Geneva, Switzerland, 13-14 Jan 1969: proceedings}}, 1969.
\newline\urlprefix\url{http://www.slac.stanford.edu/spires/find/books/www?cl=QCD175:I5:1969}

\bibitem{Kopp:2006nq}
S.~Kopp, et~al., {Secondary beam monitors for the NuMI facility at FNAL}, Nucl.
  Instrum. Meth. A568 (2006) 503--519.
\newblock \href {http://arxiv.org/abs/physics/0607229}
  {\path{arXiv:physics/0607229}}, \href
  {http://dx.doi.org/10.1016/j.nima.2006.07.062}
  {\path{doi:10.1016/j.nima.2006.07.062}}.

\bibitem{Abe:2011ks}
K.~Abe, et~al., {The T2K Experiment}, Nucl. Instrum. Meth. A659 (2011)
  106--135.
\newblock \href {http://arxiv.org/abs/1106.1238} {\path{arXiv:1106.1238}},
  \href {http://dx.doi.org/10.1016/j.nima.2011.06.067}
  {\path{doi:10.1016/j.nima.2011.06.067}}.

\bibitem{Fukuda:2002uc}
Y.~Fukuda, et~al., {The Super-Kamiokande detector}, Nucl. Instrum. Meth. A501
  (2003) 418--462.
\newblock \href {http://dx.doi.org/10.1016/S0168-9002(03)00425-X}
  {\path{doi:10.1016/S0168-9002(03)00425-X}}.

\bibitem{Abe:2017cpv}
K.~Abe, et~al., {Combined Analysis of Neutrino and Antineutrino Oscillations at
  T2K}, Phys. Rev. Lett. 118 (2017) 151801.

\bibitem{Abe:2017vif}
K.~Abe, et~al., {Measurement of neutrino and antineutrino oscillations by the
  T2K experiment including a new additional sample of $\nu_e$ interactions at
  the far detector}, Phys. Rev. D 96 (2017) 092006.
\newblock \href {http://arxiv.org/abs/1707.01048} {\path{arXiv:1707.01048}}.

\bibitem{Ichikawa:2012vif}
A.~K. Ichikawa, {Design concept of the magnetic horn system for the T2K
  neutrino beam}, Nucl. Instrum. Meth. A690 (2012) 27--33.
\newblock \href {http://dx.doi.org/10.1016/j.nima.2012.06.045}
  {\path{doi:10.1016/j.nima.2012.06.045}}.

\bibitem{Sekiguchi:2015vif}
T.~Sekiguchi, et~al., {Development and operational experience of magnetic horn
  system for T2K experiment}, Nucl. Instrum. Meth. A789 (2015) 57--80.
\newblock \href {http://dx.doi.org/10.1016/j.nima.2015.04.008}
  {\path{doi:10.1016/j.nima.2015.04.008}}.

\bibitem{Abe:2011xv}
K.~Abe, et~al., {Measurements of the T2K neutrino beam properties using the
  INGRID on-axis near detector}, Nucl. Instrum. Meth. A694 (2012) 211--223.
\newblock \href {http://arxiv.org/abs/1111.3119} {\path{arXiv:1111.3119}},
  \href {http://dx.doi.org/10.1016/j.nima.2012.03.023}
  {\path{doi:10.1016/j.nima.2012.03.023}}.

\bibitem{Bhadra:2013vif}
S.~Bhadra, et~al., {Optical transition radiation monitor for the T2K
  experiment}, Nucl. Instrum. Meth. A703 (2013) 45--58.
\newblock \href {http://dx.doi.org/10.1016/j.nima.2012.11.044}
  {\path{doi:10.1016/j.nima.2012.11.044}}.

\bibitem{Matsuoka:2010jf}
K.~Matsuoka, et~al., {Design and Performance of the Muon Monitor for the T2K
  Neutrino Oscillation Experiment}, Nucl. Instrum. Meth. A624 (2010) 591--600.
\newblock \href {http://arxiv.org/abs/1008.4077} {\path{arXiv:1008.4077}},
  \href {http://dx.doi.org/10.1016/j.nima.2010.09.074}
  {\path{doi:10.1016/j.nima.2010.09.074}}.

\bibitem{Higuchi:2003vif}
T.~Higuchi, et~al., {Development of a PCI Based Data Acquisition Platform for
  High Intensity Accelerator Experiments}, Computing in High Energy and Nuclear
  Physics\href {http://arxiv.org/abs/0305088} {\path{arXiv:0305088}}.

\bibitem{Suzuki:2014jyd}
K.~Suzuki, et~al., {Measurement of the muon beam direction and muon flux for
  the T2K neutrino experiment}, Prog. Theor. Exp. Phys. 2015~(5) (2015) 053C01.
\newblock \href {http://arxiv.org/abs/1412.0194} {\path{arXiv:1412.0194}},
  \href {http://dx.doi.org/10.1093/ptep/ptv054}
  {\path{doi:10.1093/ptep/ptv054}}.

\bibitem{Abe:2016vif}
K.~Abe, et~al., {Sensitivity of the T2K accelerator-based neutrino experiment
  with an Extended run to $20 \times 10^{21}$ POT}\href
  {http://arxiv.org/abs/1607.08004} {\path{arXiv:1607.08004}}.

\bibitem{Hamamatsu:ver3a}
H.~P.~K. K. (Ed.),
  \href{https://www.hamamatsu.com/resources/pdf/etd/PMT_handbook_v3aE.pdf}{{Photomultiplier
  Tubes Basic and Applications Third Eddition (Eddition 3a)}}, 2007.
\newline\urlprefix\url{https://www.hamamatsu.com/resources/pdf/etd/PMT_handbook_v3aE.pdf}

\bibitem{Winn:2012vif}
R.~D. Winn, Y.~Onel, {Secondary Emission Calorimeter Sensor Development},
  Journal of Physics: Conference Series 404 (2012) 012021.
\newblock \href {http://dx.doi.org/10.1088/1742-6596/404/1/012021}
  {\path{doi:10.1088/1742-6596/404/1/012021}}.

\bibitem{Abe:2013vif}
K.~Abe, et~al., {The T2K Neutrino Flux Prediction}, Phys. Rev. D 87 (2013)
  012001.
\newblock \href {http://arxiv.org/abs/1211.0469} {\path{arXiv:1211.0469}}.

\end{thebibliography}


\begin{thebibliography}{100}
%
\bibitem{pattison}
  J. B. M. Pattison, C. A. Ramm, and W. A. Venus, Neutrino Meeting, 
  CERN, Geneva, Switzerland, 13-14 Jan., 1969, proceedings (1969).
%
\bibitem{kopp}
  S. Kopp et al., 
  Nucl. Instr. Meth. Phys. Res. A 568, 503-519 (2006).
%
\bibitem{t2k}
  K. Abe et al., 
  Nucl. Instr. Meth. Phys. Res. A 659, 106-135 (2011).
%
\bibitem{superk}
  Y. Fukuda et al., 
  Nucl. Instr. Meth. Phys. Res. A 501, 418-462 (2003).
%
\bibitem{oa2016short}
  K. Abe et al., 
  Phys. Rev. Lett., 118, 151801 (2017).
%
\bibitem{oa2016long}
  K. Abe et al., 
  Phys. Rev. D, 96, 092006 (2017).
%
\bibitem{ichikawa}
  A. K. Ichikawa, 
  Nucl. Instr. Meth. Phys. Res. A 690, 27-33 (2012).
%
\bibitem{sekiguchi}
  T. Sekiguchi et al., 
  Nucl. Instr. Meth. Phys. Res. A 789, 57-80 (2015).
%
\bibitem{ingrid}
  K. Abe et al., 
  Nucl. Instr. Meth. Phys. Res. A 694, 211-223 (2012).
%
\bibitem{bhadra}
  S. Bhadra et al., 
  Nucl. Instr. Meth. Phys. Res. A 703, 45-58 (2013).
%
\bibitem{matsuoka}
  K. Matsuoka et al., 
  Nucl. Instr. Meth. Phys. Res. A 624, 591-600 (2010).
%
\bibitem{higuchi}
  T. Higuchi et al., 
  Computing in High Energy and Nuclear Physics (2003).
%
\bibitem{suzuki}
  K. Suzuki et al., 
  Prog. Theor. Exp. Phys., 053C01 (2015).
%
\bibitem{t2k2}
  K. Abe et al., 
  arXiv:1607.08004 (2016).
%
%
%
\bibitem{hamamatsu}
  Hamamatsu Photonics K. K., 
  Photomultiplier Tubes Basic and Applications Third Eddition (2007).
%
\bibitem{winn}
  R. D. Winn and Y. Onel, 
  Journal of Physics: Conference Series, 404, 012021 (2012).
%
\bibitem{flux}
  K. Abe et al., 
  Phys. Rev. D, 87, 012001 (2013).
%
\end{thebibliography}
